%% file: thesisfinal.tex
\documentclass[12pt]{report}
\usepackage[german]{babel}
\usepackage {longtable, amsmath, amsfonts, feynmf, ams,
graphicx, geometry}
\usepackage [latin1]{inputenc}
\def\beq{\begin{equation}}
\def\eeq{\end{equation}}
\def\bea{\begin{eqnarray}}
\def\eea{\end{eqnarray}}
\def\bq{\begin{quote}}
\def\eq{\end{quote}}

\title{Grenzen der supersymmetrischen $R$-Paritätsverletzung aus
Mesonenzerfällen}
\author{Margarete Herz}
\begin {document}
\hyphenation{par-ticu-lar}
\input{instituts_deckblatt.tex}
\pagestyle{plain}
\pagenumbering{Roman}
\sloppy
\tableofcontents
\listoffigures
\listoftables
\newpage
\pagestyle{headings}
\chapter*{Überblick}
\addcontentsline{toc}{chapter}{Überblick} \thispagestyle{plain}
\pagenumbering{arabic} Das Ziel dieser Arbeit ist eine Übersicht
über die wichtigsten $R$-paritätsverletzenden Prozesse in
Mesonenzerfällen. Dies geschieht in einem allgemeineren Rahmen: Es
werden Mesonenzerfälle durch die sog. Leptoquarks\footnote{Da LQs
bisher noch nicht an Hochenergiebeschleunigern über Kollisionen
nachgewiesen worden sind, wird damit der Möglichkeit nachgegangen,
zumindest aus indirekten Quellen (wie Mesonenzerfälle,
Leptonenzerfälle u. A.) Bedingungen an ihre Existenz zu stellen.}
(LQ) betrachtet und Schranken an LQ-Kopplungskonstantenprodukte in
der Form
$$\lambda_{LQ}\lambda_{LQ}^*<Zahl\left(\frac{m_{LQ}}{100GeV}\right)^2\; ,$$ berechnet.
Aus diesen Schranken können dann direkt die entsprechenden
Schranken an SUSY
(\textbf{SU}per\textbf{SY}mmetrie)-Kopplungskonstantenprodukte
extrahiert werden.\\Kapitel 1 führt in die Themenstellung ein.
Ausgehend von der in Kapitel 2 berechneten leptonischen
Mesonenzerfallsrate $\Gamma_M$ werden sowohl Zerfälle durch LQs,
als auch Zerfälle durch SUSY-Teilchen, betrachtet. Soweit möglich
werden auch semileptonische Zerfälle miteinbezogen, die dazu
notwendigen Formeln befinden sich in den jeweiligen Kapiteln.\\
\\Ausgangspunkt meiner Überlegungen ist die LQ-Lagrangefunktion
(Buchmüller-Rückl-Wyler-Modell), Kapitel 2.1 . In Kapitel 2.2 und
2.3 werden die verwendeten Konventionen sowie benutzte Näherungen
und Nebenrechnungen kurz zusammengefasst. Für Zerfälle der Art
$M\rightarrow \ell^{i}\overline{\ell}^{m}$ ($M$ ist ein Meson, die
$\ell^i$ sind Leptonen) wird in Kapitel 2.4 die Zerfallsrate
$\Gamma_M$ für nicht verschwindende LQ-Kopplungen berechnet. In
den Kapiteln 3, 4, 5 und 6 werden $\pi$-, $K$-, $D$- und $B$-Meson
im Hinblick auf mögliche LQ-Wechselwirkungen untersucht. Ergänzend
dazu wird in Kapitel 7 auf die Systeme der neutralen Mesonen
$K^{0}-\overline{K^0}$, $D^{0}-\overline{D^0}$ und
$B^{0}-\overline{B^0}$ eingegangen. In den jeweiligen Kapiteln
werden neben Schranken an LQ-Kopplungskonstantenprodukte auch die
Schranken an die entsprechenden SUSY-Kopplungskonstantenprodukte
aufgeführt. Abschließend werden die berechneten Schranken nochmals
auf ihre Aussagekraft hin überprüft (Kapitel 8).  \\ \\
Im Anhang befinden sich Tabellen mit den verwendeten Bezeichnungen
(Tabelle 1), benutzten Werten (Tabelle 2), in die Berechnungen
eingehenden Pauli-,Dirac- und \nopagebreak[4]$\gamma$-Matrizen
\nopagebreak[4](Tabelle 3)\nopagebreak[4],\nopagebreak[4]
sämtlichen \nopagebreak[4]LQ-4-Fermionenvertizes
\nopagebreak[4](Tabelle 4 und 5) sowie \nopagebreak[4]eine
\nopagebreak[4]Zusammenfassung \nopagebreak[4]der berechneten
Kopplungskonstantenprodukte (Tabelle 6 und 7).\pagebreak
\chapter{Einleitung}
Im Standardmodell (kurz SM) der Teilchenphysik können fast alle
bisherigen experimentellen Resultate \cite {pdg} (Ausnahmen: siehe
Seite 3) mit sehr hoher Genauigkeit theoretisch berechnet werden.
Dies gewährleisten die Quantenchromodynamik (QCD), die
Quantenelektrodynamik (QED) und die Theorie der schwachen
Wechselwirkung, die Bestandteile des SM sind. Die Schwächste der
vier fundamentalen Wechselwirkungen, die Gravitation, wird durch
die allgemeine Relativitätstheorie beschrieben. Trotzdem bleiben
einige theoretische Fragen und Probleme unbeantwortet bzw.
ungelöst; z.B.:
\begin{itemize}
\item Warum gibt es drei Generationen von Fermionen? \item Wie
kann man sich die experimentell bestimmten Massen der
Elementarteilchen erklären? Im SM sind dies freie Parameter.\item
Das SM selbst beinhaltet keine Quantenfeldtheorie für die
Gravitation. Bei heute in Beschleunigern erreichbaren Energien
sind die durch die Gravitation verursachten Effekte aufgrund ihrer
relativen Schwäche im Vergleich mit den drei anderen fundamentalen
Wechselwirkungen komplett vernachlässigbar. Trotzdem bleibt die
Frage nach einer Quantentheorie der Gravitation ein interessantes
Problem.\item \textbf{Das Hierarchieproblem}: Die im Experiment
messbare Higgsmasse setzt sich aus der "`reinen"' Higgsmasse und
einem $\delta {M_H^2}$-Term zusammen, der aus Schleifenkorrekturen
resultiert: \beq M_H^2=M_{Hbare}^{2}+\delta {M_H^2}_f \; .\eeq Die
Ein-Schleifen-Korrekturen der (Higgsmasse)$^2$ zeigen eine
quadratische Abhängigkeit vom Cut-off $\Lambda$\footnote {Aus
\cite {rich}, für eine detailliertere Beschreibung siehe \cite
{martin}. Hier wurde nur die Korrektur für Fermionenschleifen
(Index f) betrachtet, $m_f$ ist die Fermionenmasse.}:
\begin{equation} \delta {M_H^2}_f = \frac{|g_f|^2}{16\pi^2}\left[
-2\Lambda^2
            +6m_f^2\ln\left(\Lambda/m_f\right) \right]\; .
\end{equation} Dabei wurden beim Übergang $\Lambda\rightarrow\infty$ endlich bleibende Terme
vernachlässigt. $g_f$ ist die Kopplung des Fermions an das Higgsfeld.\\
Diese Abhängigkeit stellt noch kein Problem dar, da die Theorie
renormierbar ist. Man nimmt allerdings an, dass das SM nur eine
bei niedrigerer Energie als der GUT\footnote{\textbf{G}rand
\textbf{U}nified \textbf{T}heorie.}- oder Planck-Skala effektive
Theorie ist. Ab dieser Skala werden neue physikalische Effekte
erwartet, der Cut-off sollte also in diesem Bereich liegen.
Dadurch erwartet man (aufgrund des Cut-offs), dass die Higgsmasse
bei $10^{14}-10^{17}GeV$ liegt. Dies ist jedoch weit entfernt von
einer oberen Grenze der Higgsmasse bei 196 GeV, wie sie sich aus
elektroschwachen Strahlungskorrekturen ergibt \cite {LEP}\footnote
{Eine kurze Übersicht der Resultate der LEP-Kollaboration befindet
sich auf deren Webseite \cite {LEP}.}. Dadurch muss, um eine dem
SM entsprechende Higgsmasse unter 196 GeV zu erhalten, eine enorme
Feinabstimmung zwischen $M_{Hbare}^{2}$ und $\delta {M_H^2}_f$
existieren, was eine entsprechende Feinabstimmung der Parameter in
der Lagrangefunktion bedeutet. Dies entzieht sich jedoch dem
theoretischen Verständnis.
\end{itemize}
Eine Ausnahme zur theoretischen Vorhersagekraft des
Standardmodells stellen die Experimente zu solaren und
atmosphärischen Neutrinooszillationen dar\footnote{Für einen
Überblick: siehe \cite {sno1}.}. Insbesondere die Resultate der
SNO-Kollaboration \cite {sno1} können innerhalb des
Standardmodells nicht erklärt werden. Es bedarf damit auch aus
experimenteller Sicht der Suche nach einer umfassenderen
theoretischen Beschreibung, die das SM als Spezialfall
beinhaltet.\\ \\Als eine mögliche Erweiterung des SM hat die
Supersymmetrie (SUSY), die bisher noch nicht experimentell
bestätigt wurde, einige interessante Vorteile:
\begin{itemize}
\item Nach dem Coleman-Mandula-Theorem \cite {cole} ist eine
Erweiterung der Poincaré-Gruppe durch neue bosonische Generatoren
verboten, da diese zu einer trivialen S-Matrix führen. Die Gruppe
kann allerdings durch Generatoren, die sich wie Fermionen
transformieren, erweitert werden. Wie in \cite {haag} gezeigt
wurde, ist die Supersymmetrie die einzige mögliche derartige
Erweiterung der Poincaré-Gruppe, die nicht zu einer trivialen
S-Matrix führt. \item Die SUSY kann das Hierarchieproblem lösen:
Man erhält eine weitere Ein-Schleifen-Korrektur zum Quadrat der
Higgsmasse durch ein neues skalares Feld (S), das mit dem
Higgsboson wechselwirkt \cite {rich} :\begin{equation} \delta
{M_H^2}_S = \frac{\lambda_S}{16\pi^{2}}
           \left(\Lambda^2-2M_S^2\ln\left(\Lambda/M_S\right)
  \right)\!.
\end{equation}Es wurden wieder alle endlichen Terme im
Grenzübergang $\Lambda\rightarrow\infty$ vernachlässigt.
$\lambda_S$ ist die Kopplung des skalaren Feldes an das Higgsfeld
($\mathcal{L}=-\lambda_S|H|^2|S|^2$, siehe \cite{martin}), $M_S$
ist die Masse des skalaren Feldes.\\
Dieser Bosonenschleifen- und der Fermionenschleifenbeitrag löschen
sich gegenseitig aus, falls $\lambda_S=|g_f|^2$ (in der SUSY hat
man für jedes Dirac-Fermion zwei komplexe skalare Felder, sodass
Gl. (1.3) mit zwei zu multiplizieren ist). Dies gilt, solange die
SUSY ungebrochen ist.
\end{itemize}
Auch wenn bislang keine experimentellen Resultate eindeutig darauf
hindeuten, dass die Supersymmetrie in der Natur realisiert ist,
lohnen sich daher weitere Studien in dieser Richtung.\\Zunächst
sollen die Grundzüge der Supersymmetrie kurz erläutert werden.
\section{Die Supersymmetrie}
Die Supersymmetrie (SUSY) ordnet jedem Standardmodellteilchen
einen supersymmetrischen Partner zu, dessen Spin ganzzahlig
(halbzahlig) ist, wenn der Spin des Standardmodellteilchens
halbzahlig (ganzzahlig) ist.\\ Eine SUSY-Transformation $Q$ ($Q$
ist ein fermionischer Operator) macht aus einem fermionischen
Zustand einen bosonischen und umgekehrt: \beq Q |{\rm
Boson}\rangle = |{\rm Fermion }\rangle; \qquad\qquad \bar{Q} |{\rm
Fermion}\rangle = |{\rm Boson }\rangle . \eeq Die SUSY-Algebra ist
\cite {bailin}:
\begin{subequations}
\begin{eqnarray}
[Q_\alpha,P_\mu] &=& [\bar{Q}_{\dot{\alpha}},P_\mu] = 0, \\
 \{ Q_{\alpha},Q_{\beta} \} &=& \{ \bar{Q}_{\dot{\alpha}},\bar{Q}_{\dot{\beta}}\}
                = 0, \\
\{ Q_{\alpha},\bar{Q}_{\dot{\beta}}\} &
=&2{\sigma^\mu}_{\alpha\dot{\beta}}P_\mu.
\end{eqnarray}
\end{subequations}
Die Generatoren $Q$ und $\overline {Q}$ sind zwei-komponentige
links- bzw. rechtshändige Weyl-Spinoren und $\sigma_\mu =
(\mathbf{1} , \sigma_i)$ ($\sigma_i$ sind die Pauli-Spinmatrizen).
Der Index $\alpha$ bzw. $\beta$ ($\dot{\alpha}$ bzw.
$\dot{\beta}$) gibt die Komponente des linkshändigen
(rechtshändigen) Spinors an.\\Da der $(Masse)^2$-Operator ($P_\mu
P^\mu$-Operator) mit $Q$ und $\overline {Q}$ kommutiert, haben ein
Standardmodellteilchen und sein SUSY-Partner dieselbe Masse. Aus
Experimenten ist klar, dass dies nicht der Fall ist (SUSY-Teilchen
wurden bisher noch nicht entdeckt). Daher muss die SUSY eine
gebrochene Symmetrie sein, falls sie in der Natur realisiert
ist.\\Damit die SUSY trotzdem das Hierarchieproblem löst, muss man
fordern, dass in der supersymmetrischen Lagrangefunktion nur
"`softe"' SUSY-brechende Terme auftreten \cite {martin}. Die
Lagrangefunktion des MSSM ist: \beq \mathcal {L} = \mathcal
{L}_{\rm SUSY} + \mathcal {L}_{\rm soft}\; . \eeq  Der Term
$\mathcal {L}_{\rm SUSY}$ verletzt die SUSY nicht. Der Term
$\mathcal {L}_{\rm soft}$ ist im allgemeinen Fall \cite {martin} :
\beq \mathcal {L}_{\rm soft}\! = \! -\frac{1}{2}(M_\lambda\,
\lambda^a\lambda^a + c.c. ) - (m^2)_j^i \phi^{j*} \phi_i
-(\frac{1}{2} b^{ij} \phi_i\phi_j + \frac{1}{6}a^{ijk}
\phi_i\phi_j\phi_k  + c.c. )+h.c. \; .\qquad\>\>\>\>\>{}
\label{lagrsoft}\eeq Die $\lambda^a$ sind Gauginofelder, die
$\phi_i$ sind skalare Felder. $M_\lambda$ sind Gauginomassen.
Diese Lagrangefunktion verletzt die SUSY, ist jedoch in allen
Ordnungen frei von quadratischen Divergenzen \cite {soft}.
\\ \\In der SUSY hat man ein renormierbares und unter
$ SU(3)_C\times SU(2)_L\times U(1)_Y$ eichinvariantes
Superpotential, aus dem man neue Wechselwirkungen ableiten kann,
die über die Möglichkeiten im SM hinausgehen. Im Folgenden soll
zunächst das Superpotential dargestellt werden (Kapitel 1.1.1), um
dann die Einführung der \textbf{$R$-Parität}, $R_P$, einer neuen
Symmetrie, im Hinblick auf die dadurch im Superpotential
verbotenen Terme zu rechtfertigen (Kapitel 1.1.3).
\subsection{Das Superpotential}
\begin{table}[h]
\begin{center}
\begin{tabular}{|cc|cc|c|c|c|}
\hline \multicolumn{2}{|c|}{Superfelder} &
 Bosonische & Fermionische & $SU(3)_C$ & $SU(2)_L$ & $Y$ \\
Feld & Typ & Felder & Felder & & & \\
\hline \multicolumn{2}{|c|}{Eich-Multiplette} & & & & &
\\\cline{1-2}
 $G^a$ & Vector       & Gluonen & Gluinos & Oktett & Singlett &
                         $\phantom{-}0$ \\
 $W^a$ & Vector       & $W$ & Winos & Singlett & Triplett    &
                         $\phantom{-}0$ \\
 $B^a$ & Vector       & $B$ & Bino & Singlett & Singlett     &
                         $\phantom{-}0$ \\
 \hline
\multicolumn{2}{|c|}{Materie Multiplette} & & & & & \\ \cline{1-2}
 $L_i$ & linkshändig  & $(\widetilde{\nu}_L,\widetilde{l}^-_L)$ & $(\nu_L,\ell_L)$ &
                 Singlett & Doublett & $-1/2$\\
 $E_i$ & rechtshändig & $\widetilde{l}^-_R$        & $\ell_R$           &
                 Singlett & Singlett & $-1$ \\
 $Q_i$ & linkshändig  & $(\widetilde{u}_L,\widetilde{d}_L)$ & $(u_L,d_L)$    &
                 Triplett & Doublett & $\phantom{-}1/6$\\
 $U_i$ & rechtshändig & $\widetilde{u}_R$ & $u_R$ &
                 Triplett & Singlett & $\phantom{-}2/3$ \\
 $D_i$ & rechtshändig & $\widetilde{d}_R$ & $d_R$ &
                 Triplett & Singlett & $-1/3$  \\
\hline \multicolumn{2}{|c|}{Higgs Multiplette} & & & & &
\\\cline{1-2}
 $H_1$ & linkshändig & $(H^1_1,H^2_1)$ & $(\widetilde{H}^0_1,\widetilde{H}^-_1)_L$ &
                     Singlett & Doublett & $-1/2$\\
 $H_2$ & linkshändig & $(H^1_2,H^2_2)$ & $(\widetilde{H}^+_2,\widetilde{H}^0_2)_L$ &
                     Singlett & Doublett & $\phantom{-}1/2$\\
\hline
\end{tabular}
\end{center}
\caption [Superfelder im MSSM]{\textit{Superfelder im MSSM} \cite
{rich}} \label{tab:superfield}
\end{table}
Im MSSM (\textbf{M}inimal \textbf{S}upersymmetric
\textbf{S}tandard \textbf{M}odel) ist die $R$-Parität $R_P$
(Definition: siehe Kapitel 1.1.3) erhalten. Das allgemeinste,
renormierbare und unter $ SU(3)\times SU(2)\times U(1)$
eichinvariante Superpotential ist \cite {wein} :
\begin{equation}
  \mathcal {W} = \mathcal{W}_{MSSM} + \mathcal {W}_{\not R_p}\; ,
\end{equation}
wobei\footnote {Die benutzten Superfelder sind in Tabelle
\ref{tab:superfield} aufgeführt, zu den Konventionen siehe nächste
Seite.} \beq
 \mathcal {W}_{MSSM}=h^E_{ij}\varepsilon_{ab}L^a_iH_1^b\overline{E}_j
            +h^D_{ij}\varepsilon_{ab}\delta^{c_1c_2}Q^a_{ic_1}H_1^b\overline{D}_{jc_2}
            +h^U_{ij}\varepsilon_{ab}\delta^{c_1c_2}Q^a_{ic_1}H_2^b\overline{U}_{jc_2}
            +\mu\varepsilon_{ab} H_1^aH_2^b
\label{eqn:MSSMsuper}\eeq und \beq \mathcal {W}_{\not R_p}=
\frac{1}{2}\lambda_{ijk}\varepsilon^{ab}L_{a}^{i}L_{b}^{j}\overline{E}^{k}
+
\lambda_{ijk}'\varepsilon^{ab}\delta^{c_1c_2}L_{a}^{i}Q_{bc_1}^{j}\overline{D}^{k}_{c_2}
+
\frac{1}{2}\lambda_{ijk}''\varepsilon^{c_1c_2c_3}\overline{U}_{c_1}^{i}
\overline{D}_{c_2}^{j}\overline{D}_{c_3}^{k} +
 \kappa_i\varepsilon^{ab}L_a^{i}H_b^2\; . \label{eqn:Rsuper1}
\eeq Es wurden folgende Konventionen verwendet:\begin{tabbing}- \=
$c_{1},c_{2}$ und $c_{3}$ sind $SU(3)_{c}$ Indizes.\\- \= $i$,
$j$, $k$ sind Generationenindizes.\\- \= $a$, $b$ sind
$SU(2)_L$-Indizes.\\- \= $\mu$ und $\kappa_i$ sind
Massenparameter.\\- \= $\overline{E}$, $\overline{D}$ und
$\overline{U}$ sind die hermitesch konjugierten Felder zu $E$,
$D$, $U$.\\- \= $\delta$ ist das Kronecker-Symbol und
$\epsilon^{xy..}$ ein total antisymmetrischer Tensor, der die
Werte +1,\\ \> 0, -1 annehmen kann.\\- \= $\lambda$, $\lambda'$,
$\lambda''$ und $h^E_{ij}$, $h^D_{ij}$, $h^U_{ij}$, sind
Yukawa-Kopplungen.\end{tabbing}\vspace{5pt}
\subsection{Der Protonenzerfall}
\vspace{10pt}
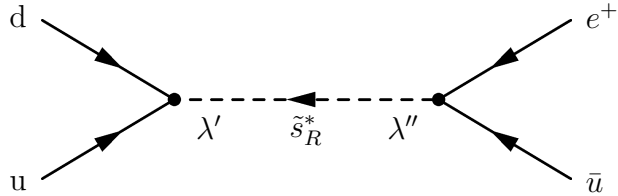
\begin{figure}[h]
\begin{center}
\begin{fmffile}{tau611}
\begin{fmfgraph*}(200,60) \fmfpen{thin}\fmfstraight \fmfleft{i2,i3} \fmfright{o2,o3}\fmf{fermion}{i2,v1}
\fmf{fermion}{i3,v1}\fmf{scalar, label=$\tilde{s}_R^*$}{v2,v1}
\fmf{fermion}{o3,v2}\fmf{fermion}{o2,v2}\fmflabel{u}{i2}\fmflabel{d}{i3}\fmflabel{$e^+$}{o3}
\fmflabel{$\bar{u}$}{o2}\fmfv{label=$ \; \; \; \;\; \; \; \;
\lambda'$,label.angle=-90}{v1}\fmfv{label=$\lambda '' \; \; \;
\;\; \; \; \;$,label.angle=-90}{v2}\fmfdotn{v}{2}
\end{fmfgraph*}\end{fmffile}\end{center}\caption[Protonenzerfall über $R$-Paritätsverletzung]
{\textit{Protonenzerfall $p\rightarrow \pi_0e^+$ über
$R$-Paritätsverletzung}}\label{proton}\end{figure} Über den
zweiten und dritten Term in $\mathcal{W}_{\not{R_P}}$ (Gl.
\ref{eqn:Rsuper1}) wird der in Abb. (\ref{proton}) dargestellte
Protonenzerfall möglich. Experimentell liegt die untere Grenze für
die Lebensdauer des Protons bei \cite{pdg}\beq\tau(p\rightarrow
\pi_{0}e^{+})>1,6\cdot 10^{33}y\; .\eeq Daraus kann man eine
Schranke für das Produkt $\lambda'_{11k}\lambda''_{11k}$
ableiten\footnote{Für eine ausführlichere Behandlung des
Protonenzerfalls verweise ich auf \cite{herbi} und
\cite{goiti}.}:\beq|\lambda'_{11k}\lambda''_{11k}|\newsymbol\lesssim
132E\lesssim
 10^{-28}\left(\frac{M_{\tilde{d}_{kR}}}{100GeV}\right)^2\; .\eeq Dies ist eine
sehr starke Einschränkung. Eine natürliche Erklärung ist daher,
dass mindestens eine der Kopplungskonstanten verschwindet. Dies
kann durch die Einführung einer neuen Symmetrie, der
\textbf{$R$-Parität}, gewährleistet werden.\\ \\ Es gibt
allerdings auch andere Lösungen, die den Protonenzerfall verbieten
\cite{herbi}:
\begin{itemize}\item Die Forderung der Erhaltung der Materie-Parität: \begin{eqnarray}
 (Q_i,\bar{U}_i,\bar{D}_i,L_i,\bar{E}_i) \rightarrow
   - (Q_i,\bar{U}_i,\bar{D}_i,L_i,\bar{E}_i),\ \ \ \ \ &
 (H_1,H_2) \rightarrow (H_1,H_2).
\end{eqnarray} Die $\mathcal{W}_{\not{R_P}}$-Terme im Superpotential
werden dadurch verboten, die MSSM-Terme bleiben
erlaubt. Die Materie-Parität ist der $R$-Parität äquivalent. \item
Erhaltung der Baryon-Parität $B_P=(-1)^{B+2S}$:
$\lambda''_{ijk}=0$ .\item Erhaltung der Lepton-Parität
$L_P=(-1)^{L+2S}$: $\lambda'_{ijk}=0 $ und $\lambda_{ijk}=0 $
.\end{itemize} Durch die letzten beiden Möglichkeiten wird zwar
der Protonenzerfall verboten, eine Verletzung der $R$-Parität ist
jedoch erlaubt.
\subsection{Die $R$-Parität}
\begin{figure}[h]
\begin{center}
\begin{fmffile}{tau501}
\begin{fmfgraph*}(170,70)
 \fmfpen{thin}\fmfleft{i1}
\fmfright{o1,o2}\fmf{scalar,label=LQ}{i1,v1}\fmf{fermion}{v1,o2}\fmf{fermion}{o1,v1}\fmflabel{q}{o1}\fmflabel{
$\ell$}{o2}\fmfdot{v1}
\end{fmfgraph*}\hspace{40pt}
\begin{fmfgraph*}(170,70)\fmfpen{thin}\fmfleft{i2}\fmfright{o3,o4}\fmf{scalar,label=$\tilde{q}$}{i2,v2}
\fmf{fermion}{v2,o4}\fmf{fermion}{o3,v2}\fmflabel{q}{o3}\fmflabel{ $\ell$}{o4}\fmfdot{v2}
\end{fmfgraph*}
\end{fmffile}\end {center}\caption[Leptoquark- und
Squark-Wechselwirkung]{\textit{Leptoquark- und
Squark-Wechselwirkung}}\label{lepto}\end{figure}
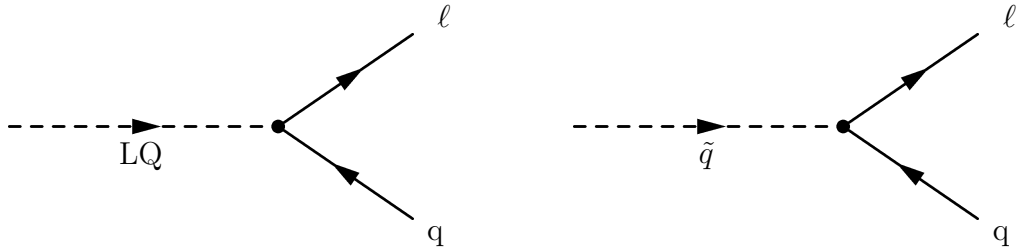

Die $R$-Parität ist eine diskrete, multiplikative Symmetrie und
definiert als
\begin{equation} R_{P}=(-1)^{3B+L+2S}\; ,\end{equation}
wobei $L$ die Leptonenzahl, $B$ die Baryonenzahl und $S$ der
Teilchenspin ist. $R_{P}$ ist +1 für Standardmodellteilchen und -1
für ihre supersymmetrischen Partner.\\Durch die Forderung der
$R$-Paritätserhaltung wird Folgendes gewährleistet:
\begin {itemize}
\item Der über $p\rightarrow \pi_{0}e^{+}$ mögliche
$R$-paritätsverletzende Protonenzerfall (siehe Kapitel 1.1.2) wird
verboten.\item Supersymmetrische Teilchen können nur paarweise
produziert werden. Dies ist eine Folge davon, dass der
Anfangszustand jedes Beschleunigerexperimentes eine gerade
$R$-Parität hat.\item Das LSP (\textbf{L}ightest
\textbf{S}upersymmetric \textbf{P}article) ist stabil.  Aus
kosmologischen Gründen \cite {LSP} ist das LSP ein neutrales
Farb-Singlett (das Neutralino $\tilde{\chi}^0$), das nur schwach
wechselwirkt. Bei Teilchenkollisionen oder durch kaskadenartigen
Zerfall instabiler supersymmetrischer Teilchen erzeugte
Neutralinos entkommen daher unbeobachtet aus dem Detektor.
Fehlende Energie und Impuls in Beschleunigerexperimenten ist somit
ein möglicher Hinweis auf das LSP. Zudem ist das LSP damit ein
Kandidat für die dunkle Materie.
\end {itemize}
Ohne die $R$-paritätsverletzenden Terme in der Lagrangefunktion
erhält man das MSSM. Die Betrachtung $R$-Paritätsverletzender
Terme gründet sich in dem Versuch, einige zurzeit ungelöste
Probleme zu bearbeiten. Dazu gehören u. A. der Ursprung der
Neutrinomassen und die kosmische Baryonenzahlverletzung. Dabei
wird versucht, die theoretischen Werte der Kopplungskonstanten
möglichst genau vorherzusagen \footnote {Eine ausführliche
Behandlung der $R$-Parität findet sich in \cite {herbi} .} .
\\ \\ Im Weiteren sollen die $R$-paritätsverletzenden Beiträge zu
Mesonenzerfällen innerhalb eines allgemeineren Rahmens untersucht
werden. Wir führen daher eine neue Klasse von Teilchen ein, die in
einigen Erweiterungen des Standardmodells, so auch der SUSY,
auftreten kann: \textbf{Die Leptoquarks} (LQs, siehe Kapitel 1.2
und 2.1). \\ \\Der zu Mesonenzerfällen beitragende LQD-Term (siehe
Gl. \ref{eqn:Rsuper1}) im $R$-paritätsverletzenden Superpotential
ist von der gleichen Form, wie der entsprechende LQ-Term (dies
wird in Kapitel 3.3 ausführlicher behandelt). Abb. \ref{lepto}
zeigt, wie ein LQ-Vertex aussieht. Der ebenfalls abgebildete
Squark-Vertex verdeutlicht den Zusammenhang zwischen LQ- und
Squark-Wechselwirkungen. LQs zerfallen ausschließlich zu Leptonen
und Quarks, während bei Squarks auch $R$-paritätserhaltende
Zerfälle der Art $\tilde{q}\rightarrow q \chi$ ($\chi$ ist ein
Neutralino oder ein Chargino) möglich sind. \vspace{20pt}
\section{Leptoquarks}
In der LQ-Lagrangefunktion (Kapitel 2.1) treten neben Termen, die
die gleiche Form haben, wie in der supersymmetrischen
Lagrangefunktion (Kapitel 3.3) weitere Wechselwirkungen auf
(pseudoskalare Wechselwirkungen, siehe Kapitel 2.1). Diese werden
in Kapitel 3.2 für das Pion behandelt. Aus den Schranken an die
Kopplungskonstantenprodukte der LQs können die entsprechenden
Schranken an die Kopplungskonstantenprodukte der
$R$-paritätsverletzenden Wechselwirkungen leicht durch einige
Einschränkungen (keine pseudoskalaren Wechselwirkungen) gewonnen
werden, wie im Falle des
Pions in Kapitel \ref{sec: R} explizit gezeigt wird.\\ \\
\textbf{LQs sind Teilchen, die an Lepton-Quark-Paare koppeln}
(siehe Abb. 1.2). Dies ist als eine Definition dieser neuen Klasse
von Teilchen zu verstehen. Sie haben ganzzahligen Spin und tragen
elektrische Ladung und Farbe. Falls LQs existieren, ergeben sich
zahlreiche neue Möglichkeiten, die innerhalb des SM's nicht
auftreten \footnote{Eine ähnliche Auflistung befindet sich in
\cite {blum} .}: \begin {itemize}\item $L$ (Leptonzahl) und/oder
$B$ (Baryonzahl) wird verletzt. Da die für $B$-Verletzung
verantwortlichen LQs einen Beitrag zum Protonenzerfall liefern,
müssen ihre Massen, aufgrund der Lebensdauer des Protons, sehr
hoch sein (bzw. das Kopplungskonstantenprodukt sehr klein), oder
diese Zerfallsart muss aufgrund einer neuen Symmetrie (wie schon
im Fall der $R$-Parität) verboten sein. \item Auch können FCNC
(\textbf{F}lavor \textbf{C}hanging \textbf{N}eutral
\textbf{C}urrents - Verletzung der Familiennummer) auftreten,
wobei die stärkste Einschränkung aus dem Zerfall $K_{L}\rightarrow
\mu e$ abgeleitet wird (Kapitel 4.2).\item Aus leptonischen
Mesonenzerfällen lassen sich relativ starke Schranken an die
Kopplungskonstantenprodukte ableiten. Insbesondere gilt dies für
LQs, die zu links- \underline{und} rechtshändigen Quarks koppeln,
wie in Kapitel 3.2 für den Pionenzerfall gezeigt wird.\item In
Kapitel 7 wird auf Meson-Antimeson-Mischungen
($K^0-\overline{K^{0}}$, $D^0-\overline{D^{0}}$ und
$B^0-\overline{B^{0}}$ eingegangen. \item Aus dem Zerfall des
$Z_0$-Bosons (dies wurde u. A. in \cite {z} behandelt) sowie
$\mu$- und $\tau$-Zerfällen \cite {myon1} lassen sich ebenfalls
Einschränkungen gewinnen.\item Eine Reihe schwächerer Bedingungen
erhält man aus
\begin {itemize}\item Messungen der atomaren Paritätsverletzung \cite {pari}, \item (g-2)-Messungen - Eine
ausführliche Behandlung des anomalen magnetischen Momentes des
Myons befindet sich in \cite {myon2}, \item
Neutrinooszillationsexperimenten \cite {neutri}. In \cite {neu}
wurde gezeigt, wie Neutrinomassen ohne Verletzung der $R$-Parität
durch Erweiterung des MSSM mit chiralen LQ-Multipletts erzeugt
werden.
\end {itemize}
\end {itemize}
Eine detaillierte Betrachtung der LQs kann \cite {comp} entnommen
werden.
\subsection{Leptoquarkmodelle}
LQs wurden zum ersten Mal in einem SU(4)-Modell von Pati und Salam
\cite{pati}, in dem $L$ als vierte Farbe behandelt wurde,
eingeführt. Der theoretische Rahmen für LQ-Kopplungen und
-Quantenzahlen wird in \cite{Buch} und \cite{Buch2} dargestellt.
Einen entsprechenden Überblick der $R$-paritätsverletzenden
Wechselwirkungen liefert \cite{butter} (siehe auch \cite{butter2}
und \cite{butter3}).
\subsection{Suche nach Leptoquarks}
An verschiedenen Beschleunigern wurde nach LQs gesucht. Hier
verweise ich insbesondere auf die Daten von HERA \cite {hera}. In
\cite {sum} werden die Ergebnisse von HERA, TEVATRON und LEP
zusammengefasst. Eine weitere Übersicht über die TEVATRON-Daten
kann \cite {teva} entnommen werden.\\ Die Möglichkeiten der
Entdeckung $R$-paritätsverletzender Wechselwirkungen am TEVATRON
(Run II) werden in \cite{herbi1} diskutiert (siehe auch
\cite{herbi2}), für die Suche an HERA verweise ich auf
\cite{schwan}. \cite{herbi3} geht auf Hadron-Hadron-Kollisionen
und die dabei mögliche Sleptonenproduktion via
$R$-Paritätsverletzung ein, die am TEVATRON (Run II) und am LHC
nachgewiesen werden kann.
\chapter{Leptonische Mesonenzerfälle}
\begin{description}
    \item[Ziel:] In dieser Arbeit sollen Schranken an LQ- und
    SUSY- Kopplungskonstantenprodukte berechnet werden. Diese
    werden in der Form
    $$\lambda_1\lambda_2<Zahl\left(\frac{m}{100GeV}\right)^2$$
    dargestellt. Mit der Division der LQ/SUSY-Masse m durch 100 GeV
    folge ich der Notation in der Literatur. Dabei wird
    angenommen, dass LQ/SUSY-Massen etwa im Bereich von 100
    GeV zu erwarten sind.
\end{description} In den folgenden Kapiteln werden neben
leptonischen Mesonenzerfällen der Art $M\rightarrow
\ell^{i}\overline{\ell}^{m}$ ($M$ ist ein Meson, die $\ell^i$ sind
Leptonen) auch semileptonische Zerfälle der Art $M\rightarrow
M'\ell^{i}\overline{\ell}^{m}$ betrachtet. Zunächst beschränken
wir uns auf leptonische Zweikörperzerfälle, da die Zerfallsrate
dann aufgrund der Kinematik einfacher zu kalkulieren ist, als bei
Dreikörperzerfällen (siehe Gln. 2.10-2.14). Ausgehend von der
LQ-Lagrangefunktion (Kap. 2.1) wird die Zerfallsrate
$\Gamma_{M\rightarrow \ell^{i}\overline{\ell}^{m}}$ berechnet
(Kap. 2.4).
\section{Die LQ-Lagrangefunktion}
Es gibt sieben renormierbare, $B$- und $L$-erhaltende
Quark-Lepton-Boson-Kopplungen, die für skalare und vektorielle LQs
in Übereinstimmung mit den $\ SU(3)\times SU(2)\times U(1)$
Symmetrien des Standardmodells sind. Die Bezeichnung
\textit{skalar} (\textit{vektoriell}) bezieht sich dabei auf den
Spin der Teilchen; Spin 0 (Spin 1). Die Lagrangefunktionen für
skalare und vektorielle Wechselwirkung ($S$- und $V$-LQs) sind
\cite {Buch} \footnote{In der Notation folge ich \cite{lepto}, der Generationenindex der Fermionen ist unterdrückt.}:\\
\begin {eqnarray} \mathcal{L}_{S}&=&
\{(\lambda_{LS_{0}}\overline{q}_{L}^{c}i\sigma_{2}\ell_{L}+
\lambda_{RS_{0}}\overline{u}_{R}^{c}e_{R})S^{\dagger}_{0}+
\lambda_{R\widetilde{S}_{0}}\overline{d}_{R}^{c}e_{R}\widetilde{S}^{\dagger}_{0}+{}\nonumber\\&&{}+
(\lambda_{LS_{1/2}}\overline{u}_{R}\ell_{L}+\lambda_{RS_{1/2}}\overline{q}_{L}i\sigma_{2}e_{R})S^{\dagger}_{1/2}+
\lambda_{L\widetilde{S}_{1/2}}\overline{d}_{R}\ell_{L}\widetilde{S}^{\dagger}_{1/2}+{}\nonumber\\&&{}+
\lambda_{LS_{1}}\overline{q}_{L}^{c}i\sigma_{2}\overrightarrow{\sigma}\ell_{L}\cdot\overrightarrow{S}_{1}^{\dagger}\}+h.c.
\end {eqnarray}
\begin {eqnarray}
\mathcal{L}_{V}&=&\{(\lambda_{LV_{0}}\overline{q}_{L}\gamma_{\mu}\ell_{L}
+ \lambda_{RV_{0}}\overline{d}_{R}\gamma_\mu
e_{R})V^{\mu\dagger}_{0}+
\lambda_{R\widetilde{V}_{0}}\overline{u}_{R}\gamma_{\mu}e_{R}\widetilde{V}^{\mu\dagger}_{0}+{}\nonumber\\&&{}+
(\lambda_{LV_{1/2}}\overline{d}_{R}^{c}\gamma_{\mu}\ell_{L}+
\lambda_{RV_{1/2}}\overline{q}_{L}^{c}\gamma_{\mu}e_{R})V^{\mu\dagger}_{1/2}+
\lambda_{L\widetilde{V}_{1/2}}\overline{u}_{R}^{c}\gamma_{\mu}\ell_{L}
\widetilde{V}^{\mu\dagger}_{1/2}+{}\nonumber\\&&{}+\lambda_{LV_{1}}\overline{q}_{L}
\gamma_{\mu}\overrightarrow{\sigma}\ell_{L}\cdot\overrightarrow{V}_{1}^{\mu\dagger}\}+h.c.
\end {eqnarray}
Die $\lambda$'s sind Kopplungskonstanten. Zur Fermionenhändigkeit
sowie der Definition von $\sigma_k$ siehe Kapitel 2.2 ; die
$\gamma_\mu$ sind 4$\times$4-Gamma-Matrizen \cite{Buch}. $q_L$ und
$\ell _L$ sind $SU(2)_L$-Dubletts, $u_R$, $d_R$ und $e_R$ sind die
entsprechenden Singletts. Der Index $c$ bedeutet
Ladungskonjugation. Aus den beiden Lagrangefunktionen ergeben sich
die in Tabelle 2.1 zusammengefassten Eigenschaften\footnote{In
einigen Arbeiten wird die Lagrangefunktion auch in der Form
$\mathcal{L}=\mathcal{L}_{F=2}+\mathcal{L}_{F=0}$ dargestellt (zur
Definition
von F siehe Tabelle 2.1).}.\\
\\Die 4-Fermion-Vertizes der vektoriellen LQs haben nach
einer Fierz-Transformation (siehe Kapitel 2.3.1) für\footnote{s
ist eine lorentzinvariante Mandelstam-Variable: $s=(\sum p_i)^2$,
$p_i$ ist der Viererimpuls der Eingangsteilchen.}
$m_{LQ}\gg\sqrt{s}$ allgemein folgende Form:
\begin{equation} \mathcal{L}_1=\pm\frac{\lambda^{ij}_{L,R}\lambda^{*mn}_{L,R}}{m^{2}_{LQ}}
(\overline{q}^{j}\gamma^{\mu}P_{L,R}q^{n})
(\overline{\ell}^{m}\gamma_{\mu}P_{L,R}\ell^{i})\; ,\label{ver1}
\end{equation}
\\und/oder
\\\begin{equation} \mathcal{L}_2=\pm\frac{\lambda^{ij}_{L,R}\lambda^{*mn}_{R,L}}{m^{2}_{LQ}}
(\overline{q}^{j}P_{R,L}q^{n})
(\overline{\ell}^{m}P_{L,R}\ell^{i})\; .\label{ver2}\end{equation}
$m_{LQ}$ ist die Leptoquarkmasse, ($i,j,m,n$) sind
Generationenindizes, $q$ und $\ell$ sind hier Dirac-Spinoren (und
nicht unbedingt SU(2)-Dubletts), $P_L$ und $P_R$ sind
Projektionsoperatoren. Das Minuszeichen muss bei
$LQ=\widetilde{V}^{\mu}_{1/2}$ oder $V^{\mu}_{1/2}$ eingesetzt
werden (es ergibt sich aus der Anwendung des
Ladungskonjugationsoperators $C$). Die Fermionen haben eine
bestimmte Chiralität, die auch den Tabellen 4 und 5 des Anhangs
entnommen werden kann. Diese enthalten sämtliche
4-Fermionen-Vertizes für die in der Lagrangefunktion auftretenden
Wechselwirkungsterme. Es wird mit der Notation
$f_{R/L}=P_{R,L}f=\frac{1}{2}(1\pm\gamma_5)f $ ($f$ ist ein
Dirac-Spinor)
gearbeitet.\begin{description}\item\textbf{Anmerkung:} In
\cite{pari} wird der Term
$\frac{\lambda_{LQ}^2}{m_{LQ}^2}(\bar{e}\gamma_\mu\gamma_5
e)(\bar{q}\gamma_\mu q)$ behandelt, der einen Beitrag zur atomaren
Paritätsverletzung durch die Wechselwirkung zwischen Quarks und
Elektronen liefert. Dieser Term wird im Weiteren nicht mehr
betrachtet.\end{description} Der Fermi-4-Fermionen-Vertex aus dem
Standardmodell (schwache Wechselwirkung) lässt sich schreiben als:
\begin{equation} \mathcal{L}_{SM}=\frac{4G_{F}}{\sqrt{2}}V_{jn}(\overline{q}^{j}\gamma^{\mu}P_{L}q^{n})
(\overline{\ell}^{m}\gamma_{\mu}P_{L}\ell ^{i})\;
.\label{sm}\end{equation} $G_{F}$ ist die Fermikonstante und
$V_{jn}$ das CKM-Matrixelement.
\newpage\setlength{\arraycolsep}{4mm}

\newcommand{\sd}[2]{\raisebox{-1.5ex}{\shortstack[r]{$\strut#1$\\ $\strut#2$}}}
\newcommand{\st}[3]{\raisebox{-3ex}{\shortstack[r]{$\strut#1$\\ $\strut#2$\\ $\strut#3$}}}
\newcommand{\stl}[3]{\raisebox{-3ex}{\shortstack[l]{$\strut#1$\\ $\strut#2$\\ $\strut#3$}}}
\begin{center}
\begin{table}[h]
\begin{tabular}{|c||c|c|c|c|c|c|l|}
  \hline
  & J & F & T & $T_3$ & Q & SU(5)-Darstellung & \multicolumn{1}{c|}{\mbox{koppelt an}} \\
  \hline\hline
         $S_0$ & 0 & 2 &   0 &              0 &           -1/3 & \textbf{5} &
         \mbox{\normalsize$e_Lu_L~~e_Ru_R~~\nu_Ld_L$} \\
  \hline
  $\tilde S_0$ & 0 & 2 &   0 &              0 &           -4/3 & \textbf{45} &
  \mbox{\normalsize$e_Rd_R$} \\
  \hline
         $S_{1/2}$ & 0 & 0 & 1/2 & \sd{-1/2}{1/2} & \sd{-5/3}{-2/3} & \textbf{45} &
         \sd{\mbox{\normalsize$e_R\bar u_L~~e_L\bar u_R$}}{\mbox{\normalsize$e_R\bar d_L~~\nu_L\bar u_R$}} \\
  \hline
  $\tilde S_{1/2}$ & 0 & 0 & 1/2 & \sd{-1/2}{1/2} & \sd{-2/3}{1/3} & \textbf{10/15} &
  \sd{\mbox{\normalsize$e_L\bar d_R$}}{\mbox{\normalsize$ \nu_L\bar d_R$}} \\
  \hline
         $S_1$ & 0 & 2 &   1 & \st{-1}{0}{1} & \st{-4/3}{-1/3}{2/3} & \textbf{45} &
         \stl{\mbox{\normalsize$e_Ld_L$}}{\mbox{\normalsize$e_Lu_L~~\nu_Ld_L$}}{\mbox{\normalsize$\nu_Lu_L$}} \\
  \hline
         $V_0^\mu$ & 1 & 0 &   0 &              0 &           -2/3 & \textbf{10} &
         \mbox{\normalsize$e_L\bar d_L~~e_R\bar d_R~~\nu_L\bar u_L$} \\
  \hline
  $\tilde V_0^\mu $ & 1 & 0 &   0 &              0 &           -5/3 & \textbf{75} & \mbox{\normalsize$e_R\bar u_R$} \\
  \hline
         $V_{1/2}^\mu$ & 1 & 2 & 1/2 & \sd{-1/2}{1/2} & \sd{-4/3}{-1/3} & \textbf{24} &
         \sd{\mbox{\normalsize$e_Rd_L~~e_Ld_R$}}{\mbox{\normalsize$e_Ru_L~~\nu_Ld_R$}} \\
  \hline
  $\tilde V_{1/2}^\mu $ & 1 & 2 & 1/2 & \sd{-1/2}{1/2} & \sd{-1/3}{2/3} & \textbf{10/15} &
  \sd{\mbox{\normalsize$e_Lu_R$}}{\mbox{\normalsize$\nu_Lu_R$}} \\
  \hline
         $V_1^\mu$ & 1 & 0 &   1 & \st{-1}{0}{1} & \st{-5/3}{-2/3}{1/3} & \textbf{40} &
  \stl{\mbox{\normalsize$e_L\bar u_L$}}{\mbox{\normalsize$e_L\bar d_L~~\nu_L\bar u_L$}}{
  \mbox{\normalsize$\nu_L\bar d_L$}} \\
  \hline
\end{tabular}
\caption[Eigenschaften der Leptoquarks]{\textit{Eigenschaften der
Leptoquarks} \cite{cuy} - \textit{Die Bezeichnungen der einzelnen
LQs wurden gemäß \cite {lepto} vorgenommen (LQs sind $SU(3)_c
$-Tripletts); F=3B+L; T ist der SU(2)-Isospin; J ist der Spin; Q
bezeichnet die elektrische Ladung: $ Q=T_3-\frac{1}{2}Y $(Y=
Hyperladung); Kopplungen wurden nur für die erste Generation
aufgelistet, außerdem wurden ergänzend die minimale
SU(5)-Darstellung angegeben, in die die LQs eingebettet werden
können} \cite{rizzo}.}\label{tab:lepto}
\end{table}\end{center}
Die Kopplungen in den beiden LQ-Vertizes tragen unterschiedliche
Händigkeitsindizes: In Gl. (\ref{ver1}) tragen beide $\lambda$s
den Index $L$ ($R$), in Gl. (\ref{ver2}) tragen die $\lambda$s
unterschiedliche Indizes ($L$, $R$ oder $R$, $L$). Desweiteren
sieht man, dass sich die Gln. (\ref{ver1}) und (\ref{sm}) nur in
den konstanten Vorfaktoren unterscheiden. Beide sind $(V\pm A)$
(vektorielle $\pm$ axialvektorielle)-Vertizes. Gl. (\ref{ver2})
ist von der Form $(S\pm P)$ (skalar $\pm$ pseudoskalar) und kann
damit die chirale Unterdrückung des Zerfalls $\pi\rightarrow
e\nu_e$ aufheben, wie in Kapitel 3 (für das Pion) dargestellt
wird. Welche LQs $(S\pm P)$-Vertizes erzeugen, kann Tabelle 2.1
und den Tabellen 4 und 5 des Anhangs entnommen werden.
\\ \\
Die effektive Lagrangefunktion für $V^{\mu}_0$ ist\footnote {Die
übrigen 4-Fermionen-Vertizes können den Tabellen 4 und 5 des
Anhangs entnommen werden.}: \begin{eqnarray}
\lefteqn{\mathcal{L}_{eff}=\frac{\lambda_{RV_{0}}\lambda^{*}_{RV_{0}}}{m^{2}_{0}}
(\overline{d}_{R}\gamma^{\mu}d_{R})
(\overline{e}_{R}\gamma_{\mu}e_{R})+\frac{\lambda_{LV_{0}}\lambda^{*}_{LV_{0}}}{m^{2}_{0}}
(\overline{u}_{L}\gamma^{\mu}u_{L})
(\overline{\nu}_{L}\gamma_{\mu}\nu_{L})+\frac{\lambda_{LV_{0}}\lambda^{*}_{LV_{0}}}{m^{2}_{0}}
(\overline{d}_{L}\gamma^{\mu}u_{L})
(\overline{\nu}_{L}\gamma_{\mu}e_{L}){}}\nonumber\\&&{}
+\frac{\lambda_{LV_{0}}\lambda^{*}_{LV_{0}}}{m^{2}_{0}}
(\overline{d}_{L}\gamma^{\mu}d_{L})
(\overline{e}_{L}\gamma_{\mu}e_{L})+\frac{\lambda_{LV_{0}}\lambda^*_{RV_{0}}}{m^{2}_{0}}
(\overline{d}_{R}u_{L})
(\overline{\nu}_{L}e_{R})+\frac{\lambda_{LV_{0}}\lambda^*_{RV_{0}}}{m^{2}_{0}}
(\overline{d}_{R}d_{L}) (\overline{e}_{L}e_{R})\;
 .{}\nonumber\\&&{}\end{eqnarray} Es treten sowohl $(V\pm A)$- als
auch $(S\pm P)$-Vertizes auf. Dies ist genauso bei den LQs
$S_{0},S_{1/2}$ und $V_{1/2}^{\mu}$. Die übrigen LQs haben nur
$(V\pm A)$-Vertizes. Aus den $(S\pm
P)$-Matrixelementen\footnote{Die dabei (nach der
Fierz-Transformation) auftretenden tensoriellen Operatoren liefern
bei semileptonischen Zerfällen nicht verschwindende Beiträge. Für
den semileptonischen $K$-Zerfall verweise ich auf Kapitel 4.1.2 .}
erhält man Schranken an das Produkt $\lambda_{L}\lambda_{R}$
(Kopplungen entgegengesetzter Chiralität an den beiden Enden des
LQ-Propagators).\\ \\Selbst wenn man nur rechts- \underline{oder}
linkshändige Kopplungen zulässt, kann der helizitätsunterdrückte
Zerfall $ \pi\longrightarrow e\nu_e $ noch starke Einschränkungen
an die Kopplungskonstanten liefern. LQ-Kopplungen mit dem
SM-Higgs-Dublett führen die nicht-chiralen Wechselwirkungsterme
wieder ein \cite {higgs}.
\section{Konventionen und Näherungen}
Im Weiteren werden einige Konventionen verwendet, die zur besseren
Übersicht hier zusammengefasst sind:
\begin {description}\item \textbf{Indizes}: In den Kopplungskonstanten
$\lambda$ sind die Generationenindizes (soweit ihre Erwähnung
notwendig ist) hochgestellt, wobei der Leptonenindex vor dem
Quarkindex kommt. Der Generationenindex der LQs ist
unterdrückt.\item \textbf{Generatoren $\sigma_k$ der SU(2)}: Die
in der LQ-Lagrangefunktion in Kapitel 2.1 auftretenden
$\sigma_{k}$ (Pauli-Matrizen), mit $\overrightarrow {\sigma}=
(\sigma_{1},\sigma_{2},\sigma_{3})$, sind in Tabelle 3 des Anhangs
definiert, sie operieren auf den Fermionenfeldern $q_L$ und
$\ell_L$ in der Lagrangefunktion.\item \textbf{Händigkeit}: Die
Fermionen in unserer Notation haben eine wie folgt definierte
Händigkeit: $\overline{f}_{R}=(P_Rf)^{\dag}\gamma_{0}$ und
$f_{L}^{c}=(P_Lf)^{c}$. Die Händigkeit der LQ-Kopplungen für einen
bestimmten 4-Fermionen-Vertex kann den Tabellen 4 und 5 des
Anhangs entnommen werden. \item \textbf{Vektorielle und skalare
LQs}: Es kann sowohl skalare, als auch vektorielle LQs geben. Der
Einfachheit halber werden sämtliche Berechnungen im Folgenden
zunächst für vektorielle LQs durchgeführt. Die Resultate für
skalare LQs erhält man so\footnote{In einigen Fällen (siehe
Kapitel 3.2.1 für das Pion) muss auch das Vorzeichen der
4-Fermionen-Vertizes (2.3) bzw. (2.4) beachtet werden.} :\begin
{equation}\lambda_{V} \longrightarrow
\frac{\lambda_{S}}{\sqrt{2}}\; .\end {equation}
\item\textbf{Berechnete Kopplungen}: Im Text werden die Kopplungen
nur für $\lambda_L\lambda_L$ bzw. $\lambda_L\lambda_R$ berechnet,
für $\lambda_R\lambda_R$ und $\lambda_R\lambda_L$ läuft die
Rechnung analog. Die Ergebnisse sind im Anhang tabellarisch
zusammengefasst.
\end {description}Es werden zudem noch einige Näherungen bzw. Abschätzungen verwendet:
\begin {description}\item \textbf{QCD-Korrekturen}: QCD-Korrekturen werden in allen Ordnungen außer Acht gelassen.
Durch den dabei entstehenden Fehler ändern sich die Schranken an
$\lambda^{2}$ nicht mehr als um einen Faktor 2 (siehe
\cite{lepto}). Bei den schwereren Mesonen erhält man aufgrund
anderer Näherungen zur Berechnung der Zerfallsraten nur ein
qualitatives Ergebnis\footnote{Insbesondere durch die teilweise
nötige Verwendung experimenteller Werte für die Zerfallskonstante
$f_{M}$, die bei den schwereren Mesonen mit einem relativ großen
Fehler behaftet sind, werden die berechneten Schranken
entsprechend ungenau.} (zur Berechnung des $B$-Mesonenzerfalls
wurde die HQE, \textbf{H}eavy \textbf{Q}uark \textbf{E}xpansion,
benutzt). Diese Näherungen werden in den entsprechenden
Abschnitten näher erläutert, siehe insbesondere Kapitel 6. \item
\textbf{Quarkmassen}: In einige Berechnungen gehen die Quarkmassen
ein, deren Werte der Tabelle 2 des Anhangs entnommen werden
können. Dabei werden die oberen Grenzwerte für die Quarkmassen
verwendet, um eine konservative Abschätzung der Schranken an die
betreffenden Kopplungskonstantenprodukte zu erhalten.
\end {description}
\section{Nebenrechnungen} \subsection{Die Fierz-Transformation}
Die Fierz-Transformation (siehe \cite{fierz}) wird verwendet, um
Berechnung mit der invarianten Amplitude $\mathcal{M}$,
insbesondere Interferenzterme mehrerer Matrixelemente, zu
vereinfachen. Die beliebigen Dirac-Spinoren $u_i$ werden dabei
umgeordnet: \beq
(\bar{u}_3\Lambda_iu_2)(\bar{u}_1\Lambda_ju_4)=\sum^5_{k,l=1}c_{ijkl}(\bar{u}_1\Lambda_ku_2)(\bar{u}_3\Lambda_lu_4)\;
,\eeq wobei $\Lambda_i=(\mathbf{1}, \gamma_\mu,
\sigma_{\mu\nu}=\frac{i}{2}[\gamma_\mu,\gamma_\nu],
\gamma_\mu\gamma_5, \gamma_5)$; $c_{ijkl}$ sind konstante
Entwicklungskoeffizienten. Für $i=j$ vereinfacht sich die
Rechnung: \beq
(\bar{u}_3\Lambda_iu_2)(\bar{u}_1\Lambda_iu_4)=\sum^5_{j=1}\lambda_{ij}(\bar{u}_1\Lambda_ju_2)(\bar{u}_3\Lambda_ju_4)\;
,\eeq wobei \beq (\lambda_{ij})=\frac{1}{4}
\left(\begin{array}{rrrrr}
1&1&1&1&1\\4&-2&0&2&-4\\6&0&-2&0&6\\4&2&0&-2&-4\\1&-1&1&-1&1
\end{array}\right)\; .\eeq
Insbesondere gilt: \beq
(\bar{u}_3\gamma_\mu(1-\gamma_5)u_2)(\bar{u}_1\gamma^\mu(1-\gamma_5)u_4)=-(\bar{u}_1\gamma_\mu(1-\gamma_5)u_2)
(\bar{u}_3\gamma^\mu(1-\gamma_5)u_4)\;
.\eeq Bei Verwendung antikommutierender Felder $\psi_i$ statt der
Spinoren $u_i$ erscheint auf der rechten Seite von Gl. (2.11) ein
zusätzliches Minuszeichen.\\Bei $(S\pm P)$-Wechselwirkungen treten
nach der Fierztransformation neben $(S\pm P)$-Termen auch
tensorielle Terme ($\bar{u}\sigma_{\mu\nu}u$) auf. Bei der
Betrachtung semileptonischer $K$-Mesonenzerfälle (Kapitel 4.1.2)
werden diese ausführlicher behandelt.
\subsection{Berechnung der Quarkströme}
In der fundamentalen Darstellung der (\textit{flavour-}) $SU(3)$,
\textbf{3}, gibt es drei Typen (\textit{flavours}) von Quarks,
$q=(u,d,s)$. Ausgehend von der PCAC- (\textbf{P}artially
\textbf{C}onserved \textbf{A}xial-vector \textbf{C}urrent)
Bedingung für das pseudoskalare Mesonenoktett\footnote{Es wird
angenommen, dass die $SU(2)$-Isospin-Symmetrie nicht gebrochen
ist; $a$ und $b$ sind Isospin-Indizes.}\beq\langle
0|j_\mu^{(5)b}(x)|\phi^a(p)\rangle
=i\delta^{ab}\frac{f_{\phi}p^{\mu}}{\sqrt{2}}e^{ip\cdot x}\; ,\eeq
kann man den Quarkstrom bei Mesonenzerfällen berechnen. Dabei ist
$f_\phi$ die Mesonenzerfallskonstante, $p^{\mu}$ der
Viererimpulsvektor des Mesons, $\phi =
\phi^a\frac{\lambda^a}{\sqrt{2}}$ und \beq
j_\mu^{(5)b}=\bar{q}\gamma^\mu\gamma^5\frac{\lambda^b}{2}q \;
.\eeq Die $\lambda^a$ (a=1,...8) sind die Gell-Mann-Matrizen
(siehe Tabelle 3 des Anhangs) mit der Normalisierung $Tr(\lambda^a\lambda^b)=2\delta^{ab}$.\\
\\Die folgende Rechnung, siehe auch \cite{peskin}, wird für das Pion durchgeführt, die Ergebnisse gelten aber
analog auch für das Kaon. Mit $\bar{u}\gamma^\mu\gamma^5
d=\bar{q}\gamma^\mu\gamma^5(\frac{\lambda^1}{2}-i\frac{\lambda^2}{2})
q$ und $\pi^-=\frac{1}{\sqrt{2}}(\phi^1+i\phi^2)$ erhält man für
$x=0$ (unter Beachtung von $\langle 0|\overline{u}\gamma_\mu
d|\pi^-\rangle =0$):
\begin{equation}\langle0|\overline{u}\gamma^{\mu}P_{L,R}d|\pi^-\rangle=\pm i\frac{f_{\pi^-}}{2}\cdot p^{\mu}
\equiv\widetilde{A}_{\pi^-}\cdot p^{\mu}\; .\end{equation}
$\widetilde{A}_{\pi^-}$ ist positiv für $P_L$ und negativ für
$P_R$. Die Ableitung von Gl. (2.12) liefert: \beq\langle
0|\partial^\mu j_\mu^{(5)b}(x)|\phi^a(p)\rangle
=-\delta^{ab}\frac{f_{\phi}m_{\phi}^2}{\sqrt{2}}e^{ip\cdot x}\;
.\eeq Unter Verwendung der Dirac-Gln. für die Quarkfelder, \beq
i\gamma^\mu\partial_\mu q=\mathbf{m}q\; ,\qquad
-i\partial_\mu\bar{q}\gamma^\mu=\bar{q}\mathbf{m}\; ,\eeq wobei
$\mathbf{m}$ eine $3\times3$-Diagonalmatrix mit den Einträgen
$m_u,m_d$ und $m_s$ ist, erhält man: \beq\partial^\mu
j_\mu^{(5)b}=i\bar{q}\gamma^5\frac{1}{2}\{\mathbf{m},\lambda^b\}q\;
.\eeq Für $(a;b)=(1;2)$ gilt \beq
tr[\{\mathbf{m},\lambda^b\}\lambda^a]=\frac{1}{2}\delta^{ab}(m_u+m_d)\;
,\eeq und damit erhält man für $x=0$:
\begin{equation}\langle 0|(\overline{u}P_{L,R}d)|\pi^-\rangle =\mp i\frac{f_{\pi^-}m_{\pi^-}}{2}
\frac{m_{\pi^-}}{m_{u}+m_{d}}\equiv\widetilde{P}_{\pi^-}\;
.\end{equation} $\widetilde{P}_{\pi^-}$ ist positiv für $P_R$ und
negativ für $P_L$.
\subsection {Die Zerfallsrate $\Gamma$}
Bei der Berechnung der Zerfallsrate aus dem Betragsquadrat der
invarianten Amplitude $\overline{|\mathcal{M}|^2}$ (über
Anfangszustandsspins wird gemittelt, über Endzustandsspins
summiert) in Kapitel 2.4 wird Fermi's Goldene Regel verwendet. Für
einen Dreikörperzerfall hat man damit \cite{rich}:
\begin{equation}
 \Gamma(a \rightarrow b+c+d) = \frac1{(2\pi)^3}\frac1{32M^3_a}
  \int^{\left(m^2_{bc}\right)_{max}}_{\left(m^2_{bc}\right)_{min}}
                            dm^2_{bc}
 \int^{\left(m^2_{cd}\right)_{max}}_{\left(m^2_{cd}\right)_{min}}
                            dm^2_{cd}\,
                \overline{|\mathcal{M}|^2}\; ,
\label{eqn:threebodyphase}
\end{equation}
wobei
\begin{itemize}
  \item $\left(m^2_{bc}\right)_{max}=(M_a-m_d)^2$,
  \item $\left(m^2_{bc}\right)_{min}=(m_b+m_c)^2$,
  \item $\left(m^2_{cd}\right)_{max}=(E^*_c+E^*_d)^2
             -\left(\sqrt{{E^*_c}^2-m^2_c}-\sqrt{{E^*_d}^2-m^2_d}\right)^2$,
\textheight 24cm
  \item $\left(m^2_{cd}\right)_{min}=(E^*_c+E^*_d)^2
             -\left(\sqrt{{E^*_c}^2-m^2_c}+\sqrt{{E^*_d}^2-m^2_d}\right)^2$,
  \item $E^*_c=\left(m^2_{bc}-m^2_b+m^2_c\right)/2m_{bc}$ und
            $E^*_d=\left(M^2_a-m^2_{bc}-m^2_d\right)/2m_{bc}$
            sind die Energien der Teilchen $c$ und $d$ im $m_{bc}$ Ruhesystem.
\end{itemize}
Für Zerfälle der Art $a\rightarrow b+c$ vereinfacht sich dies
zu:\begin{equation}\Gamma_{a\rightarrow
b+c}=\frac{|\overline{\mathcal{M}(a\rightarrow b+c)}|^{2}k}{8\pi
M_{a}^{2}}
\; .\end{equation} $k$ ist dabei der Betrag des
Leptonen-Dreierimpulsvektors im Schwerpunktsystem des zerfallenden
Mesons, $M_a$ ist die Mesonenmasse und $m_b,m_c$ sind die Massen
der Zerfallsteilchen $b$ und $c$. \beq p_{b}\cdot p_{c}=
\frac{1}{2}(M_a^2-m_b^2-m_c^2)\; \eeq benutzt. Desweiteren gilt
(im Schwerpunktsystem des zerfallenden Mesons): \beq
k^2=\frac{1}{4M_a^2}[M_a^2-(m_b+m_c)^2][M_a^2-(m_b-m_c)^2]\; .\eeq
\section {Berechnung der Zerfallsrate $\Gamma(M\rightarrow \ell^{i}\overline{\ell}^{m})$}
Es soll nun aus der invarianten Amplitude $\mathcal{M}$ die
Zerfallsrate $\Gamma$ für den Zerfall eines Mesons $M$
(gebundender Zustand zweier Quarks $\overline{q}^{j}q^{n}$) in
zwei Leptonen $ \ell^{i}\overline{\ell}^{m}$ berechnet werden.
Allgemein berechnet man die invariante Amplitude, indem man die
effektiven Hamiltonfunktionen der relevanten Zerfallsprozesse
angibt und über $\langle Endzustand\mid\mathcal{H}\mid
Anfangszustand\rangle$ integriert\footnote{Dies wurde für das Pion
in \cite{thor} durchgeführt.}, wobei zusätzlich die
Viererimpulserhaltung gewährleistet sein muss. Für das Pion sieht
dies so aus: \beq
\mathcal{M}_{ijmn}\delta^4(p_\pi-p_\ell-p_\nu)=\frac{1}{2\pi
i}\int\langle
\ell^i;\bar{\nu}^m|\mathcal{H}_{total}|\pi^-_{jn}\rangle\; .\eeq
Die Quarks sind im Anfangszustand in einem Meson gebunden. Nach
dem Einsetzen der effektiven Hamiltonfunktion erhält man daher die
Matrixelemente $\langle
0|(\overline{q}^{j}\gamma^{\mu}P_{L,R}q^{n})|M\rangle $ und
$\langle 0|(\overline{q}^{j}P_{L,R}q^{n})|M\rangle $, die in den
Gln. (2.14) und (2.19) angegeben sind. \\ \\Aus $\mathcal{L}_{SM}$
(Gl. \ref{sm}), $\mathcal{L}_1$ (Gl. \ref{ver1}) und
$\mathcal{L}_2$ (Gl. \ref{ver2}) erhält man die invariante
Amplitude\footnote{In dieser Notation entsprechen $\bar{\ell}^m$
und $\ell^i$ den Spinoren $\bar{v}_{\ell^m}$ und $u_{\ell^i}$ der,
in der Literatur \cite{halzen} verwendeten Schreibweise.}
$\mathcal{M}_{L,R}=
\mathcal{M}_{1;L,R}+\mathcal{M}_{2;L,R}+\mathcal{M}_{SM;L,R}$ :
\bea
\lefteqn{\mathcal{M}_{L,R}=\frac{(\lambda^{ij}_{L,R}\lambda^{*mn}_{L,R})}{m^{2}_{LQ}}
\langle 0|(\overline{q}^{j}\gamma^{\mu}P_{L,R}q^{n})|M\rangle
(\overline{\ell}^{m}\gamma_{\mu}P_{L,R}\ell^{i}){}}\nonumber\\&&{}+
\frac{(\lambda^{ij}_{L,R}\lambda^{*mn}_{R,L})}{m^{2}_{LQ}} \langle
0|(\overline{q}^{j}P_{R,L}q^{n})|M\rangle
(\overline{\ell}^{m}P_{L,R}\ell^{i}){}\nonumber\\&&{}+\frac{4G_FV_{jn}}{\sqrt{2}}
\langle 0|(\overline{q}^{j}\gamma^{\mu}P_{L}q^{n})|M\rangle
(\overline{\ell}^{m}\gamma_{\mu}P_{L}\ell^{i})\; .\eea Dieses
Matrixelement muss nun quadriert werden; außerdem ist eine
Summation über die Endzustandsspins durchzuführen. Über die
Eingangszustandsspins wird gemittelt (für ein Spin 0 -Meson ist
der Vorfaktor 1). Die einzelnen Beiträge werden mit den
Vollständigkeitsrelationen und unter Verwendung der Rechenregeln
für Spuren von $\gamma$-Matrizen ausgerechnet. Dabei wird nur
$\overline{|\mathcal{M}_{L}|^2}$ berechnet,
für $\overline{|\mathcal{M}_{R}|^2}$ ist das Verfahren analog.\\ \\
Im SM-Term kann man das $\gamma_\mu$ aus dem Leptonenstrom mit dem
$p^\mu$ des Matrixelements (Gl. 2.14) kontrahieren und erhält
damit (unter Verwendung der Dirac-Gln.): \bea
\sum_{Ausgangsspins}|p^\mu(\overline{\ell}^{m}\gamma_{\mu}P_{L}\ell^{i})|^2&=&
\sum_{Ausgangsspins}|m_{\ell^i}(\overline{\ell}^{m}P_{R}\ell^{i})-m_{\ell^m}(\overline{\ell}^{m}P_{L}\ell^{i})|^2
{}\nonumber\\{}&=&
\frac{1}{2}(m_{\ell^m}^2+m_{\ell^i}^2)Tr(\not{p_{\ell^m}}\not{p_{\ell^i}})+4m_{\ell^m}^2m_{\ell^i}^2\;
.\eea Es ergibt sich:
\begin{eqnarray}|\mathcal{M}_{SM}|^{2} &=& 8G_{F}^{2}|V_{jn}|^2|\tilde{A}|^{2}\cdot
\left[\frac{1}{2}(m_{\ell^m}^2+m_{\ell^i}^2)Tr(\not{p_{\ell^m}}\not{p_{\ell^i}})+
4m_{\ell^m}^2m_{\ell^i}^2\right]{}\nonumber\\{}&=&
8G_{F}^{2}|V_{jn}|^2|\widetilde{A}|^{2}\left[(m_{\ell^m}^2+m_{\ell^i}^2)
(m_M^2-m_{\ell^m}^2-m_{\ell^i}^2)+4m_{\ell^m}^2m_{\ell^i}^2\right]\;
.\end{eqnarray} Im letzten Schritt wurde Gl. (2.22) benutzt. Für
$|\mathcal{M}_1|^{2}$ und $|\mathcal{M}_2|^{2}$ erhält man:
\begin {equation}|\mathcal{M}_1|^{2}= \frac{|\lambda_{L}^{ij}\lambda_{L}^{*mn}|^{2}}{m_{LQ}^{4}}
|\widetilde{A}|^{2}\left[(m_{\ell^m}^2+m_{\ell^i}^2)(m_M^2-
m_{\ell^m}^2-m_{\ell^i}^2)+4m_{\ell^m}^2m_{\ell^i}^2\right]\;
,\end {equation} und \begin {equation}|\mathcal{M}_2|^{2}=
\frac{|\lambda_{L}^{ij}\lambda_{R}^{*mn}|^{2}}{m_{LQ}^{4}}|\widetilde{P}|^{2}(m_M^2-m_{\ell^m}^2-m_{\ell^i}^2)\;
.\end {equation} Die drei Interferenzterme haben folgende Form:
\beq 2\Re(\mathcal{M}^*_1\mathcal{M}_2)=
-2\frac{\Re[(\lambda_{L}^{ij}\lambda_{L}^{*mn})^*(\lambda_{L}^{ij}\lambda_{R}^{*mn})]}
{m_{LQ}^{4}}\widetilde{P}\widetilde{A}^*
m_{\ell^m}(m_M^2-m_{\ell^m}^2+m_{\ell^i}^2)\; ,\eeq
\begin{equation}2\Re(\mathcal{M}^*_{SM}\mathcal{M}_2)=-2\frac{\Re[(\lambda_{L}^{ij}
\lambda_{R}^{*mn})V_{jn}^*]}{m_{LQ}^{2}}\frac{4G_{F}}{\sqrt{2}}\widetilde{P}
\widetilde{A}^*m_{\ell^m}(m_M^2-m_{\ell^m}^2+m_{\ell^i}^2)\; ,
\end{equation} und
\begin{equation}2\Re(\mathcal{M}^*_{SM}\mathcal{M}_1)=2\frac{\Re[(\lambda_{L}^{ij}
\lambda_{L}^{*mn})V_{jn}^*]}{m_{LQ}^{2}}\frac{4G_{F}}{\sqrt{2}}
|\widetilde{A}|^2\left[(m_{\ell^m}^2+m_{\ell^i}^2)(m_M^2-m_{\ell^m}^2-m_{\ell^i}^2)+4m_{\ell^m}^2m_{\ell^i}^2\right]\;
.\end{equation}
\\Die Zerfallsrate $\Gamma_M=\Gamma_{M\rightarrow\l^{i}\overline{l}^{m}}$ ist damit (Gl. 2.21):
\begin{eqnarray}\lefteqn{\Gamma_M=\frac{k}{8\pi m_{M}^{2}}
\Bigg\{\frac{|\lambda_{L}^{ij}
\lambda_{R}^{*mn}|^{2}}{m_{LQ}^{4}}|\widetilde{P}|^{2}(m_M^2-m_{\ell^m}^2-m_{\ell^i}^2)
{}}\nonumber\\&&{}+
\left[\frac{\Re[(\lambda_{L}^{ij}\lambda_{R}^{*mn})^*(\lambda_{L}^{ij}\lambda_{L}^{*mn})]}{m_{LQ}^{2}}
-\frac{2\sqrt{2}G_{F}\Re[(\lambda_{L}^{ij}\lambda_{L}^{*mn})V_{jn}^*]}{m_{LQ}^{2}}\right]2\widetilde{P}\widetilde{A}^*
m_{\ell^m}(m_M^2-m_{\ell^m}^2+m_{\ell^i}^2) {}\nonumber\\&&{}
+\left|2\sqrt{2}G_{F}V_{jn}+\frac{(\lambda_{L}^{ij}\lambda_{L}^{*mn})^*}{m_{LQ}^2}\right|^2
|\widetilde{A}|^{2}\left[(m_{\ell^m}^2+m_{\ell^i}^2)(m_M^2-m_{\ell^m}^2-m_{\ell^i}^2)+
4m_{\ell^m}^2m_{\ell^i}^2\right]\Bigg\}\;,
{}\nonumber\\&&{}\label{dec}\end{eqnarray}wobei $k$ in Gl. (2.23)
angegeben ist. Man hat nun die Zerfallsrate für den leptonischen
Zerfall eines Mesons im Standardmodell mit zusätzlichen, über das
SM hinausgehenden, $(V\pm A)$ und $(S\pm P)$ -Wechselwirkungen.
\\ \\Diese Berechnung wurde in ähnlicher Form auch in \cite {lepto}
durchgeführt. Es wurde mit Kopplungskonstantenprodukten
$e_{L,R}\cdot e_{L,R}$ (bzw. $e_{L,R}\cdot e_{R,L}$) gerechnet,
die im Falle gleicher Händigkeit ($e_L \cdot e_L$ oder $e_R\cdot
e_R$) die SM-Kopplung und die LQ-Kopplung (als $|Summe\; der\;
Kopplungskonstantenprodukte|^2$), und im Fall unterschiedlicher
Händigkeit reine LQ-Kopplungen beinhalten. Außerdem wurden
$\mathcal{M}_L$ und $\mathcal{M}_R$ in einer einzigen Zerfallsrate
als Summe dargestellt und angenommen, dass die Masse $m_{\ell^i}$
verschwindet.
\chapter{Das Pion}
Innerhalb des  SMs zerfallen geladene Pionen hauptsächlich über
schwache Wechselwirkung in ein Myon und ein Myonneutrino: \beq
BR(\pi^{+}\rightarrow\mu^{+}\nu_{\mu})=\frac{\Gamma_{\pi^{+}\rightarrow\mu^{+}
\nu_{\mu}}}{\Gamma_\pi}=(99,98770\pm0,00004)\cdot 10^{-2}\; .\eeq
Der Zerfall $\pi^{+}\rightarrow e^{+}\nu_{e}$ ist etwa um den
Faktor $(\frac{m_{e}}{m_{\mu}})^{2}=2,34\cdot 10^{-5}$
unterdrückt. Das $\pi^{+}$ hat Spin 0, d.h. da das Neutrino
linkshändig ist, muss das Positron bzw. positiv geladene Myon auch
linkshändig sein (Spinerhaltung). Im Grenzübergang $m_e\rightarrow
0$ ($m_\mu\rightarrow 0$) koppelt der Strom der schwachen WW
allerdings nur an rechtshändige Positronen (Antimyonen), d.h. die
linkshändige Kopplung ist stark
unterdrückt. \\
Wie schon erwähnt existiert diese chirale Unterdrückung bei
Mesonenzerfällen über LQs nicht, da diese links- und rechtshändige
Kopplungen haben. Daraus ergeben sich starke Schranken für die
entsprechenden Kopplungskonstantenprodukte, die im Folgenden
(Kapitel 3.2) aus der allgemein berechneten Zerfallsrate
$\Gamma_M$ (Gl. 2.33) abgeleitet werden.
\section {Das Verhältnis $R$}
Um die experimentelle Ungenauigkeit in der Zerfallskonstante
$f_{M}$ \cite{pdg}, insbesondere bei schwereren Mesonen, zu
umgehen wird, statt mit der Zerfallsrate, mit dem Verhältnis \beq
R\equiv\frac{\Gamma(\pi^{+}\rightarrow\overline{e}\nu_e)}{\Gamma(\pi^{+}\rightarrow\overline{\mu}\nu_\mu)}\eeq
gearbeitet. Die Zerfallskonstante kürzt sich bei dieser
Vorgehensweise aus den Gleichungen heraus. Die
Standardmodellvorhersage für $R$ mit Strahlungskorrekturen ist
\cite{fink}: \beq
R_{th}=\frac{m^{2}_{e}}{m^{2}_{\mu}}\frac{(m_{\pi}^{2}-m_{e}^{2})^{2}}
{(m_{\pi}^{2}-m_{\mu}^{2})^{2}}(1+\Delta)=(1,2354\pm0,0004)\times10^{-4}\;
.\eeq Die Standardabweichung ist in 2$\sigma$ angegeben. Der
experimentelle Wert für $R$ (mit 2$\sigma$-Standardabweichung) ist
\cite{pdg}\footnote{$ \frac{\Gamma(\pi^{+}\rightarrow
\overline{e}\nu)}{\Gamma_\pi}=(1,230\pm 0,008)\times 10^{-4}$
(2$\sigma$-Bereich).} \beq R_{exp}=\frac{\Gamma(\pi^{+}\rightarrow
\overline{e}\nu)}{\Gamma(\pi^{+}\rightarrow
\overline{\mu}\nu)}=(1,230\pm0,008)\times10^{-4}\; , \eeq und
steht damit innerhalb der Fehlergrenzen in Übereinstimmung mit der
theoretischen Vorhersage.
\section {LQ-Wechselwirkungen}
\subsection {$\pi^{+}\rightarrow\overline{e}\nu_e$ und $\pi^{+}\rightarrow\overline{\mu}\nu_\mu$}
Die pseudoskalare Wechselwirkung durch den Vertex (2.4) liefert
den größten Beitrag zur gesamten LQ-induzierten WW, da die
Helizitätsunterdrückung fehlt. Die anderen Beiträge werden daher
zunächst vernachlässigt. Wenn man aus der Zerfallsrate (\ref{dec})
den Beitrag $I_{SM-LQ_2}$ der Interferenz zwischen Standardmodell
und den als vergleichsweise klein angenommenen LQ-induzierten
pseudoskalaren Operatoren (2.31) benutzt und seinen Beitrag zu $R$
berechnet, erhält man\footnote{Indizierung: Elektron -
$m=j=n=i=1$; Myon - $m=i=2, j=n=1$. Es wird angenommen, dass
$V_{ud}$ reell ist.}: \bea
R_{SM+I_{SM-LQ_2}}&=&\frac{k_e}{k_\mu}\frac{(|\mathcal{M}_{SM}|^2
+ 2\Re(\mathcal{M}^*_{SM}\mathcal{M}_2) )_{\pi^{+}\rightarrow
\overline{e}\nu_e}}{(|\mathcal{M}_{SM}|^2 +
2\Re(\mathcal{M}^*_{SM}\mathcal{M}_2) )_{\pi^{+}\rightarrow
\overline{\mu}\nu_\mu}}{}\\{}&=&
-\frac{\widetilde{P}}{\widetilde{A}}\left(\frac{\Re(\lambda_{L}^{11}\lambda_{R}^{*11})}
{\sqrt{2}G_{F}V_{ud}m_{LQ}^{2}}
\frac{1}{m_{e}}-\frac{\Re(\lambda_{L}^{21}\lambda_{R}^{*21})}{\sqrt{2}
G_{F}V_{ud}m_{LQ}^{2}}\frac{1}{m_{\mu}}\right)R_{th}+R_{th} \;
.{}\nonumber\\{}\eea \\ Die Neutrinomassen wurden vernachlässigt.
Die Taylorentwicklung des Nenners wird nach dem linearen Term
abgebrochen, da die LQ-Kopplungen klein gegenüber dem Term aus der
schwachen Wechselwirkung sind.
\begin{description}
    \item[Anmerkung zum Vorzeichen in (3.6):] Gemäß Gl. (2.14)
    und Gl. (2.19) ist $$\tilde{A}_{\pi^-}=\pm i \frac{f_{\pi^-}}{2}
    \qquad\textrm{und}\qquad \tilde{P}_{\pi^-}=\mp i\frac{f_{\pi^-}}{2}\frac{m_{\pi^-}^2}{m_u+m_d}\; .$$ Es muss
    allerdings hier das unterschiedliche Vorzeichen der
    Projektionsoperatoren $P_L$ bzw. $P_R$ beachtet werden, was zu
    einem relativen Minuszeichen zwischen $\tilde{A}$ und
    $\tilde{P}$ führt, d.h. zu dem positiven Vorfaktor
    $-(\widetilde{P}/\widetilde{A})$ .
\end{description}
Der gesamte Beitrag zu $R$ muss kleiner sein als es die
Abweichungen zwischen SM und Experiment innerhalb der
Fehlergrenzen zulassen. Nimmt man an, dass jeweils nur zwei
LQ-Kopplungskonstanten nicht verschwinden, erhält man folgende
Bedingungen:
\begin{equation}-\frac{\widetilde{P}}{\widetilde{A}}\left(\frac{\Re(\lambda_{L}^{11}\lambda_{R}^{*11})}
{\sqrt{2}G_{F}V_{ud}m_{LQ}^{2}}\frac{1}{m_{e}}\right)
<\frac{R_{exp}}{R_{th}}-1+\Delta\left(\frac{R_{exp}}{R_{th}}\right)\equiv
R_{max}=2,1\times 10^{-3}\; ,\end{equation} und
\begin{equation}-\frac{\widetilde{P}}{\widetilde{A}}\left(\frac{\Re(\lambda_{L}^{21}\lambda_{R}^{*21})}
{\sqrt{2}G_{F}V_{ud}m_{LQ}^{2}}\frac{1}{m_{\mu}}\right)<-\left(\frac{R_{exp}}{R_{th}}-1-\Delta
\left(\frac{R_{exp}}{R_{th}}\right)\right)\equiv R_{min}=1,1\times
10^{-2}\; ,\end{equation}wobei, unter Verwendung von
$\frac{\Delta(R_{th})} {R_{th}}\ll \frac{\Delta(R_{exp})}
{R_{exp}}$ ,
\beq\Delta\left(\frac{R_{exp}}{R_{th}}\right)=\frac{R_{exp}}{R_{th}}\sqrt{\left(\frac{\Delta(R_{exp})}
{R_{exp}}\right)^2+\left(\frac{\Delta(R_{th})}{R_{th}}\right)^2}\simeq
\frac{\Delta(R_{exp})}{R_{th}}=6,5\times 10^{-3}\; .\eeq \\Damit
gelten folgende Einschränkungen für die
Kopplungskonstantenprodukte:
\begin{equation}\Re(\lambda_{L}^{11}\lambda_{R}^{*11})<R_{max}\frac{\sqrt{2}G_{F}V_{ud}
m_{LQ}^{2}m_{e}}{m_{\pi}^2}(m_u+m_d)=9,8\times10^{-8}\left(\frac{m_{LQ}}{100GeV}\right)^{2}\;
,\end{equation} und
\begin{equation}\Re(\lambda_{L}^{21}\lambda_{R}^{*21})<R_{min}\frac{\sqrt{2}G_{F}V_{ud}
m_{LQ}^{2}m_{\mu}}{m_{\pi}^2}(m_u+m_d)=1,0\times10^{-4}\left(\frac{m_{LQ}}{100GeV}\right)^{2}\;
.\end{equation}
Der relativ große Unterschied zwischen diesen beiden Werten
resultiert aus dem Unterschied in den Leptonenmassen
($m_e/m_\mu\simeq 0,005$) und dem Verhältnis der $R$-Werte,
$R_{max}/R_{min}\simeq 0,20$. Die verwendeten Konstanten können
der Tabelle 3 des Anhangs entnommen werden. \\ \\ Der Wert aus Gl.
(3.11) stimmt mit dem in \cite{lepto}
 berechneten Wert überein, der Wert aus Gl. (3.10) ist um etwas weniger
 als eine Größenordnung niedriger und stimmt mit dem aus Gl.
 (3.13) überein. In \cite{lepto} wurde die Differenz zwischen $R_{min}$ und
 $R_{max}$ vernachlässigt und nur mit $R_{min}$ gearbeitet.
\\
\\Die LQs, die zu diesen Termen beitragen können (aus den
Tabellen 4 und 5 des Anhangs ersichtlich), sind $V_{0}$ und
$V_{1/2}$ ( sowie $S_{1/2}$ und $S_{0}$ mit der Ersetzung 2.7),
d.h.:
\begin{equation}\Re(\lambda_{LV_{0}}^{11}\lambda_{RV_{0}}^{*11})<
9,8\times10^{-8}\left(\frac{m_{LQ}}{100GeV}\right)^{2} \;
,\label{bound2}\end{equation}
und\begin{equation}\Re(\lambda_{LV_{1/2}}^{11}\lambda_{RV_{1/2}}^{*11})<
5,0\times10^{-7}\left(\frac{m_{LQ}}{100GeV}\right)^{2}\;
.\end{equation} Der Unterschied in den beiden Werten resultiert
aus dem unterschiedlichen Vorzeichen der 4-Fermionen-Vertizes
(siehe Tabelle 4 des Anhangs). Um den Wert (3.13) zu erhalten muss
in Gl. (3.10) $R_{max}$ durch $R_{min}$ ersetzt werden. Für
$\lambda_{LV_{0}}^{21}\lambda_{RV_{0}}^{*21}$und
$\lambda_{LV_{1/2}}^{21}\lambda_{RV_{1/2}}^{*21}$ ist Gl. (3.11)
zu verwenden (wieder unter Berücksichtigung der Vorzeichen). \\
\\Aus den Termen (2.27) (SM-Term) und (2.32) kann man
ebenfalls Schranken ableiten, wenn man annimmt, dass die
pseudoskalaren Terme verschwinden. Dieses Verbot von Kopplungen
der Art $\lambda_L\lambda_R$ kann durch eine zur $R$-Parität
analoge Symmetrie erzielt werden. Die Terme (2.27) und (2.32)
werden entsprechend in (3.5) eingesetzt und man erhält (mit
gleichem $R_{max}$, $R_{min}$) analog zu Gl. (3.7) und Gl. (3.8):
\begin{equation}\frac{1}{\sqrt{2}G_{F}V_{ud}m_{LQ}^{2}}\Re(\lambda_{L}^{11}
\lambda_{L}^{*11})<R_{max}\; ,\end{equation} und
\begin{equation}\frac{1}{\sqrt{2}G_{F}V_{ud}m_{LQ}^{2}}\Re(\lambda_{L}^{21}\lambda_{L}^{*21})
<R_{min}\; , \end{equation}woraus sich ergibt:\begin{equation}
\Re(\lambda_{L}^{11}\lambda_{L}^{*11})<3,4\times10^{-4}\left(\frac{m_{LQ}}{100GeV}\right)^{2}\;
,\label{bound1}
\end{equation}bzw.:
\begin{equation}|\lambda_{L}^{11}|<1,8\times10^{-2}\left(\frac{m_{LQ}}{100GeV}\right)\;
,\label{bound2}
\end{equation}und \beq|\lambda_{L}^{21}|<4,2\times10^{-2}\left(\frac{m_{LQ}}{100GeV}\right)\; .\eeq
Die hier wechselwirkenden LQs sind $V_{0}$ und $V_{1}$ (bzw.
$S_{0}$ und $S_{1}$ unter Verwendung von Gl. 2.7 und Beachtung der
Vorzeichen). Im Vergleich mit \cite{lepto} wirkt sich auch hier
die Verwendung unterschiedlicher $R_{max}$ und $R_{min}$ aus. Die
Schranke (3.16) ist niedriger, stimmt aber bei Verwendung von
$R_{min}$ statt $R_{max}$
mit \cite{lepto} überein.\\ \\
Im Folgenden werden die wechselwirkenden LQs nicht mehr
aufgeführt; dies kann Tabelle 6 des Anhangs entnommen werden. Es
werden nur noch Schranken wie in Gl. (3.16) angegeben.
\subsection {$\pi^{+}\rightarrow\overline{\mu}\nu_{e}$}Der Zerfall
$\pi^{+}\rightarrow\overline{\mu}\nu_{e}$, bei dem $L$ verletzt
wird, ist im SM verboten. Unter Verwendung von
$BR(\pi^{+}\rightarrow\overline{\mu}\nu_{e})<8\times 10^{-3}$
\cite{pdg} ergibt sich: \beq
\Gamma(\pi^{+}\rightarrow\overline{\mu}\nu_{e})<8\times10^{-3}\tau_\pi^{-1}=2,0\times
10^{-16}\; MeV\; . \eeq \pagebreak \\ Daraus werden Schranken an
mögliche LQ-Wechselwirkungen berechnet, die diesen Zerfall
ermöglichen:
\begin{itemize} \item Aus dem ($S\pm P$)-Term (das ist der reine LQ-Term,
 der sich aus Gl. 2.29 ergibt) der Zerfallsrate
$\Gamma_M$ (Gl. 2.33) erhält man die Bedingung
\begin{equation}\frac{|\lambda_{L}^{11}\lambda_{R}^{*21}|^2}{m_{LQ}^4}\frac{m_\pi
f_\pi^2(m_\pi^2-m_\mu^2)^2}{64\pi(m_u+m_d)^2}<\Gamma(\pi^{+}\rightarrow\overline{\mu}\nu_{e})\;
,\end{equation} \item und aus dem ($V\pm A$)-Term (der reine
LQ-Term, der aus Gl. 2.28 hervorgeht) der Zerfallsrate $\Gamma_M$
(Gl. 2.33) erhält man
\begin{equation}\frac{|\lambda_{L}^{11}\lambda_{L}^{*21}|^2}{m_{LQ}^4}\frac{f_\pi^2m_\mu^2
(m_\pi^2-m_\mu^2)^2}{64\pi m_\pi^3}
<\Gamma(\pi^{+}\rightarrow\overline{\mu}\nu_{e})\;
.\label{bound5}\end{equation}\end{itemize} Für die
Kopplungskonstantenprodukte ergibt sich damit: \beq
|\lambda_{L}^{11}\lambda_{R}^{*21}|<1,7\times10^{-3}\left(\frac{m_{LQ}}{100GeV}\right)^{2}\;
,\quad \textrm{und}\quad|\lambda_{L}^{11}\lambda_{L}^{*21}|<
2,9\times10^{-2}\left(\frac{m_{LQ}}{100GeV}\right)^{2}\; .\eeq Mit
der Zerfallskonstante $f_\pi = (130,7\pm 0,4) MeV$ zu rechnen,
verursacht keine großen Ungenauigkeiten. Der größte Fehler ist in
den Werten für die Quarkmassen (siehe Tabelle 2 des Anhangs) zu
finden. Diese Schranken für die Produkte von Kopplungskonstanten
sind nicht sehr niedrig und aus den in Kapitel 3.2.1 berechneten
Werten, Gln. (3.17) und (3.18), besser ableitbar:
\beq|\lambda_{L}^{11}\lambda_{L}^{*21}|\leq
|\lambda_{L}^{11}||\lambda_{L}^{21}|<7,7\times10^{-4}\left(\frac{m_{LQ}}{100GeV}\right)^{2}\;
.\label{bound10} \eeq In \cite{lepto} wurde statt mit
$\tau_\pi^{-1}$ in Gl. (3.19) mit
$\Gamma(\pi^+\rightarrow\mu\nu_\mu)$ gerechnet. Aufgrund der in
Gl. (3.20) durch $\widetilde{P}$ auftretenden Quarkmassen kommt
der Hauptteil der Differenz zwischen dem hier berechneten Wert zu
\cite{lepto} zustande (dabei wurde die tabellarische Auflistung
aus \cite{lepto} verwendet, der im Text angegebene Wert weicht
davon um mehr als die Hälfte nach unten ab). Sie beträgt etwa
10\%. In \cite{lepto} wurde $\widetilde{P}=7f_\pi m_\pi$
verwendet, was mit dem aktuellen Wert für $f_\pi$ die erwähnte
Abweichung von etwa 10\% des hier verwendeten Wertes nach oben
erklärt. Die Schranke für $|\lambda_{L}^{11}\lambda_{L}^{*21}|$
stimmt mit der aus \cite{lepto} überein.\\ \\Die Gln. (3.20) und
(3.21) lassen sich auch auf die Zerfälle $\pi^+\rightarrow
e\nu_\tau$, $\pi^+\rightarrow e\nu_\mu$ (mit $m_e$ statt $m_\mu$)
und $\pi^+\rightarrow \mu\nu_\tau$ anwenden.
\begin{center}---------------------------------------------------------------------\end{center}
Das neutrale Pion zerfällt zu $(98,798\pm 0,032)$ \% \cite {pdg}
durch die elektromagnetische Wechselwirkung in zwei Photonen.
Damit hat es eine um acht Größenordnungen geringere Lebensdauer
als die über schwache Wechselwirkung zerfallenden, geladenen
Pionen. Da die Wurzel der Lebensdauer in die Schranken an die
Kopplungskonstantenprodukte eingeht können hieraus keine starken
Bedingungen abgeleitet werden. Dies gilt insbesondere für den
Zerfall $\pi_0 \rightarrow \mu e$, aus dem man gute Schranken für
nicht verschwindende LQD- und LLE-Terme erwarten könnte (siehe
\cite{thor}).\newpage
\section {Squark- und Slepton- Wechselwirkungen}\label {sec:
R} \vspace{20pt}
\begin{figure}[h]
\begin{center}
\begin{fmffile}{tau206}
\begin{fmfgraph*}(150,100)
 \fmfpen{thin}\fmftop{i1,i2}
\fmfbottom{o1,o2}\fmf{fermion}{v1,i1}\fmf{fermion}{v1,i2}\fmf{scalar,label=$\tilde{d}_R^k$}{v2,v1}
\fmf{fermion}{o2,v2}\fmf{fermion}{o1,v2}
\fmflabel{u}{i1}\fmflabel{$\ell$}{i2}\fmflabel{d}{o1}\fmflabel{$\nu$}{o2}\fmfdotn{v}{2}
\end{fmfgraph*}\hspace{20pt}
\begin{fmfgraph*}(200,100)
 \fmfpen{thin}\fmfleft{i1,i2}
\fmfright{o1,o2}\fmf{fermion}{v1,i1}\fmf{fermion}{i2,v1}\fmf{scalar,label=$\tilde{l}_L^k$}{v1,v2}
\fmf{fermion}{o1,v2}\fmf{fermion}{v2,o2}
\fmflabel{u}{i1}\fmflabel{d}{i2}\fmflabel{$\ell$}{o2}\fmflabel{$\nu$}{o1}\fmfdotn{v}{2}
\end{fmfgraph*}
\end{fmffile}\end {center}\caption[Squark- und Slepton-Wechselwirkung im Pionenzerfall]
{\textit{Squark- und Slepton-Wechselwirkung im
Pionenzerfall}}\label{squark}\end{figure} Wie in \cite{han} und
ausführlicher in \cite{thor} durchgeführt, kann man auch die
$R$-paritätsverletzenden Beiträge zum Pionenzerfall durch Squarks
und Sleptonen berechnen. Die dafür relevanten Terme aus dem
Superpotential sind $L_{L}Q_{L}\overline{D}_{R}$ und
$L_{L}L_{L}\overline{E}_{R}$. Die daraus folgenden
Kopplungskonstantenprodukte lassen sich einfach aus den
Berechnungen für LQs extrahieren; es handelt sich gewissermaßen um
einen Spezialfall der obigen Berechnungen. In vier-komponentiger
Dirac-Schreibweise erhält man für den Lagrangian aus dem LQD- und
dem LLE- Term:
\beq\mathcal{L}_{LQD}=\lambda_{ijk}'[\tilde{\nu}^i_L\bar{d}^k_Rd^j_L+
\tilde{d}^j_L\bar{d}^k_R\nu^i_L+(\tilde{d}^k_R)^*(\bar{\nu}^i_L)^cd^j_L
-\tilde{e}^i_L\bar{d}^k_Ru^j_L-
\tilde{u}^j_L\bar{d}^k_Re^i_L-(\tilde{d}^k_R)^*(\bar{e}^i_L)^cu^j_L]+h.c.\;
.\eeq
\beq\mathcal{L}_{LLE}=\lambda_{ijk}[\tilde{\nu}^i_L\bar{e}^k_Re^j_L+
\tilde{e}^j_L\bar{e}^k_R\nu^i_L+(\tilde{e}^k_R)^*(\bar{\nu}^i_L)^ce^j_L-(i\leftrightarrow
j)]+h.c.\eeq Zum Pionenzerfall können zwei Prozesse beitragen,
deren 4-Fermionen-Vertizes so aussehen:
\beq\mathcal{L}_{\not{R}_P1}=\frac{1}{2}\frac{\lambda'{}_{ijk}^*\lambda'{}_{mnk}}
{m^2_{\tilde{d}_R^k}}[\bar{\ell}^i\gamma^\mu
P_L\nu^m][\bar{u}^j\gamma_\mu P_L d^n]\; ,\eeq und \beq
\mathcal{L}_{\not{R}_P2}=-\frac{\lambda'{}_{kjn}^*\lambda_{mki}}{m^2_{\tilde{l}_L^k}}[\bar{l}^i
P_L\nu^m][\bar{u}^j P_R d^n ]\; .\eeq Beide Terme finden sich (bis
auf Konstanten) auch bei den 4-Fermionen-Vertizes der LQs (siehe
Tabelle 5 des Anhangs). Aus Gl. (\ref{bound2}) und Gl. (3.18)
lassen sich daher direkt folgende Schranken ableiten: \beq
 |\lambda'_{11k}|<0,026\left(\frac{m_{\tilde{d}_R^k}}{100GeV}\right)\;
,\eeq\beq
 |\lambda'_{21k}|<0,059\left(\frac{m_{\tilde{d}_R^k}}{100GeV}\right)\;
,\eeq und\beq|\lambda'_{11k}\lambda'_{21k}|<1,5\cdot
10^{-3}\left(\frac{m_{\tilde{d}_R^k}}{100GeV}\right)^2\; .\eeq
Dabei mussten die berechneten Kopplungen $\lambda$ noch durch
$\sqrt{2}$ geteilt werden um die entsprechenden Squark-Kopplungen
zu erhalten (Faktor 1/2 in Gl. 3.26). Diese Schranken wurden in
\cite {han} und \cite{thor} ebenfalls berechnet; die Ergebnisse
stimmen, abgesehen von
marginalen Abweichungen (die Werte aus \cite {han} wurden leicht verbessert) aufgrund
neuerer experimenteller Werte, überein.\\ \\
Das Pion kann auch durch Slepton-Austausch (LLE- \textbf{und} LQD-
Term des Superpotentials) zerfallen (siehe Abbildung
\ref{squark}). Zur Berechnung von Schranken an
$\Re(\lambda'_{k11}\lambda_{1k1}^*)$ und
$\Re(\lambda'_{k11}\lambda_{2k2}^*)$ werden die Gln. (3.10) und
(3.11) benutzt. Es muss hier folgende Modifikation vorgenommen
werden: $R_{max}$ wird durch $R_{min}$ ersetzt und umgekehrt (dies
folgt aus dem Vorzeichen in Gl. 3.27). Es ergibt sich:
\begin{equation}\Re(\lambda'_{k11}\lambda_{1k1}^*)<5,0\times10^{-7}\left(\frac{m_{\widetilde{l}^{k}_{R}}}{100GeV}
\right)\; ,\end{equation} und mit analoger Rechnung aus (3.11):
\begin{equation}\Re(\lambda'_{k11}\lambda_{2k2}^*)<2,0\times10^{-5}\left(\frac{m_{\widetilde{l}^{k}_{R}}}{100GeV}
\right)\; .\end{equation} Die Berechnung dieser
Kopplungskonstantenprodukte wurde ebenfalls in \cite{thor}
durchgeführt, die Ergebnisse stimmen in etwa überein. Weiterhin
ergeben sich aus Kapitel 3.2.2 folgende Schranken:
\beq|\lambda'_{k11}\lambda_{1k2}^*|<1,7\times10^{-3}\left(\frac{m_{\widetilde{l}^{k}_{R}}}{100GeV}\right)\;
,\eeq und
\beq|\lambda'_{21k}\lambda'{}_{11k}^*|<5,8\times10^{-2}\left(\frac{m_{\tilde{d}^k_R}}{100GeV}\right)\;.\eeq
Die Schranke an das Kopplungskonstantenprodukt in (3.33) kann
besser aus Leptonenzerfällen gewonnen werden (siehe \cite{alla}).
Die Schranke (3.34) stimmt mit der aus \cite{thor} überein, das
Produkt der einzelnen Kopplungskonstanten (Gl. 3.23) liefert
allerdings eine stärkere Einschränkung:
\beq|\lambda'_{21k}\lambda'{}_{11k}^*|<1,5\times10^{-3}\left(\frac{m_{\tilde{d}^k_R}}{100GeV}\right)\;
.\eeq
\chapter {Der $K$-Mesonenzerfall}
\section {Zerfälle geladener Kaonen}
Die geladenen Kaonen ($K^{+}=u\overline{s}$ und
$K^{-}=\overline{u}s$) zerfallen zu (63,51$\pm$0,18)\% in
$\overline{\mu}\nu_{\mu}$ (bzw. $\mu\overline{\nu}_{\mu}$).
Daneben existieren zahlreiche Zerfallskanäle, hauptsächlich solche
mit einem oder mehreren Pionen im Endzustand.
\subsection {Leptonische Zerfälle}
Wie im Fall der Pionen ist auch bei den Kaonen der Zerfall in
$e\nu_{e}$ im SM durch die Drehimpulserhaltung stark unterdrückt.
Die Rechnungen sind analog zum Pionenzerfall: Die Gln. (3.7) und
(3.8) bzw. (3.14) und (3.15) mit den entsprechenden Ersetzungen
für den $K$-Zerfall. Die Quarkströme können Kapitel 2.3.2 mit den
entsprechenden Ersetzungen für das Kaon entnommen werden. Daraus
ergeben sich obere Schranken an
$\Re(\lambda_{L}^{12}\lambda_{R}^{11*})$ und
$\Re(\lambda_{L}^{22}\lambda_{R}^{21}{}^*)$ (LQs $V_{0}$, und
$V_{1/2}$ unter Berücksichtigung des Vorzeichens) und an
$\Re(\lambda_{L}^{12}\lambda_{L}^{11}{}^*)$ und
$\Re(\lambda_{L}^{22}\lambda_{L}^{21}{}^*)$ (LQs $V_{0}$ und
$V_{1}$), aus denen wiederum die entsprechenden Schranken an
SUSY-Kopplungskonstantenprodukte abgeleitet werden können. Aus
\cite{pdg} und \cite{fink} hat man: \beq R_{exp}=(2,44\pm
0,22)\cdot 10^{-5}\; ,\quad \textrm{und} \quad R_{th}=(2,472\pm
0,002)\cdot 10^{-5}\; ,\eeq die übrigen Größen können Tabelle 2
des Anhangs entnommen werden. Für die Kopplungskonstantenprodukte
erhält man damit:
\begin{equation}\Re(\lambda_{L}^{12}\lambda_{R}^{11}{}^*)<9,2\times10^{-7}\left(\frac{m_{LQ}}{100GeV}\right)^{2}\; ,
\qquad\Re(\lambda'_{k12}\lambda_{1k1}^*)<1,2\times10^{-6}\left(\frac{m_{\tilde{l}^k_R}}{100GeV}\right)^{2}\;
,\end{equation}
\begin{equation}\Re(\lambda_{L}^{22}\lambda_{R}^{21}{}^*)<2,6\times10^{-4}\left(\frac{m_{LQ}}{100GeV}\right)^{2}\; ,
\qquad\Re(\lambda'_{k12}\lambda_{2k2}^*)<1,9\times10^{-4}\left(\frac{m_{\tilde{l}^k_R}}{100GeV}\right)^{2}\;
,\end{equation}
\begin{equation}\Re(\lambda_{L}^{12}\lambda_{L}^{11}{}^*)<2,8\times10^{-3}\left(\frac{m_{LQ}}{100GeV}\right)^{2}\; ,
\qquad\Re(\lambda'_{11k}\lambda'{}_{12k}^*)<5,5\times10^{-3}\left(\frac{m_{\tilde{d}^k_R}}{100GeV}\right)^{2}\;
,\end{equation}
\begin{equation}\Re(\lambda_{L}^{22}\lambda_{L}^{21}{}^*)<3,7\times10^{-3}\left(\frac{m_{LQ}}{100GeV}\right)^{2}\; ,
\qquad\Re(\lambda'_{21k}\lambda'{}_{22k}^*)<7,4\times10^{-3}\left(\frac{m_{\tilde{d}^k_R}}{100GeV}\right)^{2}\;
.\end{equation}Die LQ-Schranken haben sich im Vergleich zu
\cite{lepto} leicht verändert, die ersten beiden Schranken sind
niedriger. Die Schranke aus Gl. (4.4) hat sich nicht verändert und
die Schranke Gl. (4.5) ist geringfügig schlechter. Dies kann nur
an den verwendeten $R$-Werten liegen, die benutzten Größen gehen
jedoch aus \cite{lepto} nicht hervor. Die SUSY-Schranken, Gl.
(4.2) und Gl. (4.3), stimmen in etwa mit denen aus \cite{thor}
überein, der Unterschied beruht auf den verwendeten Werten für die
Quarkmassen (die Schranken aus \cite{thor} sind etwas niedriger).
Hier wurden die von der \textit{particle data group} \cite{pdg}
angegebenen oberen Grenzen für die Quarkmassen eingesetzt (siehe
Tabelle 2 des Anhangs), um eine möglichst
konservative Abschätzung zu erhalten. \\
\\Auch beim $K$-Meson ist der $L$-verletzende Zerfall in ein Myon und
ein Elektronneutrino via LQ-Austausch möglich (siehe Kapitel
3.2.2). Laut \cite {pdg} ist:\beq \frac{\Gamma(K \rightarrow \mu
\nu_e)}{\Gamma(K \rightarrow \mu \nu_{\mu})}< 6,31 \times
10^{-3}\; . \eeq Durch die Verwendung dieses Verhältnis wird das
Erscheinen der Zerfallskonstante $f_K$ vermieden. Nun werden
einmal nur die pseudoskalaren LQ-Beiträge ($|\mathcal{M}_2|^2$,
siehe Gl. 2.29) und dazu separat nur die SM-ähnlichen LQ-Beiträge
($|\mathcal{M}_1|^2$, siehe Gl. 2.28) zur Zerfallsrate $\Gamma_M$
für $K \rightarrow \mu \nu_e$ betrachtet. Wird für den Zerfall $K
\rightarrow \mu \nu_{\mu}$ nur der SM-Term (für
$|\mathcal{M}_{SM}|^2$: siehe Gl. 2.27) berücksichtigt, so erhält
man für die beiden LQ-Beiträge zu $K \rightarrow \mu \nu_e$
folgende Bedingungen:
  \beq
  \frac{|\lambda_L^{12}\lambda_L^{21}{}^*|^2}{8 m_{LQ}^4 G_F^2|V_{us}|^2} ~, ~ \frac{|\lambda_L^{12}
  \lambda_R^{21}{}^*|^2m_K^4}{8m_{LQ}^4 G_F^2|V_{us}|^2 m_{\mu}^2(m_u+m_s)^2} < 6,31 \times
10^{-3}\; .
  \eeq
Daraus ergibt sich:
\begin{equation}|\lambda_{L}^{12}\lambda_{L}^{21}{}^*|<5,8\times10^{-3}\left(\frac{m_{LQ}}{100GeV}\right)^{2}\; ,
\qquad
|\lambda'_{11k}\lambda'{}_{22k}^*|<1,2\times10^{-2}\left(\frac{m_{\tilde{d}^k_R}}{100GeV}\right)^{2}\;
,\end{equation}
\begin{equation}|\lambda_{L}^{12}\lambda_{R}^{21}{}^*|<4,0\times10^{-4}\left(\frac{m_{LQ}}{100GeV}\right)^{2}\; ,
\qquad
|\lambda'_{k12}\lambda_{1k2}^*|<4,0\times10^{-4}\left(\frac{m_{\tilde{d}^k_R}}{100GeV}\right)^{2}\;
.\end{equation} Die LQ-Schranken haben sich im Vergleich zu
\cite{lepto} verbessert (beide haben sich etwa halbiert). Die
SUSY-Schranke (4.9) ist gegenüber \cite{thor} um etwas über die
Hälfte niedriger. Aus leptonischen $K$-Zerfällen berechenbare
Schranken an das Produkt $|\lambda'_{ijk}\lambda'{}_{lmn}^*|$ sind
jedoch deutlich schwächer als die in \cite {alla} aus
Leptonenzerfällen berechneten Schranken.
\subsection {Semileptonische Zerfälle} Das Kaon kann semileptonisch in ein
Pion und zwei geladene Leptonen zerfallen ($K\rightarrow
  \pi e\overline{e},\pi\mu\overline{\mu},\pi\mu\overline{e}$ und
  $\pi e\overline{\mu}$, siehe Abb. 4.1). Innerhalb des Standardmodells ist dies in niedrigster Ordnung nicht
  erlaubt (aufgrund der FCNC-Unterdrückung).\\ \\
  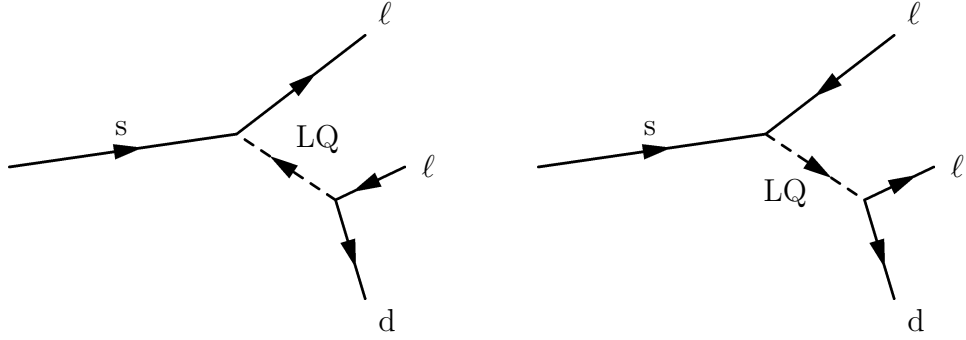
\begin{figure}[h]
\begin{center}
\begin{fmffile}{tau98}
\begin{fmfgraph*}(150,100)
\fmfpen{thin}\fmfleft{i1}\fmfright{o1,o2,o3}\fmf{fermion,label=s}{i1,v1}\fmf{fermion}{v1,o3}
\fmf{scalar,label=LQ}{v2,v1}\fmf{fermion}{o2,v2}\fmf{fermion}{v2,o1}\fmflabel{d}{o1}
\fmflabel{$\ell$}{o2}\fmflabel{$\ell$}{o3}
\end{fmfgraph*}\hspace{50pt}\begin{fmfgraph*}(150,100)
\fmfpen{thin}\fmfleft{i1}\fmfright{o1,o2,o3}\fmf{fermion,label=s}{i1,v1}\fmf{fermion}{o3,v1}
\fmf{scalar,label=LQ}{v1,v2}\fmf{fermion}{v2,o2}\fmf{fermion}{v2,o1}\fmflabel{d}{o1}\fmflabel{$\ell$}{o2}
\fmflabel{$\ell$}{o3}
\end{fmfgraph*}\end{fmffile}\end{center}\caption[$s$-Quarkzerfall via LQ-Austausch]{
\textit{$s$-Quarkzerfall via LQ-Austausch}}\label{tau100}\end{figure}
\\Zunächst soll nur eine Abschätzung der $|\lambda_L\lambda_L^*|$
-Schranke gewonnen werden, ehe auf pseudoskalare Beiträge
eingegangen wird, d.h. es wird zunächst angenommen, dass sämtliche
$(S\pm P)$-Beiträge verschwinden. Dazu wird von ungebrochener
Isospinsymmetrie ausgegangen: \beq
  \langle K^+ | \tilde{O} | \pi^o\rangle = \frac{1}{\sqrt{2}}  \langle K^+ | \tilde{O} |
\pi^+\rangle\; ,
  \eeq
  wobei $\tilde{O}$ ein Isospin-1/2-Operator ist. Laut \cite{pdg} ist \beq\frac{BR(K
\rightarrow \pi^+ \mu \bar{\mu})}{BR(K \rightarrow \pi^o \bar{\mu}
\nu_\mu)}<2,3\times 10^{-6},\;\;\;\textrm{und}\;\;\;\frac{BR(K
\rightarrow \pi^+ e \bar{e})}{BR(K \rightarrow \pi^o \bar{e}
\nu_e)}<5,9\times 10^{-6} .\eeq Die beiden Zerfallsraten lassen
sich als Produkt der Kopplungen, der Quarkströme und kinetischer
Terme $\mathcal{T}$ darstellen: \beq \Gamma(K \rightarrow \pi^+
\ell \bar{\ell})=
\frac{|\lambda_L^{i2}\lambda_L^{m1}{}^*|^2}{m_{LQ}^4}\langle
K^+|\bar{s}\gamma^\mu\gamma^5 d|\pi^+\rangle^2 \mathcal{T}_{K
\rightarrow \pi^+ \ell \bar{\ell}}\; ,\eeq und \beq \Gamma(K
\rightarrow \pi^o \bar{\ell} \nu)=8G_F^2|V_{su}|^2\langle
K^+|\bar{s}\gamma^\mu\gamma^5 u|\pi^0\rangle^2 \mathcal{T}_{K
\rightarrow \pi^o \bar{\ell} \nu}\; .\eeq Vernachlässigt man
sämtliche Leptonenmassen, was einen Fehler von weniger als einem
Faktor zwei bedeutet, da $BR(K \rightarrow \pi^o \bar{e} \nu)
\simeq 1.5 ~BR(K \rightarrow \pi^o \bar{\mu} \nu)$ \cite{pdg}, so
ist $\mathcal{T}_{K \rightarrow \pi^+ \ell \bar{\ell}}$ gleich
$\mathcal{T}_{K \rightarrow \pi^o \bar{\ell} \nu}$. Damit erhält
man unter Benutzung von Gl. (4.10) und Gl. (4.11):
  \beq
  \frac{|\lambda_L^{22}\lambda_L^{21}{}^*|^2}{ 4 G_F^2 |V_{su}|^2m_{LQ}^4} <2,3\times 10^{-6},\;\;\textrm{und}\;\;
   \frac{|\lambda_L^{12}\lambda_L^{11}{}^*|^2}{ 4 G_F^2 |V_{su}|^2m_{LQ}^4} < 5,9\times
   10^{-6}\; ,\eeq
  woraus sich ergibt:
\begin{equation}|\lambda_{L}^{22}\lambda_{L}^{21}{}^*|<7,8\times10^{-5}\left(\frac{m_{LQ}}{100GeV}
\right)^{2}\; ,\qquad
|\lambda'_{2k2}\lambda'{}_{2k1}^*|<1,6\times10^{-4}\left(\frac{m_{\tilde{u}^k_L}}{100GeV}\right)^{2}
\; ,\end{equation} und
\begin{equation}|\lambda_{L}^{12}\lambda_{L}^{11}{}^*|<1,5\times10^{-4}\left(\frac{m_{LQ}}{100GeV}
\right)^{2}\; ,\qquad
|\lambda'_{1k2}\lambda'{}_{1k1}^*|<3,0\times10^{-4}\left(\frac{m_{\tilde{u}^k_L}}{100GeV}\right)^{2}
\; .\end{equation} Die LQ-Schranke (4.15) beträgt etwa 25\% der in
 \cite{lepto} berechneten Schranke, (4.16) hat sich gegenüber \cite{lepto} etwa halbiert. \\
\\Abgesehen von diesem Beitrag, haben wir allerdings noch skalare
und tensorielle Beiträge. Diese tauchen nur bei den skalaren
Wechselwirkungen auf, sie werden dort durch die
Fierz-Transformation (Kapitel 2.3.1) generiert. Um sie zu
berücksichtigen, muss zunächst etwas Stromalgebra betrieben
werden.\\
\subsubsection{Stromalgebra beim semileptonischen K-Zerfall}  Mit \cite{weak} \beq\langle
\pi^0(p_\pi)|\bar{s}\gamma_\mu u|K^+(p_K)\rangle
=\frac{1}{\sqrt{2}}(f_+p_\mu +f_-q_\mu)\; ,\eeq wobei $f_+$ und
$f_-$ nur von $q^2$ abhängige Formfaktoren sind und \beq
p=p_K+p_\pi\; , \qquad q=p_K-p_\pi\; ,\eeq ergibt sich: \beq
\langle \pi^0(p_\pi)|\bar{s}u|K^+(p_K)\rangle =\frac{p\cdot
q}{m_u-m_s}\frac{f_0}{\sqrt{2}}\; . \eeq Dabei wurde Gl. (4.17)
mit $q^\mu$ mulitpliziert und die Dirac-Gl. angewendet. Außerdem
wurde der Formfaktor $f_0$ eingeführt: \beq f_0
=f_++\frac{q^2}{p\cdot q}f_-\; .\eeq Für das Tensor-Matrixelement
gilt allgemein \cite{form}: \beq \langle
\pi^0(p_\pi)|\bar{s}\sigma_{\mu\nu}u|K^+(p_K)\rangle
=\frac{-i}{\sqrt{2}}\mathcal{B}\cdot(p_\mu q_\nu -p_\nu q_\mu)\;
,\eeq mit
\beq\mathcal{B}=c_-(q^2)\frac{f_0(q^2)}{m_s-m_u}+(m_s+m_u)\frac{f_-(q^2)}{p\cdot
q}\;.\eeq $c_-=c_-(q^2)$ wird durch\beq\langle
\pi^0(p_\pi)|i\big[\bar{s}(\partial_\mu u)-(\partial_\mu\bar{s})
u\big]|K^+(p_K)\rangle = (c_+p_\mu+c_-q_\mu)\langle
\pi^0(p_\pi)|\bar{s}u|K^+(p_K)\rangle\eeq definiert, wobei
gilt:\beq c_+=-\frac{1}{p\cdot q}(m_s^2-m_u^2+c_-q^2)\; .\eeq
\subsubsection {Die invariante Amplitude $\mathcal{M}(K^+\rightarrow
\pi^0\ell^+\nu_\ell)$} Die invariante Amplitude unter
Berücksichtigung nicht verschwindender skalarer, vektorieller und
tensorieller Formfaktoren ist in allgemeiner Form \cite{stein}:
\beq\mathcal{M}(K^+\rightarrow
\pi^0\ell^+\nu_\ell)=G_FV_{su}\left[-(\bar{\nu}_L\gamma_\mu\ell_L)(f_+p^\mu
+f_-q^\mu)+2m_K(\bar{\nu}_L\ell_R)f_S+i\frac{f_T}{m_K}(\bar{\nu}_L\gamma_\mu\gamma_\nu\ell_R)p^\mu
q^\nu\right].\eeq Der skalare und der tensorielle Formfaktor sind
dabei so definiert:\beq
f_S=\frac{1}{\sqrt{2}m_K}\langle\pi^0(p_\pi)|\bar{s}
u|K^+(p_K)\rangle\;\; \textrm{und}\;\; f_T=-m_K\mathcal{B}\; .\eeq
\begin{description}\item[Anmerkung:] Im SM sind $f_S$
und $f_T$ in führender Ordnung gleich null, während dies bei
LQ-Kopplungen nicht der Fall sein muss. Durch die Kontraktion des
vektoriellen Leptonenstromes (1. Summand in Gln. 4.25) unter
Verwendung der Dirac-Gln. (die Neutrinomasse wird vernachlässigt)
erhält man $-f_-(\bar{\nu}_L\gamma_\mu\ell_L) q^\mu =
f_-m_\ell(\bar{\nu}_L\ell_R)$, und somit wieder einen skalaren
Term: \beq f_S^{SM}=\frac{m_\ell}{2m_K}f_-\; .\eeq
\end{description}
Um eine Abschätzung der skalaren und tensoriellen Beiträge zu
erhalten wird hier ein einfaches Verfahren gewählt (siehe nächster
Abschnitt): Es wird das Verhältnis der Formfaktoren
$\frac{f_S}{f_+}$ und $\frac{f_T}{f_+}$ gebildet, ohne die
Zerfallsrate $\Gamma_M$ konkret zu berechnen. Separat werden
jeweils nur die $f_S$ und die $f_T$ -Beiträge (übernächster
Abschnitt) betrachtet. Die Leptonenmassen werden dabei
vernachlässigt.\begin{description}
    \item[Einschub:]
Um die Zerfallsrate $\Gamma_M$ zu berechnen muss das Matrixelement
(4.25) quadriert werden: \bea
|\mathcal{M}|^2&=&G_F^2|V_{su}|^2\{|\mathcal{M}_{f_+\&f_-}|^2+|\mathcal{M}_{f_S}|^2+|\mathcal{M}_{f_T}|^2
{}\nonumber\\&&{}+2\Re(\mathcal{M}_{f_+\&f_-}^*\mathcal{M}_{f_S})+2\Re(\mathcal{M}_{f_+
\&f_-}^*\mathcal{M}_{f_T})+2\Re(\mathcal{M}_{f_S}^*\mathcal{M}_{f_T})\}.\eea
Die einzelnen Terme lassen sich mit der Dirac-Gln. und den
Spurregeln für Gamma-Matrizen berechnen ($p_\nu$ ist der
Viererimpuls des Neutrinos, für $p_\ell$ gilt Entsprechendes, die
Leptonenmassen werden vernachlässigt):
\bea|\mathcal{M}_{f_+\&f_-}|^2&=& \frac{1}{4}Tr[\not{p}_\nu
(f_+\not{p}+f_-\not{q})(1-\gamma^5)\not{p}_\ell
(1+\gamma^5)(f_+\not{p}+f_-\not{q})]{}\nonumber\\{}&=&
2f_+^2(2(p_\nu\cdot p)(p_\ell\cdot p)-p^2(p_\nu\cdot p_\ell))\;
,\eea \bea|\mathcal{M}_{f_S}|^2&=& 4m_K^2f_S^2\cdot
\frac{1}{4}Tr[\not{p}_\nu (1+\gamma^5)\not{p}_\ell
(1-\gamma^5)]=4m_K^2f_S^2q^2\; ,\eea $|\mathcal{M}_{f_T}|^2$,
$2\Re(\mathcal{M}_{f_+\&f_-}^*\mathcal{M}_{f_S})$,
$2\Re(\mathcal{M}_{f_+ \&f_-}^*\mathcal{M}_{f_T})$ und
$2\Re(\mathcal{M}_{f_S}^*\mathcal{M}_{f_T})$ sind analog
berechenbar. Diese Terme können gemäß Kapitel 2.3.3 im
dreidimensionalen Phasenraum integriert werden, um die
Zerfallsrate $\Gamma_M$ zu erhalten.\end{description}
\textbf{Zuerst der skalare Term}: Mit den Gln. (4.19), (4.20) und
(4.26) erhält man (bei gleichen Kopplungen im $f_S$ und im $f_+$
 -Faktor des Matrixelements):\beq\left|\frac{f_S}{f_+(0)}\right|=\frac{m^2_K-m^2_\pi}{2(m_s-m_u)m_K}\;
.\eeq Dabei wurden Terme proportional zu $q^2$ vernachlässigt.
Unter Verwendung von \cite{pdg}
\beq\left[\frac{f_S}{f_+(0)}\right]_{exp}=0,045\pm 0,033\eeq kann
nun eine Abschätzung der LQ-Kopplungen gewonnen werden, der
Standardmodellbeitrag $f_S^{SM}$ wird dabei vernachlässigt. Laut
\cite{pdg} ist $\frac{f_-(0)}{f_+(0)}=-0,096\pm 0,043$ und da
zusätzlich das Verhältnis $\frac{m_\ell}{m_K}$ eingeht, ist diese
Näherung gerechtfertigt.\\ \\Um die invariante Amplitude für
LQ-Wechselwirkungen zu erhalten muss in Gl.(4.25) $G_FV_{su}$
durch $\frac{\lambda_{LQ}\lambda_{LQ}^*}{2\sqrt{2}m_{LQ}^2}$
ersetzt werden. Für den ersten Term aus (4.25) tragen beide
$\lambda$s den Index L, beim $f_S$ und $f_T$ -Summanden ist die
Händigkeit unterschiedlich. Folglich kann man (mit Gl. 4.31)
fordern:
\beq\frac{|\lambda_L^{i2}\lambda_R^{m1}{}^*|}{m_{LQ}^2}\left(\frac{4G_FV_{su}}{\sqrt{2}}\right)^{-1}\frac{m^2_K-m^2_\pi}
{2(m_s-m_u)m_K}\leq\left[\frac{f_S}{f_+(0)}\right]_{exp} \; .\eeq
Daraus ergibt sich: \beq |\lambda_L^{i2}\lambda_R^{m1}{}^*|\leq
3,8\times 10^{-3}\left(\frac{m_{LQ}}{100GeV}\right)^2\; .\eeq In
\cite{form} wurde diese Rechnung ebenfalls durchgeführt. Die
Ergebnisse unterscheiden sich um einen Faktor 2, da in \cite{form}
der 4-Fermionen-Vertex der LQ-Wechselwirkung einen Faktor
$\frac{1}{2}$ im Vergleich zu Gl. (2.4) aufweist (dort wurden
skalare LQs betrachtet).
\\ \\ \\\textbf{Nun der tensorielle Term}: Dieser Term wurde in \cite{form} ausführlich
behandelt. Hier werden die Ergebnisse kurz zusammengefasst.
$c_-(q^2)$ (siehe Gl. 4.22) ist im Modell der Mesonendominanz
(siehe Abb. 4.2): \beq
c_-=\frac{f_+}{f_0}\frac{m_s-m_u}{m_{K^*}}-\frac{f_-}{f_0}\frac{m_s^2-m_u^2}{p\cdot
q}\; .\eeq Es wurde dabei der Beitrag angeregter K-Zustände ($K^*$
und $K_0^*$) zu $K^+\rightarrow \pi^0\ell^+\nu_\ell$ betrachtet.
\begin{figure}[t]
\begin{center}
\begin{fmffile}{tau782}
\begin{fmfgraph*}(250,100)
\fmfpen{thin}\fmfleft{i1}\fmfright{o1,o2}\fmf{vanilla}{i1,v1}\fmf{vanilla}{v1,o1}
\fmf{vanilla,label=$
K^*$}{v1,v2}\fmf{photon}{v2,o2}\fmflabel{$\pi$}{o1}\fmflabel{j}{o2}\fmflabel{K}{i1}\fmfdot{v1}\fmfdot{v2}
\end{fmfgraph*}\hspace{30pt}
\end{fmffile}\end{center}\caption[Semileptonischer K-Zerfall im Modell der Mesonendominanz]
{\textit{Beitrag angeregter K-Zustände zum hadronischen
Matrixelement des Stromes j (faktorisiert aus dem leptonischen
Teil des Zerfalls $K^+\rightarrow\pi^0\ell^+\nu_\ell$). Das Symbol
$K^*$ steht für $K^*,\, K_0^*(0^+)$.}}\label{tau}\end{figure}
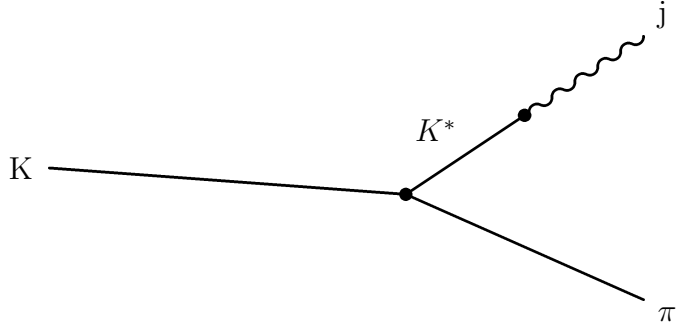 Man
erhält dabei für das hadronische Matrixelement: \beq \langle
\pi^0(p_\pi)|\bar{s}\sigma_{\mu\nu}u|K^+(p_K)\rangle
=\frac{-i}{\sqrt{2}}\frac{f_+(q^2)}{m_{K^*}}(p_\mu q_\nu -p_\nu
q_\mu)\approx\frac{-i}{\sqrt{2}}\frac{f_+(0)}{m_{K^*}}(p_\mu q_\nu
-p_\nu q_\mu)\; ,\eeq wobei im letzten Schritt die Abhängigkeit
von $f_+$ von $q^2$ vernachlässigt wurde. Aus den Gln. (4.26) und
(4.36) erhält man damit analog zur Berechnung der Schranke beim
skalaren Formfaktor (Gl. 4.33):
\beq\frac{|\lambda_L^{i2}\lambda_R^{m1}{}^*|}{m_{LQ}^2}\left(\frac{4G_FV_{su}}{\sqrt{2}}\right)^{-1}
\frac{m_{K}}{4m_K^*}\leq \left[\frac{f_T}{f_+(0)}\right]_{exp}\;
.\eeq Mit \cite{pdg}
\beq\frac{f_T}{f_+(0)}\left(K^+_{e3}\right)=0,31\pm 0,25\;
,\qquad\frac{f_T}{f_+(0)}\left(K^+_{\mu 3}\right)=0,02\pm 0,12\;
,\eeq folgt daraus:\beq |\lambda_L^{12}\lambda_R^{11}{}^*|\leq
0,29\left(\frac{m_{LQ}}{100GeV}\right)^2\; ,\qquad
|\lambda_L^{22}\lambda_R^{21}{}^*|\leq
0,074\left(\frac{m_{LQ}}{100GeV}\right)^2\; .\eeq Diese Schranken
sind schwächer als die Schranke aus Gl. (4.34). Sie werden daher
im Weiteren nicht mehr betrachtet. Sie stimmen, bis auf den
erwähnten Faktor zwei, mit den in \cite{form} berechneten
Schranken überein. Die aus den skalaren Beiträgen zu
semileptonischen K-Zerfällen berechenbaren Schranken an
$\not{R_P}$-Beiträge zur supersymmetrischen Lagrangefunktion sind
mit der Schranke aus Gl. (4.34) identisch: \beq|\lambda
'_{k12}\lambda_{ikj}^*|\leq 3,8\times
10^{-3}\left(\frac{m_{\tilde{l}_k}}{100GeV}\right)^2\; \eeq Diese
Schranke ist allerdings vergleichsweise schwach und aus anderen
Prozessen besser berechenbar (siehe \cite{alla} und \cite{thor} ).
\section {Zerfälle neutraler Kaonen} Wenn man die CP-Verletzung im neutralen $K$-System vernachlässigt,
dann sind die neutralen $CP$-Eigenzustände im Kaonensystem $K^0_L$
und $K^0_S$ ($K^0=\bar{s}d$): \beq
|K^0_S\rangle=\sqrt{\frac{1}{2}}\left(|K^0\rangle
+|\bar{K}^0\rangle\right)\; ,\qquad
|K^0_L\rangle=\sqrt{\frac{1}{2}}\left(|K^0\rangle
-|\bar{K}^0\rangle\right)\; .\eeq Im SM sind Zerfälle der Art
$K_{L}\rightarrow\mu\overline{\mu}, e\overline{e},
\mu\overline{e}$ über FCNC (und Verletzung von $L$) in niedrigster
Ordnung verboten (Letzterer ist in allen Ordnungen verboten),
wohingegen LQs diese Art von Zerfällen erlauben. Die
experimentellen Schranken an diese Zerfälle sind \cite{pdg}: \bea
&&BR(K_L \rightarrow e^{\pm}\mu^{\mp})<4,7\cdot 10^{-12}\;
,\;\;BR(K_L \rightarrow e\overline{e})<9^{+6}_{-4}\cdot 10^{-12}\;
, {}\nonumber\\&&{} BR(K_L \rightarrow \mu\overline{\mu})<(7,25\pm
0,16)\cdot 10^{-9}\; .\eea  Mit $\tau_K
  =5,17\cdot 10^{-8}s$ \cite{pdg} gilt für den Zerfall $\Gamma(K_L
\rightarrow e^{\pm} \mu^{\mp})$ (mit Gl. \ref{dec}, unter
ausschließlicher Berücksichtigung der $(V\pm A)$-Beiträge):
  \beq \Gamma(K_L \rightarrow e^{\pm} \mu^{\mp})_A=\frac{(m_K^2 -
m_{\mu}^2)^2}{64 \pi m_K^3}
\frac{|\lambda_L^{i2}\lambda_L^{*m1}|^2}{m_{LQ}^4} f_K^2 m_{\mu}^2
< 5,98 \times 10^{-26} {\rm ~MeV}\; . \label{nkao}
  \eeq
  Dabei ist: $i=(1;2)$ und $m=(2;1)$.
  Die pseudoskalaren Beiträge (unter Annahme einer möglichen, diese Beiträge verbietenden, Symmetrie)
  sowie die Elektronenmasse wurden
  vernachlässigt.
  Genauso lassen sich die Zerfälle in zwei Elektronen bzw. zwei
  Myonen mit Gl. (\ref{dec}) berechnen:
  \beq
\Gamma(K_L \rightarrow e\overline{e})_A=\frac{1}{32 \pi}
\frac{|\lambda_L^{12}\lambda_L^{*11}|^2}{m_{LQ}^4} f_K^2
m_{e}^2m_{K}<1,15\times 10^{-25}MeV\; .\eeq
  Aus diesen Überlegungen erhält man folgende Bedingungen an die
  Kopplungskonstantenprodukte:
  \beq
 K_L \rightarrow \mu^{\pm}e^{\mp}:\qquad|\lambda_{L}^{i2}\lambda_{L}^{*m1}|<9,7\times10^{-8}
 \left(  \frac{m_{LQ}}{100 {\rm ~GeV}}
\right)^2\; ,
  \eeq   \beq
 K_L \rightarrow e\overline{e}:\qquad|\lambda_{L}^{12}\lambda_{L}^{*11}|<1,9\times10^{-5}
 \left(  \frac{m_{LQ}}{100 {\rm ~GeV}}
\right)^2
  \; ,\eeq und \beq
 K_L \rightarrow \mu\overline{\mu}:\qquad|\lambda_{L}^{22}\lambda_{L}^{*21}|<2,7\times10^{-6}
 \left(  \frac{m_{LQ}}{100 {\rm ~GeV}}
\right)^2 \; .\eeq Im Vergleich zu \cite{lepto} haben sich die
Schranken leicht verbessert. \\ \\Die entsprechenden Schranken an
Produkte supersymmetrischer Kopplungskonstanten erhält man durch
Multiplikation der LQ-Schranken (Gln. 4.44-6) mit zwei, das
Austauschteilchen ist ein u-Squark:
  \beq
 K_L \rightarrow \mu^{\pm}e^{\mp}:\qquad|\lambda '_{1k(1/2)}\lambda '{}_{2k(2/1)}^*|<1,9\times10^{-7}
 \left(  \frac{m_{SUSY}}{100 {\rm ~GeV}}
\right)^2\; ,
  \eeq   \beq
 K_L \rightarrow e\overline{e}:\qquad|\lambda '_{2k1}\lambda '{}_{1k1}^*|<3,7\times10^{-5}
 \left(  \frac{m_{SUSY}}{100 {\rm ~GeV}}
\right)^2
  \; ,\eeq und \beq
 K_L \rightarrow \mu\overline{\mu}:\qquad|\lambda '_{2k2}\lambda '{}_{1k2}^*|<5,4\times10^{-6}
 \left(  \frac{m_{SUSY}}{100 {\rm ~GeV}}
\right)^2 \; .\eeq
Die Schranke (4.47) ist mit der in \cite{thor} berechneten Schranke vergleichbar.\\
\\\textbf{Nun wird der pseudoskalare Term P betrachtet}: Die
Berechnungen sind analog zu denen im vorangegangenen Abschnitt und
man erhält: \beq \Gamma(K_L \rightarrow e^{\pm} \mu^{\mp})_P=
\frac{|\lambda_L^{i2}\lambda_R^{*m1}|^2}{m_{LQ}^4} \frac{(m_K^2 -
m_{\mu}^2)^2f_K^2 m_{K}}{64 \pi(m_s+m_d)^2} < 5,98 \times 10^{-26}
{\rm ~MeV} \; ,\label{nkao1}
  \eeq und \beq \Gamma(K_L \rightarrow e\bar{e})_P= \frac{|\lambda_L^{12}\lambda_R^{*11}|^2}{m_{LQ}^4}\frac{
m_K^5f_K^2}{64 \pi(m_s+m_d)^2} <1,15 \times 10^{-25} {\rm ~MeV}
\label{nkao2}
  \; .\eeq Daraus erhält man wiederum Schranken an die
  Kopplungskonstantenprodukte:
\beq
 K_L \rightarrow \mu^{\pm}e^{\mp}:\qquad|\lambda_{L}^{i2}\lambda_{R}^{*m1}|\; ,\;
 |\lambda '_{k12}\lambda_{k12/k21}^*|<6,9\times10^{-9} \left(  \frac{m_{LQ/\tilde{\nu}_k}}{100 {\rm ~GeV}}
\right)^2\; ,
  \eeq   \beq
 K_L \rightarrow e\overline{e}:\qquad|\lambda_{L}^{12}\lambda_{R}^{*11}|\; ,\;
 |\lambda '_{k12}\lambda_{k11}^*|<9,1\times10^{-9} \left(  \frac{m_{LQ/\tilde{\nu}_k}}{100 {\rm ~GeV}}
\right)^2
  \; ,\eeq und \beq
 K_L \rightarrow \mu\overline{\mu}:\qquad|\lambda_{L}^{22}\lambda_{R}^{*21}|\; ,\;
 |\lambda '_{k12}\lambda_{k22}^*|<2,8\times10^{-7} \left(  \frac{m_{LQ/\tilde{\nu}_k}}{100 {\rm ~GeV}}
\right)^2 \; .\eeq Die LQ-Schranken haben sich deutlich gegenüber
\cite{lepto} verbessert (bei Gl. 4.54. um etwa eine
Grössenordnung). Die SUSY-Schranke (4.52) stimmt mit der Schranke
aus \cite{thor} überein.
\chapter{Der $D$-Mesonenzerfall}
$D$-Mesonen haben $C=\pm 1$ (C=\textit{charm}): $D^+=c\bar{d}$
($D^+_s=c\bar{s}$) und $D^0=c\bar{u}$ . Die geladenen $D$-Mesonen
haben sowohl (semi-)leptonische Zerfallskanäle, als auch rein
hadronische, wobei mit $(59\pm 7)\%$ \cite{pdg} der inklusive
Zerfall in $\bar{K}^0$ bzw. $K^0$ dominiert. Beim $D^0$ dominiert
der inklusive Zerfall in $K^-$ mit $(53\pm 4)\%$ \cite{pdg}.
\section {Zerfälle geladener $D$-Mesonen}
\begin{description}
    \item[Vorab:] Da $(V\pm A)$-LQ-Wechselwirkungen (Gl. 2.3) bis auf die unterschiedlichen Konstanten dem
     Standardmodellterm (Gl. 2.5) entsprechen, können unter der Annahme, dass die übrigen
LQ-Wechselwirkungen durch eine Symmetrie verboten sind, einfach
Schranken an die zugehörigen Kopplungskonstantenprodukte gewonnen
werden: \\Aus dem Streuprozess $\nu_\mu +d\rightarrow \mu +c$ kann
$V_{cd}$ bestimmt werden, aus $D^0\rightarrow K^-e^+\nu_e$ das
CKM-Matrixelement $V_{cs}$. Unter der einfachen (und
konservativen) Annahme, dass die LQ-Beiträge geringer sind, als
die SM-Beiträge, kann man daher für c, d, $\mu$ und $\nu_{\mu}$
-Vertizes fordern: \beq |\lambda_L^{21}\lambda_L^{22}{}^*| <
\frac{4 G_F}{\sqrt{2}} |V_{cd}| m_{LQ}^2= 7,4\times 10^{-2}
 \left(\frac{m_{LQ}}{100GeV}\right)^2\eeq und für c, s, e, $\nu_e$
 -Vertizes:
 \beq |\lambda_L^{22}\lambda_L^{22}{}^*| < \frac{4 G_F}{\sqrt{2}}
 |V_{cs}|m_{LQ}^2=3,3\times 10^{-1}\left(\frac{m_{LQ}}{100GeV}\right)^2
\; .\eeq Diese Schranken wurden ebenfalls in \cite{lepto}
berechnet, die Ergebnisse stimmen, abgesehen von einer
Verbesserung der Werte der CKM-Matrixelemente (dies führt zu
keiner starken Abweichung), überein und können für beliebige
Leptonenindizes verallgemeinert werden.\end{description} Beim
$D$-Meson ist, wie bei $\pi^{+}\rightarrow e^{+}\nu$, der Zerfall
$D^{+}\rightarrow \mu^{+}\nu$ durch die Drehimpulserhaltung
unterdrückt \cite{pdg}: \beq BR(D^{+}\rightarrow
\mu^{+}\nu)=(8^{+17}_{-5})\times 10^{-4}\; .\eeq Mit $\Gamma_M$
(Gl. \ref{dec}) sind die LQ-Beiträge (unter Verwendung der
Interferenzterme I zwischen SM und A($V\pm A$)- bzw. P($S\pm
P$)-LQ-Wechselwirkung und unter der Annahme, dass $V_{cd} $ reell
ist): \beq \Gamma (D^+\rightarrow \mu^+
\nu_\mu)_{A_I}=\Gamma_{SM}+m_\mu^2
(m_D^2-m_\mu^2)^2|\tilde{A}|^2G_FV_{cd}\frac{\sqrt{2}}{4\pi
m_D^3}\frac{\Re(\lambda_L\lambda_L^*)}{m_{LQ}^2}\; ,\eeq und\beq
\Gamma (D^+\rightarrow \mu^+ \nu_\mu)_{P_I}=\Gamma_{SM}+m_\mu
(m_D^2-m_\mu^2)^2\tilde{P}\tilde{A}^*G_FV_{cd}\frac{\sqrt{2}}{4\pi
m_D^3}\frac{\Re(\lambda_L\lambda_R^*)}{m_{LQ}^2}\; .\eeq Der
Interferenzterm (5.4) ist, wie der SM-Term, durch die Forderung der Drehimpulserhaltung unterdrückt.\\
\\
Für $BR_{SM}$ gilt (ohne Korrekturen höherer Ordnung,
Berechnungsgrundlage ist der reine SM-Term aus Gl. \ref{dec}) in
guter Übereinstimmung mit experimentellen Daten \cite{pdg}: \beq
BR_{SM}=8^{+14}_{-10}\times 10^{-4}\qquad
(BR_{exp}=8^{+17}_{-5}\times 10^{-4})\;\; .\eeq Die Lebensdauer
des $D$-Mesons kann Tabelle 2 des Anhangs entnommen werden. Der
Fehler des SM-Term wurde mit dem Gaußschen
Fehlerfortpflanzungsgesetz aus der Unsicherheit in $f_{D^+}$
(siehe Tabelle 2 des Anhangs) berechnet:
$\Delta_{SM}=BR_{SM}\frac{2\Delta f_{D^+}}{f_{D^+}}$. Eine obere
Schranke für die Größe des zweiten (LQ)-Summanden in den Gln.
(5.4) bzw. (5.5) ist daher:
\beq\Delta_{D^+}(max)=\frac{BR_{exp}}{\tau_{D^+}}-\frac{BR_{SM}}{\tau_{D^+}}+\frac{1}
{\tau_{D^+}}\sqrt{(\Delta_{SM}(max))^2+
(\Delta_{exp}(max))^2}=1,4\times 10^{-12}MeV\; .\eeq  Es wurden
jeweils die maximalen Abweichungen $\Delta_{SM}(max)$
($\Delta_{exp}(max)$) aus (5.6) eingesetzt, um eine möglichst
konservative Abschätzung zu erhalten. Für die LQ-Kopplungen
impliziert dies:
\beq\Re(\lambda_L^{21}\lambda_L^{22}{}^*)<1,0\times
10^{-1}\left(\frac{m_{LQ}}{100GeV}\right)^2\; ,\eeq und
\beq\Re(\lambda_L^{21}\lambda_R^{22}{}^*)<4,3\times
10^{-3}\left(\frac{m_{LQ}}{100GeV}\right)^2\; .\eeq Die Schranke
(5.9) kann auch aus der reinen LQ-Kopplung (der auf Gl. 2.29
basierende Teil von Gl. \ref{dec}) statt dem zweiten Summanden in
(5.5) abgeleitet werden:
\beq|\lambda_L^{21}\lambda_R^{22}{}^*|<\frac{8\sqrt{\pi}(m_c+m_d)}{f_D\sqrt{m_D}(m_D^2-m_\mu^2)}\sqrt{\Delta_{D^+}(max)
}m_{LQ}^2=5,2\times 10^{-3}\left(\frac{m_{LQ}}{100GeV}\right)^2\;
.\eeq Dieser Wert weicht nur geringfügig von dem aus Gl. (5.9) ab,
da der reine ($P\pm S$)-LQ-Term keine Unterdrückung durch die
Forderung der Drehimpulserhaltung aufweist. Auf eine kombinierte
Bestimmung des Kopplungskonstantenproduktes aus dem reinen LQ-Term
und dem Interferenzterm wird hier verzichtet. Gegenüber
\cite{lepto} liegen sämtliche Schranken geringfügig niedriger. Es
wurde hier jedoch eine wesentlich konservativere Fehlerabschätzung
vorgenommen (in \cite{lepto} wurde die damalige obere Schranke an
die Zerfallsrate für $D^+\rightarrow \mu^+\nu_\mu$ statt des hier verwendeten $\Delta_{D^+}(max)$ eingesetzt).\\
\\Für die Schranken an $\Re(\lambda '_{22k}\lambda '{}_{21k}^*)$ und
$|\lambda '_{k21}\lambda_{2k2}^*|$ verweise ich auf Tabelle 6 und
7 des Anhangs, die Schranke an Letzteres ist identisch mit der
zugehörigen LQ-Schranke (Gl. 5.10 bzw. 5.9). \\ \\Für das
$D_s^+$-Meson lassen sich auf verschiedene Weisen Schranken
ableiten: \begin{description}
    \item[Analog zum Pionenzerfall, Kapitel 3.2.1 :] Da beim $D_s^+$-Meson experimentelle
    Werte für $BR(D_s^+\rightarrow \mu\nu_\mu)$ und $BR(D_s^+\rightarrow
    \tau\nu_\tau)$ existieren ($BR(D_s^+\rightarrow \mu\nu_\mu)=(5,1\pm 1,9)\times 10^{-3}$ und $BR(D_s^+\rightarrow
    \tau\nu_\tau)=(6,4\pm 1,5)\% $, siehe \cite{pdg}), kann man
    auch analog zu Kapitel 3.2.1 vorgehen. Der aus Gl.
    (3.3) ableitbare SM-$R$-Wert ohne Korrekturen höherer Ordnung weicht um
    fast $30\%$ vom experimentellen R-Wert \cite{pdg}
    ab:\beq R_{th}=1,0\times 10^{-1}\quad \textrm{und}\quad R_{exp}=(8,0\pm 3,5)\times 10^{-2}\quad\; .\eeq
Die Standardabweichung von $R_{th}$ ist gegenüber der von
$R_{exp}$ vernachlässigbar. Analog
    zu den Gln. (3.10-11) und (3.14-15) erhält man für die Schranken an die Kopplungskonstantenprodukte:
    \bea \lefteqn{}\Re(\lambda_{L}^{22}\lambda^*{}_{R}^{22})<8,3\times
    10^{-4}\left(\frac{m_{LQ}}{100GeV}\right)^2 ,
    \quad \Re(\lambda_{L}^{32}\lambda^*{}_{R}^{32})<6,6\times 10^{-2}\left(\frac{m_{LQ}}{100GeV}\right)^2 ,\eea
 und \beq|\lambda_{L}^{22}|<0,14\left(\frac{m_{LQ}}{100GeV}\right)\; ,\;\;
 |\lambda_{L}^{32}|<0,30\left(\frac{m_{LQ}}{100GeV}\right)\; .\eeq
 Die zweite Schranke aus Gln. (5.12) ist aufgrund der hohen
 $\tau$-Masse, die linear eingeht, schwächer als die erste Schranke.
\item[Analog zu den $D^+$-Berechnungen:] Unter Verwendung der
entsprechenden Werte für das $D_s^+$-Meson können analog zu den
Gl. (5.8) und (5.9) Schranken an die entsprechenden
LQ-Kopplungskonstantenprodukte bestimmt werden. $BR_{SM}$ (ohne
Korrekturen höherer Ordnung) weicht allerdings um mehr als
$1\sigma_{exp}$ von $BR_{exp}$ \cite{pdg} ab: \beq
BR_{SM}=7,1\times 10^{-3}\;\;\textrm{und}\;\; BR_{exp}=(5,1\pm
1,9)\times 10^{-3}\; .\eeq Es wird daher mit der
$2\sigma_{exp}$-Fehlergrenze gearbeitet, d.h.:
\beq\Delta_{D^+_s}(max)=\frac{1}{\tau_{D_s^+}}(BR_{exp}-BR_{th}+2\sigma_{exp})=2,4\times
10^{-12}MeV\; .\eeq Für die Kopplungskonstantenprodukte erhält
man: \beq\Re(\lambda_L^{22}\lambda_L^{22}{}^*)<4,2\times
10^{-2}\left(\frac{m_{LQ}}{100GeV}\right)^2\; ,\qquad
 \Re(\lambda_L^{22}\lambda_R^{22}{}^*)<1,8\times
10^{-3}\left(\frac{m_{LQ}}{100GeV}\right)^2\; , \eeq und für
$D^+_s\rightarrow \tau\nu_\tau$ erhält man entsprechend:
\beq\Re(\lambda_L^{32}\lambda_L^{32}{}^*)<2,3\times
10^{-2}\left(\frac{m_{LQ}}{100GeV}\right)^2\; ,\qquad
 \Re(\lambda_L^{32}\lambda_R^{32}{}^*)<1,7\times
10^{-2}\left(\frac{m_{LQ}}{100GeV}\right)^2\; . \eeq Hier wurde
mit der $1\sigma_{exp}$-Standardabweichung gearbeitet, da
$BR_{SM}=6,9\% $ gut innerhalb der Fehlergrenzen des
experimentellen Wertes $BR_{exp}=(6,4\pm 1,5)\% $ \cite{pdg}
liegt.
\end{description}
Die aus diesen beiden Methoden berechneten Werte weichen leicht
voneinander ab: Die Werte aus (5.17) sind besser als die aus
(5.12-13), da dort die 2$\sigma_{exp}$-Fehlergrenze benutzt wurde.
Die Werte aus (5.16) sind schlechter als die entsprechenden Werte
aus (5.12-13). Dies gründet sich in dem Wert für $R_{min}$ (siehe
Gln. 3.7-8), der in (5.12-13) eingeht: Dieser ist aufgrund der
relativ starken Abweichung von $R_{th}$ und $R_{exp}$ (Gl. 5.11)
sehr hoch. Die Abweichung des theoretischen vom experimentellen
$BR$-Wert, die in (5.17) eingeht, ist vergleichsweise niedrig.\\ \\
Es soll aber noch besonders darauf hingewiesen werden, dass die
Zerfallskonstante $f_{D_s}$, mit ihrer hohen experimentellen
Ungenauigkeit, bei der Berechnung des Verhältnisses $R$ nicht
eingeht, da sie sich aus den Gleichungen herauskürzt. Diese
Schranken erhalten dadurch ein stärkeres Gewicht. Die relative
Übereinstimmung in den beiden Werten kommt dadurch zustande, dass
die Zerfallskonstante aus den betrachteten leptonischen Zerfällen
berechnet wird (Fehlerfortpflanzung der experimentellen Fehler in
den $BR$-Werten). Ein Fehler im $f_{D_s}$-Wert wird daher durch
den ebenfalls in die Berechnung von (5.16-17) eingehenden
$BR$-Wert (dieser fließt linear in $\Delta_{D_s^+}(max)$ ein), der
einen entsprechenden Fehler aufweist, in etwa kompensiert.
\section{semileptonische $D^+$-Zerfälle}Wie beim
Kaonenzerfall (Kapitel 4.1.2, Gln. 4.11 und 4.14) erhält man,
unter Verwendung der Isospin-Symmetrie, d.h.
 \beq \langle D^+| \bar{u}
\gamma^{\mu}P_L c|\pi^+\rangle ~= ~\langle D^o| \bar{d}
\gamma^{\mu}P_L c|\pi^-\rangle
  \; ,\eeq
ohne Berücksichtigung skalarer und pseudoskalarer Beiträge und
unter Vernachlässigung der Leptonenmassen Schranken an
LQ-Kopplungskonstantenprodukte. Laut \cite{pdg} ist: \bea
\lefteqn{}BR(D^+ \rightarrow \pi^+ \mu \bar{\mu})<1,5\times
10^{-5}\quad ,\quad  BR(D^+ \rightarrow \pi^+ e \bar{e})<5,2\times
10^{-5}\; ,{}\eea und \beq BR(D^0 \rightarrow \pi^- \nu_e
\bar{e})=(3,6\pm 0,6)\times 10^{-3}\; .\eeq Für das $D$-Meson
lässt sich analog zu Gl. (4.14) Folgendes ableiten:
  \beq
\frac{|\lambda_L^{\ell 2}\lambda_L^{\ell 1}{}^*|}{m_{LQ}^2}
\frac{\sqrt{2}}{4 G_F |V_{cd}|} =
\sqrt{\frac{\tau_{D^0}}{\tau_{D^+}}\frac{BR(D^+ \rightarrow \pi^+
\ell \bar{\ell})} {BR(D^0 \rightarrow \pi^- \nu \bar{e})}}\; ,
\eeq also:\beq
 |\lambda_L^{22}\lambda_L^{21}{}^*|< 3,0 \times 10^{-3} \left(
\frac{m_{LQ}}{100 {\rm ~GeV}}\right)^2\quad \textrm{und}\quad
|\lambda_L^{12}\lambda_L^{11}{}^*|< 5,6 \times 10^{-3} \left(
\frac{m_{LQ}}{100 {\rm ~GeV}}\right)^2\; .\eeq Diese Schranken
sind um etwa eine Größenordnung niedriger als die in \cite{lepto}
berechneten Schranken.\\ \\ Beiträge von über das SM
hinausgehenden Wechselwirkungen zu Zerfällen der Art
$D^\pm\rightarrow K\pi\ell^\pm \nu$ wurden in \cite{Dz}
betrachtet.
\section {Zerfälle neutraler $D$-Mesonen}
Wie im Fall der neutralen Kaonen können LQs auch beim
$D^{0}$-Meson Zerfälle über FCNC, die in niedrigster Ordnung im SM
verboten sind, verursachen. In den Formeln aus Kapitel 4.2 müssen
nur die notwendigen Veränderung für das $D$-Meson vorgenommen
werden und für die Kopplungskonstanten folgt (analog zu den Gln.
4.47-49): \beq
 |\lambda_L^{i2} \lambda_L^{*m1}| < 1,2 \times 10^{-2} \left(
\frac{m_{LQ}}{100 {\rm ~GeV}}\right)^2  \qquad(\textrm{für}\;\;
D^0 \rightarrow \mu \bar{e})\; ,
  \eeq\beq
 |\lambda_L^{22} \lambda_L^{*21}| < 6,3\times 10^{-3} \left(\frac{m_{LQ}}{100 {\rm ~GeV}}
\right)^2 \qquad(\textrm{für}\;\; D^0 \rightarrow \mu \bar{\mu})
  \; ,\eeq
  \beq
 |\lambda_L^{12} \lambda_L^{*11}| < 1,6  \qquad\quad\left(\frac{m_{LQ}}{100 {\rm ~GeV}}
\right)^2 \qquad(\textrm{für}\;\; D^0 \rightarrow e \bar{e})
  \; .\eeq
Die letzten beiden Schranken sind schwächer als in Gl. (5.22). Aus
den pseudoskalaren Beiträgen ergibt sich (analog zu Gl. 4.52-54):
\beq
 |\lambda_L^{i2}\lambda_R^{*m1}| < 5,3 \times 10^{-4} \left(
\frac{m_{LQ}}{100 {\rm ~GeV}}\right)^2  \qquad(\textrm{für}\;\;
D^0 \rightarrow \mu \bar{e})\; ,
  \eeq\beq
 |\lambda_L^{22} \lambda_R^{*21}| < 3,8\times 10^{-4} \left(\frac{m_{LQ}}{100 {\rm ~GeV}}
\right)^2 \qquad(\textrm{für}\;\; D^0 \rightarrow \mu \bar{\mu})
  \; ,\eeq
  \beq
  |\lambda_L^{12} \lambda_R ^{*11}|< 4,6 \times 10^{-4} \left(  \frac{m_{LQ}}{100 {\rm ~GeV}}
\right)^2  \qquad(\textrm{für}\;\; D^0 \rightarrow e \bar{e})
  \; .\eeq
Diese Schranken wurden auch in \cite{lepto} berechnet, die dort
berechneten Schranken sind durchschnittlich um etwa eine
Größenordnung schwächer als die hier aus aktuelleren
experimentellen Daten berechneten Werte.
\chapter{Der $B$-Mesonenzerfall}
Die $B$-Mesonen ($B^+=u\bar{b}$, $B^0=d\bar{b}$, $B^0_s=s\bar{b}$,
$B^+_c=c\bar{b}$) haben $B=\pm 1$. Die $b$-Physik ist ein
Schwerpunkt theoretischen Interesses, da hier durch
Präzisionsmessungen ein Einblick in Niedrigenergie-Prozesse
gewährt werden kann, der indirekte Effekte (LQ-Wechselwirkung) in
höheren Energieskalen offenbart. Zudem kann hier das SM der
Teilchenphysik auf seine Vorhersagekraft hin überprüft werden
\cite{babar}\footnote{In \cite{babar} findet sich eine
ausführliche Beschreibung des theoretischen Hintergrundes der
$B$-Physik.}.
\section{Zerfälle geladener $B$-Mesonen}
\subsection{Leptonische $B^+$-Zerfälle}
Für die Zerfälle $B^+\rightarrow e^+\nu_e$, $B^+\rightarrow
\mu^+\nu_\mu$ und $B^+\rightarrow \tau^+\nu_\tau$ existieren in
der Literatur \cite{pdg} nur obere Schranken: \beq
BR(B^+\rightarrow e^+\nu_e)<1,5\times 10^{-5} ,\;BR(B^+\rightarrow
\mu^+\nu_\mu)<2,1\times 10^{-5} , \; BR(B^+\rightarrow
\tau^+\nu_\tau)<5,7\times 10^{-4}. \eeq Aus dem reinen SM-Term von
Gl. (\ref{dec}) können theoretische Werte für $\Gamma_{SM}$ ohne
Korrekturen höherer Ordnung berechnet werden. Diese weichen -
abgesehen vom Zerfall $B^+\rightarrow e^+\nu_e$ - um mehrere
Größenordnungen von der aus (6.1) und $\tau_{B^+}$ (Tab. 2 des
Anhangs) berechenbaren oberen Grenze $\Gamma_{exp}$ nach oben ab.
Dies zeigt die Unzulänglichkeit dieser Berechnungen bei Verwendung
im $B$-System. Die auf der Grundlage von (\ref{dec}) berechneten
Werte können daher nur als grobe Abschätzung dienen. Eine weitere
Schwierigkeit ergibt sich aus den nur schlecht bekannten Größen
$f_B$ und $|V_{ub}|$. Zu $f_B$ verweise ich auf \cite{bernie}, im
Folgenden wird die auf mehreren Experimenten beruhende Abschätzung
$f_B=200(30)MeV$ (siehe
\cite{bernie}) benutzt.\\ \\
Zunächst werden die LQ-Kopplungskonstantenprodukte unter
Verwendung der reinen LQ-Terme Gln. (2.28) und (2.29) abgeschätzt
(dabei können Gl. 3.20 und 3.21 mit den entsprechenden Änderungen
für das $B$-Meson verwendet werden). Der SM-Term und sämtliche
Interferenzterme werden vernachlässigt. Man erhält: \beq
 \textrm{Für}\; B^+\rightarrow e^+\nu_e \; :\; |\lambda_L^{13} \lambda_R^{*11}| < 1,2\times 10^{-4}
 \left(\frac{m_{LQ}}{100 {\rm ~GeV}}
\right)^2  ,\quad |\lambda_L ^{13}\lambda_L^{*11}| < 1,5
\left(\frac{m_{LQ}}{100 {\rm ~GeV}} \right)^2 ,\;\;\;\;\;\;\; \eeq
\beq \textrm{Für}\; B^+\rightarrow \mu^+\nu_\mu \; :\;
|\lambda_L^{23} \lambda_R^{*21}| < 1,4\times 10^{-4}
\left(\frac{m_{LQ}}{100 {\rm ~GeV}} \right)^2 ,\quad
|\lambda_L^{23} \lambda_L^{*21}| < 8,4\times 10^{-3}
\left(\frac{m_{LQ}}{100 {\rm ~GeV}} \right)^2 , \eeq \beq
\textrm{Für}\; B^+\rightarrow \tau^+\nu_\tau \; :\;
|\lambda_L^{33} \lambda_R^{*31}| < 8,4\times 10^{-4}
\left(\frac{m_{LQ}}{100 {\rm ~GeV}} \right)^2 ,\quad
|\lambda_L^{33} \lambda_L^{*31}| < 2,9\times 10^{-3}
\left(\frac{m_{LQ}}{100 {\rm ~GeV}} \right)^2. \eeq Es sei hier
noch einmal darauf hingewiesen, dass diese Resultate aufgrund der
erwähnten Unsicherheit in den experimentellen Daten nur grobe
Abschätzungen darstellen. Sie gelten direkt entsprechend auch für
beliebigen Neutrino-\textit{flavour}. Die
$\lambda_L\lambda_R$-Schranken sind direkt identisch mit den
entsprechenden SUSY-Schranken und die Ergebnisse aus \cite{thor}
stimmen mit denen aus den Gln. (6.2-4) in etwa überein.
\begin{description}
    \item[Anmerkung: ]Um die Unsicherheit in $f_B$ zu umgehen und die theoretische SM-Vorhersage miteinzubeziehen
   kann auch mit der theoretischen
Vorhersage \cite{cleo} \beq BR(B^+\rightarrow
\tau^+\nu_\tau)=(4,08\pm 0,24)\times
10^{-4}\left|\frac{V_{ub}}{V_{td}}\right|^2\; ,\eeq gearbeitet
werden (siehe \cite{thor} für Schranken an
SUSY-Kopplungskonstantenprodukte). In der
Wolfenstein-Parameterisierung \cite{wolfen} \beq
\frac{V_{ub}}{V_{td}}=\frac{\bar{\rho}-i\bar{\eta}}{1-\frac{\lambda^2}{2}-\bar{\rho}-i\bar{\eta}}\;
,\eeq können die miteinander korrelierten Unsicherheiten in
$V_{ub}$ und $V_{td}$ berücksichtigt werden\footnote{Laut
\cite{neuber} ist $\bar{\rho}=0,21\pm 0,12$, $\bar{\eta}=0,38\pm
0,11$ und $\lambda = 0,222\pm 0,004$.}. Die daraus berechenbaren
Werte für $|\lambda'{}_{313}^*\lambda_{233}|$ und $|\Re [
\lambda'{}_{213}^*\lambda_{323}]|$ sind mit der Schranke aus Gl.
(6.4) vergleichbar.
\end{description}
\subsection{Semileptonische $B^+$-Zerfälle}
Den Überlegungen aus \cite{lepto} folgend werden hier
\textit{Flavour}-ändernden Zerfälle der Art $B^+\rightarrow
\ell_1\bar{\ell}_2X$ (X symbolisiert ein oder mehrere Hadronen)
betrachtet. \begin{description}
    \item[Vorab:]
Analog zu der Abschätzung beim $D^+$, Anfang Kapitel 5.1 , kann
auch hier verfahren werden. Für $X=D^0$ ist
\beq|\lambda_L^{\ell_13}\lambda_L^{*\ell_22}|<\frac{4G_F}{\sqrt{2}}|V_{cb}|m_{LQ}^2=1,4\times
10^{-2}\left(\frac{m_{LQ}}{100GeV}\right)^2\; ,\eeq und für
$X=\pi^0$ erhält man:
\beq|\lambda_L^{\ell_13}\lambda_L^{*\ell_21}|<\frac{4G_F}{\sqrt{2}}|V_{ub}|m_{LQ}^2=1,2\times
10^{-3}\left(\frac{m_{LQ}}{100GeV}\right)^2\; .\eeq  Diese
Resultate verbessern die Ergebnisse aus \cite{lepto} leicht. Dass
die LQ-Kopplungen bei $b$-Zerfällen kleiner als die SM-Kopplungen
sind, ist allerdings nicht mehr als eine unbegründete -wenn auch
plausibel erscheinende- Annahme.
\end{description} Das $B^+$ zerfällt mit \cite {pdg} $(10,2\pm
0,9)\%$ in ein Lepton $\ell^+$, ein Neutrino $\nu_\ell$ und X. Wie
in Kapitel 4.1.2 für das Kaon soll nun auch hier zunächst aus
relativ einfachen Überlegungen zur Struktur der Zerfallsraten eine
Abschätzung an das Produkt $|\lambda_L\lambda_L^*|$ gewonnen
werden. Zunächst lässt sich unter Vernachlässigung der
kinematischen Unterschiede (unter Vernachlässigung der
Leptonenmassen) zwischen den Zerfällen
$B^+\rightarrow\ell\bar{\ell}X$ und $B^+\rightarrow\nu\bar{\ell}X$
($\ell = e,\; \mu$) analog zu den Gln. (4.12-14), fordern
($V_{jn}=\; V_{cb}\; \textrm{oder}\; V_{ub}$): \beq
 \frac{|\lambda_L^{i3}\lambda_L^{*mn}|^2}{m_{LQ}^4} \simeq  8 G_F^2 |V_{jn}|^2 \frac{BR(B^+
\rightarrow \ell \bar{\ell} X)}{BR(B^+ \rightarrow \nu \bar{\ell}
X)} \; ,\label{BB2}
  \eeq und für $BR(B^+\rightarrow \pi^0e^+\nu_e)=(9,0\pm 2,8)\times 10^{-5}$ und
$BR(B^+\rightarrow \pi^+e^+e^-)<3,9\times 10^{-3}$ erhält man:
\beq |\lambda_L^{13}\lambda_L^{*11}|<7,8\times
10^{-3}\left(\frac{m_{LQ}}{100GeV}\right)^2\; .\eeq Dieses
Resultat stimmt mit dem in \cite{lepto} berechneten Wert aufgerundet überein. \\
\\Desweiteren kann der Zerfall $B^+\rightarrow \tau^\pm e^\mp X$ betrachtet
werden, wobei das $\tau$ sehr schnell weiter zerfällt,
$\tau\rightarrow \nu +\textrm{Hadronen}$ oder $\tau\rightarrow
e\nu \bar{\nu}$, und zu $B^+\rightarrow \nu \bar{e}X$ oder
$B^+\rightarrow e\bar{e}\nu\bar{\nu}X$ im Endzustand führt. Die
Phasenraumunterdrückung $PS(x)$ (Ursache hierfür ist die
$\tau$-Masse) ist gegeben durch \cite{koy}\beq
PS(x)=1-8x+8x^3-x^4-12x^2\ln x=3,2\times 10^{-1}\; , \qquad\quad
x=\frac{m_\tau^2}{m_b^2}\; .\eeq  Daraus erhält man ($n= u$ oder
$c$ und $(i;m)=(1;3)$ oder $(3;1)$): \beq BR(B^+ \rightarrow
\tau^\pm e^\mp X) \simeq PS(x)
\frac{|\lambda_L^{i3}\lambda_L^{*mn}|^2}{m_{LQ}^4} \frac{BR(B
\rightarrow e \nu X)}{8 G_F^2 |V_{nb}|^2}~~. \eeq Diese Gleichung
stimmt mit den entsprechenden Ersetzungen und unter
Berücksichtigung des PS-Faktors, der die kinematischen
Unterschiede in den beiden Zerfällen darstellt, mit Gl. (6.9)
überein. Daraus folgt mit der Abschätzung (s.o.)\beq BR(B^+
\rightarrow \tau^- e^+ X)<\frac{BR(B^+ \rightarrow e^+ \nu
X)}{BR(\tau^-\rightarrow \nu X')}\; ,\eeq für $X=\pi$ mit
$BR(\tau\rightarrow\nu_\tau\pi)=(11,06\pm0,11)\%$, siehe
\cite{pdg} folgende Schranke: \beq
\frac{|\lambda_L^{33}\lambda_L^{*11}|}{m_{LQ}^2} < 2
\sqrt{\frac{2}{PS}} G_F \frac{V_{ub}}{[BR(\tau \rightarrow \nu
X')]^{1/2}}~~.\label{tau} \eeq  Man erhält:
\beq|\lambda_L^{33}\lambda_L^{*11}|<6,3\times
10^{-3}\left(\frac{m_{LQ}}{100GeV}\right)^2\; .\eeq Dies
verbessert den entsprechenden Wert aus \cite{lepto} um etwa eine
Größenordnung.
\begin{description}
    \item[Anmerkung:]Hier wurde der Grenzübergang
$\rho_q=\frac{m_q^2}{m_b^2}\rightarrow 0$ betrachtet. Außerdem
wurden (wie auch im Folgenden) sämtliche nicht-perturbativen
Korrekturen vernachlässigt. Laut \cite{koy} vermindern diese
Korrekturen (bis zur Ordnung $\frac{1}{m_b^2}$) die Zerfallsrate
um 6-10\%, abhängig von der Quarkmasse im Endzustand. Die
Berechnungen basieren auf HQE (\textit{Heavy Quark Expansion}) und
der Operatorproduktentwicklung. In \cite{koy} findet sich eine
ausführlichere Behandlung zu inklusiven Zerfällen der Art
$H_b\rightarrow \tau\bar{\nu}X$ .
\end{description}
Nimmt man an, dass das $\tau$ in $ e\nu\bar{\nu}$ zerfällt, so
kann der Zerfall $B^+\rightarrow\tau\bar{e}X$ einen Beitrag bei
der experimentellen Suche nach $B^+\rightarrow e \bar{e}\pi^+$
liefern ($BR(B^+\rightarrow e\bar{e}\pi^+)<3,9\times 10^{-3}$
\cite{pdg}). Mithilfe von \beq PS\cdot BR(B \rightarrow \tau
\bar{e} X)\cdot BR(\tau \rightarrow e
\nu\bar{\nu})<BR(B^+\rightarrow e\bar{e}X)_{exp}\; ,\eeq und Gl.
(6.9) erhält man: \beq
\frac{|\lambda_L^{33}\lambda_L^{*11}|}{m_{LQ}^2} < \frac{4
G_FV_{ub}}{\sqrt{2}}
 \sqrt{\frac{BR(B \rightarrow e \bar{e} \pi^+)}
{BR(B \rightarrow e \nu \pi^0)BR(\tau \rightarrow e \nu
\bar{\nu})PS(x)}}\; . \eeq Mit $BR(B \rightarrow e \nu
\pi^0)=(9,0\pm 2,8)10^{-5}$ und $BR(\tau \rightarrow e\nu
\bar{\nu})=(17,84\pm0,06)\%$ (siehe \cite{pdg} ) ergibt sich
daraus:\beq|\lambda_L^{33}\lambda_L^{*11}|<3,2\times
10^{-2}\left(\frac{m_{LQ}}{100GeV}\right)^2\; .\eeq Dieses
Ergebnis ist nur geringfügig besser als in \cite{lepto}, die
Berechnungsgrundlagen sind jedoch unterschiedlich: In \cite{lepto}
wurde der PS-Faktor nicht berücksichtigt. Das Ergebnis ist
allerdings um etwa eine Größenordnung schlechter als (6.15). Das
Elektron kann auch durch ein Myon ersetzt werden, um weitere Werte
zu erhalten.
\\ \\Abschließend zu diesen Überlegungen wird nun noch der Zerfall
$B^+\rightarrow \tau\bar{\tau} X$ betrachtet. Nimmt man an, dass
beide $\tau$'s in $e$'s zerfallen, so erhält man analog zu den
vorherigen Berechnungen (siehe Gl. 6.17):\beq
\frac{|\lambda_L^{33}\lambda_L^{*31}|}{m_{LQ}^2} < \frac{4
G_F}{\sqrt{2}} \frac{V_{ub}}{BR(\tau \rightarrow e \nu \bar{\nu})}
\sqrt{\frac{BR(B \rightarrow e \bar{e} X)} {BR(B \rightarrow e \nu
X)~PS(x)}} \; .\eeq  Der Phasenraumunterdrückungsfaktor PS kann
mithilfe von \cite{koy} berechntet werden. Dabei wird
$\rho_q=\rho_\tau=m_\tau^2/m_b^2$ gesetzt und man erhält: \beq
PS(\rho_\tau)=\sqrt{1-4\rho_\tau}[1-14\rho_\tau-2\rho_\tau^2-12\rho_\tau^3]+24\rho_\tau^2(1-\rho_\tau^2)
ln\frac{1+\sqrt{1-4\rho_\tau}}{1-\sqrt{1-4\rho_\tau}}=0,030\;
.\eeq In \cite{lepto} wurde mit \beq
PS=\left[1-4\frac{m_\tau^2}{m_b^2}\right]^{2,48}=0,089\eeq
gerechnet. Mit \cite{pdg} $BR(\tau \rightarrow e \nu
\bar{\nu})=(17,84\pm 0,06)\%$, $BR(B \rightarrow e
\bar{e}\pi)<3,9\times 10^{-3}$ und $BR(B \rightarrow e \nu \pi)
=(9,0\pm 2,8)\times 10^{-5}$ erhält man: \beq
|\lambda_L^{33}\lambda_L^{*31}|<2,5\times
10^{-1}\left(\frac{m_{LQ}}{100GeV}\right)^2\; .\eeq Dieser Wert
ist um mehr als eine Größenordnung schlechter als der in
\cite{lepto} berechnete Wert. Da die in \cite{lepto} eingesetzten
$BR$-Werte nicht angegeben waren, kann hier keine genaue
Begründung für diese Abweichung gegeben werden. Das Pion kann auch
durch ein D-Meson ersetzt werden. Dabei sind bessere Schranken zu
erwarten, die experimentellen Daten (BR-Werte) sind allerdings
unvollständig, sodass zusätzliche Näherungen gemacht werden
müssen.
\\ \\Nun werden kurz die exklusiven Zerfälle der Art $B^+\rightarrow
\ell_1\bar{\ell}_2X$ ($\ell_1\bar{\ell}_2=e\bar{e},\;
\mu\bar{\mu}\;\textrm{oder}\; \mu^\pm e^\mp$ und X ist ein Kaon
oder ein Pion) betrachtet. Dazu muss das Matrixelement $\langle
X^+|\bar{s}\gamma^\mu b|B^+\rangle$ berechnet werden (Gl. 4.17):
\beq\langle X^+|\bar{s}\gamma^\mu b|B^+\rangle =
\frac{1}{\sqrt{2}}(f_+p^\mu +f_-q^\mu)\; .\eeq Einen
entsprechenden Ausdruck kann man für das $D^0\rightarrow X^+$
-Matrixelement finden und die Formfaktoren für $B$ und $D$ können
mit der \textit{heavy quark symmetry} \cite{wise} zueinander in
Beziehung gebracht werden. Unter der Annahme einer
$SU(2)$-\textit{flavour}-Symmetrie der schweren $c$ und $b$ Quarks
kann man nähern \cite{koy}:\beq(f_+-f_-)^{B\rightarrow
K}=\left(\frac{m_b}{m_c}\right)^{\frac{1}{2}}\left(\frac{\alpha_s(m_b)}{\alpha_s(m_c)}\right)^{\frac{-6}{25}}
(f_+-f_-)^{D\rightarrow K}\; ,\eeq wobei $f_+-f_-\simeq 2f_+$ und
$\alpha_s$ ist die starke Kopplungskonstante. Nun können, unter
Vernachlässigung der kinematischen Unterschiede (Vernachlässigung
der Massen aller Endzustandsteilchen), $B$ und $D$-Zerfälle
zueinander in Beziehung gesetzt werden. Zusätzlich muss wieder von
ungebrochener Isospinsymmetrie ausgegangen werden. Man erhält
(analog zu Gl. 6.9, die Zerfallsrate ist proportional zum Quadrat
der Zerfallskonstante und $m_{B/D}^5$): \beq
\frac{|\lambda_L^{i3}\lambda_L^{*m1}|^2}{m_{LQ}^4}\simeq
8G_F^2|V_{13}|^2\frac{m_D^5}{m_B^5}\frac{m_c}{m_b}\left(\frac{\alpha_s(m_b)}{\alpha_s(m_c)}\right)^{\frac{12}{25}}
\frac{\Gamma(B^+\rightarrow K\ell^+\ell^-)}{\Gamma(D^0\rightarrow
K\ell\nu)}\; .\eeq Es muss allerdings betont werden, dass diese
Berechnung auf vielen Näherungen basiert. So wurde auch die
Abhängigkeit der Formfaktoren vom Impulsübertrag vernachlässigt.
Gl. (6.25) gilt eigentlich nur für maximalen Impulsübertrag, bei
kleinem Impulsübertrag $q^2$ ändert sich die Schranke, siehe
\cite{lepto}, zu \beq
\frac{|\lambda_L^{i3}\lambda_L^{*m1}|^2}{m_{LQ}^4}\simeq
8G_F^2|V_{13}|^2\frac{m_D^3}{m_B^3}\frac{m_c}{m_b}\left(\frac{\alpha_s(m_b)}{\alpha_s(m_c)}\right)^{\frac{12}{25}}
\frac{\Gamma(B^+\rightarrow K\ell^+\ell^-)}{\Gamma(D^0\rightarrow
K\ell\nu)}\; .\eeq Die aus diesen Näherungen berechenbaren
Schranken können Tabelle 6 und 7 des Anhangs entnommen werden (es
wurde jeweils die schwächste Schranke angenommen). Als
Zerfallsteilchen kann statt des $K$-Mesons ein Pion eingesetzt
werden. Es muss aber betont werden, das es sich dabei nur um grobe
Abschätzungen handelt. Die Werte verbesserten sich gegenüber
\cite{lepto} um etwa eine Größenordnung. Das Verhältnis der
starken Kopplungskonstanten $\alpha_s(m_b)$
 und $\alpha_s(m_c)$ wurde mit eins angenähert.\\
\\Zerfälle der Art $B\rightarrow D\pi\ell\nu$ wurden in \cite{Bz}
ausführlich betrachtet, sie werden hier daher nicht mehr
behandelt.
\section{Zerfälle neutraler $B$-Mesonen}
Den Rechnungen für den Zerfall neutraler Kaonen aus Kapitel 4.2
folgend lässt sich fordern (die Zahlen wurden \cite{pdg}
entnommen, zur Lebensdauer des $B$-Mesons siehe Tabelle 2 des
Anhangs): \beq \Gamma(B_0 \rightarrow e^{\pm}
\mu^{\mp})_A=\frac{(m_{B_0}^2 - m_{\mu}^2)^2}{64 \pi m_{B_0}^3}
\frac{|\lambda_L^{i3}\lambda_L^{*m1}|^2}{m_{LQ}^4} f_{B_0}^2
m_{\mu}^2 < 5,98 \times 10^{-26} {\rm ~MeV} \label{nkao10}
  \; ,\eeq und \beq
\Gamma(B_0 \rightarrow e\overline{e})_A=\frac{1}{32 \pi}
\frac{|\lambda_L^{i3}\lambda_L^{*i1}|^2}{m_{LQ}^4} f_{B_0}^2
m_{e}^2\sqrt{m_{B_0}^{2}-4m_{e}^{2}}<1,15\times 10^{-25}MeV\;
.\eeq Die pseudoskalaren Beiträge liefern \cite{pdg}: \beq
\Gamma(B_0 \rightarrow e^{\pm} \mu^{\mp})_P=
\frac{|\lambda_L^{i3}\lambda_R^{*m1}|^2}{m_{LQ}^4}
\frac{(m_{B_0}^2 - m_{\mu}^2)^2f_{B_0}^2 m_{B_0}}{64
\pi(m_b+m_d)^2} < 5,98 \times 10^{-26} {\rm ~MeV} \label{nkao1}
  \; ,\eeq und \beq \Gamma(B_0 \rightarrow e\bar{e})_P= \frac{|\lambda_L^{i3}\lambda_R^{*i1}|^2}{m_{LQ}^4}\frac{
(m_{B_0}^2 - 2m_{e}^2)f_{B_0}^2 m_{B_0}^2}{64
\pi(m_b+m_d)^2}\sqrt{m_{B_0}^2-4m_e^2} <1,15 \times 10^{-25} {\rm
~MeV} \label{nkao2}
  \; .\eeq Daraus erhält man für die Kopplungskonstantenprodukte (unter Verwendung der Werte $BR(B^0\rightarrow
  \mu\bar{\mu})<6,1\cdot 10^{-7}$, $BR(B^0\rightarrow
  e^\pm\tau^pm)<5,3\cdot 10^{-4}$ und $BR(B^0\rightarrow
  \mu^\pm\tau^pm)<8,3\cdot 10^{-4}$, siehe \cite{pdg}) folgende Werte:
\beq
 B^0
\rightarrow e \bar{e}\; :\quad |\lambda_L^{13} \lambda_L^{*11}| <
2,5 \times 10^{-1} \left( \frac{m_{LQ}}{100 {\rm ~GeV}}\right)^2,
\quad |\lambda_L \lambda_R^*| < 3,0 \times 10^{-5} \left(
\frac{m_{LQ}}{100 {\rm ~GeV}}\right)^2 ,\;\;\;
  \eeq
  \beq
  B^0
\rightarrow \mu \bar{\mu}\; :\quad |\lambda_L^{23}
\lambda_L^{*21}| < 1,1 \times 10^{-3} \left( \frac{m_{LQ}}{100
{\rm ~GeV}}\right)^2\; , \quad |\lambda_L \lambda_R^*| < 2,5
\times 10^{-5} \left( \frac{m_{LQ}}{100 {\rm ~GeV}}\right)^2
,\;\;\;
  \eeq\beq
  B^0
\rightarrow e^\mp \mu^\pm\; :\quad |\lambda_L^{i3}
\lambda_L^{*m1}| < 2,3 \times 10^{-3} \left( \frac{m_{LQ}}{100
{\rm ~GeV}}\right)^2\; , \quad |\lambda_L \lambda_R^*| < 4,0
\times 10^{-5} \left( \frac{m_{LQ}}{100 {\rm ~GeV}}\right)^2  ,
  \eeq\beq
  B^0
\rightarrow e^\mp \tau^\pm\; :\quad |\lambda_L^{i3}
\lambda_L^{*m1}| < 2,9 \times 10^{-3} \left( \frac{m_{LQ}}{100
{\rm ~GeV}}\right)^2, \quad |\lambda_L^{i3} \lambda_R^{*m1}| < 8,5
\times 10^{-4} \left( \frac{m_{LQ}}{100 {\rm ~GeV}}\right)^2 ,
  \eeq\beq
  B^0
\rightarrow \mu^\mp \tau^\pm\; :\quad |\lambda_L^{i3}
\lambda_L^{*m1}| < 3,7 \times 10^{-3} \left( \frac{m_{LQ}}{100
{\rm ~GeV}}\right)^2 , \quad |\lambda_L^{i3} \lambda_R^{*m1}| <
1,1 \times 10^{-3} \left( \frac{m_{LQ}}{100 {\rm ~GeV}}\right)^2
.\eeq Diese Schranken sind durchgehend niedriger (um bis zu eine
Größenordnung) als die in \cite{lepto} auf ähnliche Weise
berechneten Werte. Die entsprechenden
SUSY-Schranken können Tabelle 7 des Anhangs entnommen werden. Diese bestätigen die Resultate aus \cite{jang}.\\ \\
Genauso können auch Schranken an den leptonischen LQ-Zerfall des
$B^0_s$-Mesons berechnet werden. Dabei wird
$f_{B^0_s}=(1,16\pm0,04)f_{B^0})$ \cite{bernie} benutzt und
außerdem \cite{pdg}: \beq BR(B^0_s\rightarrow e \bar{e})<5,4\cdot
10^{-5}\; ,\;\;BR(B^0_s\rightarrow \mu \bar{\mu})<2,0\cdot
10^{-6}\;\textrm{und}\;\; BR(B^0_s\rightarrow e^\pm
\mu^\pm)<6,1\cdot 10^{-6}\; .\eeq Man erhält folgende Schranken:
\beq
 B^0_s
\rightarrow e \bar{e}:\quad |\lambda_L^{13} \lambda_L^{*12}| < 1,8
\times 10^{0} \left( \frac{m_{LQ}}{100 {\rm ~GeV}}\right)^2, \;
|\lambda_L^{13} \lambda_R^{*12}| < 2,0 \times 10^{-4} \left(
\frac{m_{LQ}}{100 {\rm ~GeV}}\right)^2 ,
  \eeq
  \beq
  B^0_s
\rightarrow \mu \bar{\mu}:\quad |\lambda_L^{23} \lambda_L^{*22}| <
1,7 \times 10^{-3} \left( \frac{m_{LQ}}{100 {\rm ~GeV}}\right)^2\;
, \; |\lambda_L^{23} \lambda_R^{*22}| < 3,9 \times 10^{-5} \left(
\frac{m_{LQ}}{100 {\rm ~GeV}}\right)^2 ,
  \eeq\beq
  B^0_s
\rightarrow e^\mp \mu^\pm :\quad |\lambda_L^{i3} \lambda_L^{*m2}|
< 4,1 \times 10^{-3} \left( \frac{m_{LQ}}{100 {\rm ~GeV}}\right)^2
, \; |\lambda_L^{i3} \lambda_R^{*m2}| < 6,8 \times 10^{-5} \left(
\frac{m_{LQ}}{100 {\rm ~GeV}}\right)^2  .
  \eeq
Die entsprechenden SUSY-Schranken sind für LL-Kopplungen doppelt
so gross und für LR-Kopplungen identisch mit den aufgelisteten
Werten. Die SUSY-Schranken für $B^0_s\rightarrow e^\pm\mu^\mp$
stimmen mit den in \cite{thor} berechneten Werten überein.\\ \\Zu
leptonischen $B$-Zerfällen existieren detailliertere
Betrachtungen. Hier verweise ich auf \cite{dedes}, wo der Zerfall
in ein Myon und ein Antimyon analysiert wurde. Weitere leptonische
$B$-Zerfälle wurden in \cite{chanko} behandelt, QCD-Korrekturen
können u. A. \cite{urban} entnommen werden.
\chapter {$K^{0}-\overline{K}^{0}$-, $D^{0}-\overline{D}^{0}$- und $B^{0}-\overline{B}^{0}$- Mischzustände }
Abbildung 8.1 zeigt den Standardmodellbeitrag zur Mischung der
neutralen Kaonen in niedrigster Ordnung. Entsprechende Diagramme
existieren für die neutralen $D$ und $B$ -Mesonen. Ohne $c$- und
$t$-Austausch würde die $CP$-Symmetrie nicht verletzt werden und
$|K^0_S\rangle$ bzw. $|K^0_L\rangle$ (Gl. 4.31) wären exakte
Eigenzustände des $CP$-Operators. Die Existenz dreier
Quarkgenerationen ist für die experimentell nachgewiesene schwache
$CP$-Verletzung verantwortlich.\\
\begin{figure}[h]
\begin{center}
\begin{fmffile}{tau413}
\begin{fmfgraph*}(160,150)
\fmfpen{thin}\fmfstraight\fmfleft{i1,i2}\fmfright{o1,o2}\fmf{fermion,tension=2.5,label=s}{i1,v1}
\fmf{fermion,tension=2.5,label=d}{v2,i2}
\fmf{dashes,label=$W^-$}{v1,v3}\fmf{dashes,label=$W^+$}{v2,v4}\fmf{fermion,label=$u,,c,,t$}{v1,v2}
\fmf{fermion,label=$u,,c,,t$}{v4,v3}\fmf{fermion,tension=2.5,label=d}{v3,o1}\fmf{fermion,tension=2.5,label=s}{o2,v4}
\end{fmfgraph*}\hspace{30pt}
\begin{fmfgraph*}(160,150)
\fmfpen{thin}\fmfstraight\fmfleft{i1,i2}\fmfright{o1,o2}\fmf{fermion,tension=2.5,label=s}{i1,v1}
\fmf{fermion,tension=2.5,label=d}{v2,i2}
\fmf{dashes,label=$W^-$}{v1,v2}\fmf{dashes,label=$W^+$}{v3,v4}\fmf{fermion,label=$u,,c,,t$}{v1,v3}
\fmf{fermion,label=$u,,c,,t$}{v4,v2}\fmf{fermion,tension=2.5,label=d}{v3,o1}\fmf{fermion,tension=2.5,label=s}{o2,v4}
\end{fmfgraph*}
\end{fmffile}\end{center}\caption[$K^0-\bar{K}^0$-Mischung im SM]{\textit{$K^0-\bar{K}^0$-Mischung im SM}}
\label{mix}\end{figure}\\Hier soll der LQ-Beitrag dieses
Diagramms zur Massendifferenz (bei ungebrochener $CPT$-Symmetrie
\cite{pdg}), \beq\Delta m_K=m_{K^0_L}-m_{K^0_S}=(3,490\pm
0,006)\times 10^{-12}\; MeV\; ,\eeq berechnet werden. Der
Standardmodellbeitrag zu dieser Massendifferenz kann unter
Verwendung der Vakuumsättigungsnäherung \cite{Cheng} \beq\langle
K|\left[\bar{d} \gamma^{\mu}P_L s\right]\left[\bar{d} \gamma_{\mu}
P_L s\right]|\bar{K}\rangle = \frac{8}{3}\frac{1}{4}\langle
K|\bar{d} \gamma^{\mu}\gamma_5 s|0\rangle\langle 0|\bar{d}
\gamma_{\mu} \gamma_5 s|\bar{K}\rangle = \frac{2}{3} \frac{f_K^2
m_K}{2} \; ,\label{K1} \eeq berechnet werden. Der Faktor
$\frac{8}{3}$ in Gl. (8.2) setzt sich zusammen aus den vier
möglichen Wick-Kontraktionen multipliziert mit einem Farbfaktor
2/3 . Den Überlegungen in \cite{Cheng} folgend, erhält man für die
Massendifferenz im Standardmodell unter Vernachlässigung der
$u$-Quark Masse:
\begin{eqnarray}\frac{\Delta m_K}{2}&=& -\Re\left[\langle K|-\mathcal
{L}^{\Delta S=2}_{eff}|\overline{K}\rangle\right] \nonumber \\&=&
\frac{G_F}{\sqrt{2}} \frac{\alpha}{4 \pi \sin^2 \theta_W}
(sin^2\theta_ccos^2\theta_c)X\langle K|(\bar{d} \gamma^{\mu} P_L
s)(\bar{d} \gamma_{\mu} P_L s)|\bar{K}\rangle\; , \label{K0}
\end{eqnarray} wobei \beq X=(sin^2\theta_ccos^2\theta_c)^{-1}\sum_{q
= c,t} \Re\left[(V_{qs} V_{qd}^*)^2 \frac{m_q^2}{m_W^2}+ V_{cs}
V_{cd}^* V_{ts} V_{td}^* \frac{2 m_c^2 m_t^2}{m_W^2 ( m_t^2 -
m_c^2)} \ln \left( \frac{m_t^2}{m_c^2} \right)\right]\; .\eeq
Dabei ist $\alpha=7,297352533(27)\times 10^{-3}$ die
Feinstrukturkonstante, $\theta_c$ der Cabibbowinkel und
$sin^2\theta_W=0,23113(15)$ das Sinusquadrat des Weinbergwinkels
der schwachen WW. \\ \\Daraus kann man für die SM-Massendifferenz
einen Wert von $(\Delta M_K)_{th}^{SM}=3,08\times 10^{-12}MeV$
berechnen\footnote{Die beiden Terme, in denen $V_{td}$ vorkommt,
wurden dabei nicht - wie in \cite{lepto} - vernachlässigt. Sie
liefern beide jeweils einen Beitrag von etwa 10\% des
Hauptbeitrages. Für $V_{td}$ und $V_{ts}$ wurden die oberen
Grenzwerte aus \cite{pdg} verwendet: $V_{td}=0,014$,
$V_{ts}=0,044$, $V_{cd}=0,224\pm 0,016$ und $V_{cs}=0,996\pm
0,013$. }, der mit dem experimentellen Wert \cite{pdg} $(\Delta
M_K)_{exp}=3,49\times 10^{-12}MeV$ recht gut übereinstimmt.
\section {LQ-($V\pm A$)-Wechselwirkungen}
Der LQ-Beitrag zu $\Delta m_{K}$ ist von der gleichen Form, wie
der SM-Beitrag (Gl. 8.3). Es müssen jedoch einige Veränderungen
vorgenommen werden:\begin{itemize}
    \item Die Quarkmassen müssen durch Leptonenmassen ersetzt werden, \item $\frac{4G_{F}V}{\sqrt{2}}$
ist durch $\frac{\lambda_{LQ}^{ij}\lambda_{LQ}^{mn*}}{m_{LQ}^2}$
zu ersetzen, \item $m_W^2$ wird $m_{LQ}^2$ und $\frac{\alpha}{4\pi
sin^2\theta_W}$ vereinfacht sich zu $\frac{1}{8\pi^2}$.
\end{itemize}      Man kann daher
fordern: \beq\frac{1}{32 \pi^2 m_{LQ}^2} \left[  |\lambda^{\ell d}
\lambda^{\ell s}{}^*|^2 \frac{m_\ell^2}{m_{LQ}^2}  + \lambda^{\ell
d} \lambda^{\ell s}{}^* \lambda^{\ell 'd}{}^* \lambda^{\ell 's}
\frac{2m_\ell^2m_{\ell '}^2}{m_{LQ}^2(m_\ell^2 - m_{\ell '}^2)}
\ln \left( \frac{m_{\ell '}^2}{m_{\ell}^2} \right) \right]< \frac{
(\Delta M_K)_{exp}-(\Delta M_K)_{th}^{SM}}{2}.\label{54} \eeq
Vernachlässigt man den zweiten Term komplett (dies impliziert die
Annahme, dass maximal zwei LQ-Kopplungskonstanten nicht
verschwinden), so erhält man eine Schranke an alle vektoriellen
Leptoquarks die ein $d$-Typ-Quark an ein massives (geladenes)
Lepton $\ell$ koppeln: \beq |\lambda^{\ell 1}\lambda^{\ell
2}{}^*|<3,9\times
10^{-2}\left(\frac{m_{LQ}}{100GeV}\right)^2\left(\frac{1GeV}{m_\ell}\right)\;
.\eeq Der in \cite{lepto} angegebene Wert dieser Schranke
verbessert sich damit um fast eine Größenordnung. Dies liegt
hauptsächlich daran, dass in \cite{lepto} nicht die Differenz
zwischen SM-Wert und experimentellem Wert als Schranke für
LQ-Kopplungen angesetzt wurde. Stattdessen wurde gefordert, dass
die LQ-Kopplungen kleiner sind als die SM-Kopplung. Diese Methode
hat den Vorteil, dass sich die Zerfallskonstante $f_K$ aus den
Gleichungen herauskürzt. Man erhält dann:\beq |\lambda^{\ell
1}\lambda^{\ell 2}{}^*|<9,9\times
10^{-2}\left(\frac{m_{LQ}}{100GeV}\right)^2\left(\frac{1GeV}{m_\ell}\right)\;
,\eeq was mit \cite{lepto} übereinstimmt. Beim $D$-Meson ist für
die Massendifferenz nur eine obere Schranke bekannt: $\Delta
M_D<4,6\times 10^{-11} MeV$ \cite{pdg}. Mit der Forderung (dabei
bediene ich mich der Vorgehensweise, die auch zu Gl. 7.6 führte)
 \beq \frac{1}{32 \pi^2} | \lambda^{\ell u} \lambda^{\ell c}{}^*|^2
\frac{m_\ell ^2} {m_{LQ}^4} < \frac{3 \Delta m_D}{2 f_D^2 m_D}\;
,\eeq erhält man:  \beq | \lambda^{\ell u} \lambda^{\ell c}{}^*| <
1,1\times 10^{-1} \left( \frac{m_{LQ}}{100 {\rm ~GeV}} \right)^2
\left( \frac{1 {\rm ~GeV}}{m_{\ell}} \right)\; . \eeq Dies ist nur
geringfügig niedriger als in \cite{lepto}. Für die
$B^0$-$\bar{B}^0$ Massendifferenz existiert der Wert \cite{pdg}
$\Delta m_B = 3.2 \times 10^{-10} MeV$ . Man kann daher fordern
(analog zu Gl. 7.8): \beq | \lambda^{\ell d} \lambda^{\ell b}{}^*|
< 2,7\times 10^{-1}
 \left( \frac{m_{LQ}}{100 {\rm ~GeV}} \right)^2
\left( \frac{1 {\rm ~GeV}}{m_{\ell }} \right)\; . \eeq Auch dieser
Wert ist nur geringfügig niedriger als in \cite{lepto}.
\section{LQ-($S\pm P$)-Wechselwirkungen}
Auch hier können $\lambda_L\lambda_R$ -Kopplungen auftreten. Diese
wurden in \cite{miri} behandelt, liefern jedoch keine stärkeren
Schranken, als die im vorangegangenen Abschnitt berechneten Werte.
Daher werden sie hier nicht weiter verfolgt. \\ \\Stärkere
Schranken erhält man, wenn man annimmt, dass das Leptoquark kein
Eichboson ist (siehe \cite{lepto}), dieser Möglichkeit wird hier
jedoch nicht nachgegangen.
\chapter {Diskussion}
In diesem Kapitel werden die Resultate aus den Kapiteln 3-7 im
Hinblick auf die Güte der berechneten Schranken an die
Kopplungskonstantenprodukte und an die einzelnen
Kopplungskonstanten analysiert. Dabei werden die vorgenommenen
Abschätzungen und Näherungen näher betrachtet. Weitere, in den
vorangegangenen Kapiteln nicht behandelte, Mesonenzerfälle werden
dahingehend überprüft, ob aus ihnen gute Schranken an
Kopplungskonstanten(produkte) erhalten werden können. Zudem werden
- sofern vorhanden - weitere Möglichkeiten, insbesondere
Leptonenzerfälle, diskutiert, um Schranken an SUSY- und LQ-
Kopplungskonstanten(produkte) zu berechnen.\section{Schranken aus
$R$-Werten} Die berechneten Schranken aus $R$-Werten haben den
Vorteil, dass die Zerfallskonstante sich aus den Berechnungen
herauskürzt. Die LR-Kopplungen liefern besonders starke Schranken,
da die Drehimpulsunterdrückung aufgehoben ist. Zudem sind
Schranken unter Berücksichtigung beliebigen
Neutrino-\textit{flavours} berechnbar (siehe \cite{thor}). Die in
diese Rechnungen eingehenden Quarkmassen sind jedoch besonders bei
den leichten Mesonen eine Fehlerquelle. Es muss weiterhin darauf
hingewiesen werden, das beim K-Meson die Schranken aus
FCNC-Prozessen bei neutralen Mesonen deutlich stärker sind, als
die Schranken aus dem $R$-Wert.\section{Schranken aus
semileptonischen Zerfällen} Die Schranken an LR-Kopplungen (siehe
Kapitel 4.1.2) sind vergleichsweise schwach. Aus den LL-Kopplungen
können zahlreiche Schranken berechnet werden, wobei allerdings
auch einige Näherungen gemacht wurden, so wurden beispielsweise
sämtliche Leptonenmassen vernachlässigt. Diese Näherung ist für
Myonen problematisch, da deren Masse von der Größenordnung der
Pionenmasse ist. Die daraus abgeleitete Schranke ist daher nur als
Abschätzung zu verstehen. Trotzdem bieten semileptonische Zerfälle
gerade bei schweren Mesonen ein weites Feld um Kopplungskonstanten
zu berechnen. Hier verweise ich auf inklusive Zerfälle des
$B$-Mesons in ein oder zwei $\tau$-Leptonen und die HQE im
Zusammenhang mit den Zerfallskonstanten $f_D$ und $f_B$. Dort
bieten sich interessante Möglichkeiten, um Schranken an
Kopplungskonstantenprodukte abzuschätzen (Kapitel 6.1.2).
\section{Schranken aus Zerfällen schwerer Mesonen}
Bei den Zerfällen schwerer Mesonen ergeben sich einige
Schwierigkeiten:
\begin{itemize}
    \item Die Zerfallskonstanten $f_D$ und $f_B$ sind nur sehr
    ungenau bekannt. Daneben ist die experimentelle Unsicherheit
    in den CKM-Matrixelementen zu beachten.
    \item Die QCD-Korrekturen wurden hier in allen Ordnungen außer
    Acht gelassen (dazu verweise ich auf die Literatur, siehe z.B.
    \cite{urban}).
    \item Die Beziehung zwischen $f_D$ und $f_B$ kann nur bei
    minimalem und maximalem Impulsübertrag auf einfache Weise
    beschrieben werden. Die Zwischenregion muss daher übergangen
    werden.
\end{itemize} Trotzdem bieten Zerfälle schwerer Mesonen einzigartige Möglichkeiten in der
Suche nach seltenen Zerfällen, FCNC und CP-Verletzung\footnote{Da
die starke Kopplungskonstante bei hohen Energien klein wird
(asymptotische Freiheit), eröffnet sich zudem die Möglichkeit
perturbativer Berechnungen auf diesem Gebiet.}. Durch zukünftige
Experimente können die hier berechneten Werte verbessert und neue
Werte hinzugewonnen werden. Für das $\Upsilon$-Meson können die
Berechnungen zu
leptonischen Zerfällen neutraler Mesonen übernommen werden.\\
\\Aus hadronischen
Zerfällen können Schranken an SUSY-Kopplungskonstantenprodukte mit
$\lambda''_{ijk}$ berechnet werden.
    \section{Sonstige Mesonenzerfälle} Es wurden im Verlauf dieser Arbeit einige Zerfälle nicht näher betrachtet.
    Die Zerfälle des $\eta$-Meson
    liefern aufgrund der niedrigen
    Lebensdauer dieses Mesons nur sehr schwache Schranken. Diese können aus anderen
    Zerfällen besser berechnet werden. Genauso verhält es sich
    auch mit einigen der berechneten D-Zerfälle. Hier kann
    allerdings noch auf eine signifikante Verbesserung der
    experimentellen Daten gehofft werden. Auch die Mischung der
    neutralen Mesonen liefert noch keine starken Schranken an die
    Kopplungskonstantenprodukte.
    \section{Alternative Methoden} Es gibt zahlreiche alternative
    Methoden, um die berechneten und weitere Schranken an
    Kopplungskonstantenprodukte zu erhalten (siehe auch Kapitel
    1.2). Ich möchte hier nur noch auf $\tau$-Zerfälle in der SUSY eingehen,
    da ich zu diesen Zerfällen einige Berechnungen durchgeführt
    habe. Diese werden im Folgenden kurz zusammengefasst.\\ \\Motiviert durch \cite{sher}
    werden Schranken durch Squark- und Slepton- Austausch in $\tau$-Zerfällen über $R$-Paritätsverletzung diskutiert.
Eine detailliertere Studie der $\tau$-Zerfälle kann \cite{black}
entnommen werden. M. Sher konzentriert sich in \cite{sher} auf
Higgs-vermittelte $\tau$ Zerfälle. Im MSSM+$\not{R_P}$ existieren
aber weitere mögliche Zerfallsmoden, die durch den LLE- und den
LQD-Term des $\not{R_P}$-Superpotentials erzeugt
werden.\\\vspace{5pt}\\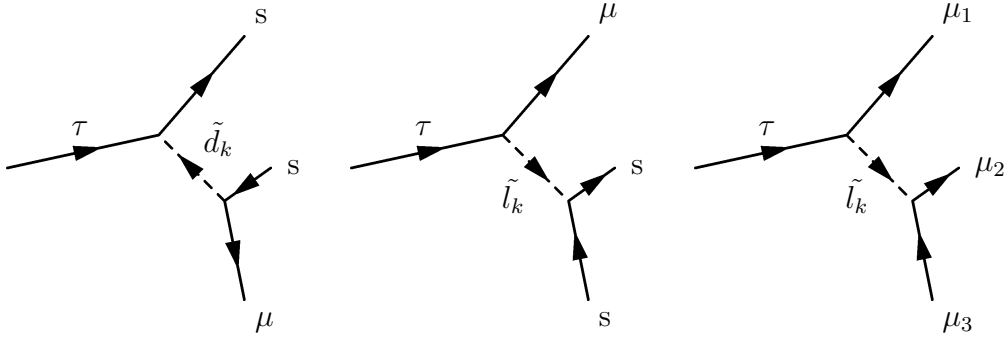
\begin{figure}[h]
\begin{center}
\begin{fmffile}{tau16}
\begin{fmfgraph*}(100,100)
\fmfpen{thin}\fmfleft{i1}\fmfright{o1,o2,o3}\fmf{fermion,label=$\tau$}{i1,v1}\fmf{fermion}{v1,o3}
\fmf{scalar,label=$\tilde{d_k}$}{v2,v1}\fmf{fermion}{o2,v2}\fmf{fermion}{v2,o1}\fmflabel{$\mu$}{o1}
\fmflabel{s}{o2}\fmflabel{s}{o3}
\end{fmfgraph*}\hspace{30pt}\begin{fmfgraph*}(100,100)\fmfpen{thin}\fmfleft{i2}\fmfright{o4,o5,o6}
\fmf{fermion,label=$\tau$}{i2,v3}
\fmf{fermion}{v3,o6}\fmf{scalar,label=$\tilde{l_k}$}{v3,v4}\fmf{fermion}{v4,o5}
\fmf{fermion}{o4,v4}\fmflabel{s}{o4}\fmflabel{s}{o5}\fmflabel{$\mu$}{o6}
\end{fmfgraph*}\hspace{30pt}\begin{fmfgraph*}(100,100)\fmfpen{thin}\fmfleft{i2}\fmfright{o4,o5,o6}
\fmf{fermion,label=$\tau$}{i2,v3}
\fmf{fermion}{v3,o6}\fmf{scalar,label=$\tilde{l_k}$}{v3,v4}\fmf{fermion}{v4,o5}
\fmf{fermion}{o4,v4}\fmflabel{$\mu_3$}{o4}\fmflabel{$\mu_2$}{o5}\fmflabel{$\mu_1$}{o6}
\end{fmfgraph*}
\end{fmffile}\end{center}\caption[$\tau$-Zerfall über Squark- oder Slepton-Austausch]{\textit{
$\tau$-Zerfall über Squark- oder Slepton-Austausch}}\label{tau}\end{figure}
\\Um die Zerfallsrate $\Gamma$ zu erhalten muss die invariante Amplitude
$\mathcal{M}$ berechnet werden, aus der mit Fermi's Goldener Regel
$\Gamma$ gewonnen werden kann. Die benötigten Matrixelemente sind
\cite{black}:
\begin{equation}
    <0|\overline{s}\gamma^{\mu}\gamma^{5}s|\eta_8> =
    i\frac{F_{\eta}^8}{\sqrt{2}}p_{\mu_{\eta_8}}\; ,
\end{equation} und \beq<0|\overline{s}\gamma^{5}s|\eta_8> =
    -i\sqrt{6}\frac{F_{\eta}^8m_{\eta_8}^2}{m_u+m_d+4m_s}\; .\eeq
    Das Quadrat der invarianten Amplitude der einzelnen Beiträge ist damit: \beq
(\mathcal{M}_{\tau\rightarrow\mu\eta})_{1}=\frac{\lambda'_{32k}\lambda'_{22k}}{
m_{\tilde{d_k}}^2}\frac{1}{2}<0|\overline{s}\gamma^{\mu}\gamma^{5}s|\eta>
\frac{1}{2}(\bar{\tau}\gamma_\mu(1-\gamma^5)\mu)\; ,\eeq \beq
(\mathcal{M}_{\tau\rightarrow\mu\eta})_{2}=\frac{\lambda_{3k2}\lambda'_{k22}}{
m_{\tilde{l_k}}^2}\frac{1}{2}<0|\overline{s}\gamma^{5}s|\eta>
\frac{1}{2}(\bar{\tau}(1-\gamma^5)\mu)\; ,\eeq \beq
(\mathcal{M}_{\tau\rightarrow\mu\eta})_{3}=\frac{\lambda_{2k3}\lambda'_{k22}}{
m_{\tilde{l_k}}^2}\frac{1}{2}<0|\overline{s}\gamma^{5}s|\eta>
\frac{1}{2}(\bar{\tau}(1-\gamma^5)\mu)\; ,\eeq und damit hat man
(für $\eta=\eta_8$):
\beq|\mathcal{M}_{\tau\rightarrow\mu\eta}|^2=\frac{1}{8}\left|\frac{\lambda'_{32k}
\lambda'_{22k}}{m_{\tilde{d_k}}^2}
-2\sqrt{3}\frac{m_{\eta_8}^2}{m_{\tau}(m_u+m_d+4m_s)}\frac{(\lambda_{3k2}
\lambda'_{k22}+\lambda_{2k3}\lambda'_{k22})}{m_{\tilde{l_k}}^2}
\right|^2 F_{\eta_8}^2m_{\tau}^{2}(m_{\tau}^{2}-m_{\eta_8}^{2})\;
. \eeq Die $\mu$ - Masse wurde vernachlässigt. Für den
Slepton-vermittelten Zerfall $\tau\rightarrow 3\mu$ erhält man:
\beq|\mathcal{M}_{\tau\rightarrow
3\mu}|^2=\frac{\left|\lambda_{3k2}\lambda_{2k2}+\lambda_{2k3}
\lambda_{2k2}\right|^2}{m_{\tilde{l_k}}^4}16(p_\tau
p_{\mu_3})
(p_{\mu_1}p_{\mu_2})=\frac{\left|\lambda_{3k2}\lambda_{2k2}+
\lambda_{2k3}\lambda_{2k2}\right|^2}{m_{\tilde{l_k}}^4}m_\tau^4
\; .\eeq Die Zerfallsraten sind damit (die $\mu$-Masse wurde
vernachlässigt, eine Summation über die Anfangszustandspins,
$\frac{1}{2s_\tau+1}=\frac{1}{2}$, wurde durchgeführt):
\bea\Gamma(\tau\rightarrow\mu\eta)=\frac{1}{2^8\pi}\left|\frac{
\lambda'_{32k}\lambda'_{22k}}{m_{\tilde{d_k}}^2}
-2\sqrt{3}\frac{m_{\eta_8}^2}{m_{\tau}(m_u+m_d+4m_s)}\frac{(
\lambda_{3k2}\lambda'_{k22}+\lambda_{2k3}\lambda'_{k22})}{m_{\tilde{l_k}}^2}
\right|^2
F_{\eta}^{2}\frac{(m_{\tau}^{2}-m_{\eta}^{2})^2}{m_\tau}\nonumber\eea
\bea =1.29\cdot
10^{-12}\left|\frac{\lambda'_{32k}\lambda'_{22k}}{m_{\tilde{d_k}}^2}
-2\sqrt{3}\frac{m_{\eta_8}^2}{m_{\tau}(m_u+m_d+4m_s)}\frac{(
\lambda_{3k2}\lambda'_{k22}+\lambda_{2k3}\lambda'_{k22})}{m_{\tilde{l_k}}^2}
\right|^2100GeV^5,\eea und \bea\Gamma(\tau\rightarrow
3\mu)&=&\frac{1}{(2\pi)^3
2^6}\frac{\left|\lambda_{3k2}\lambda_{2k2}+\lambda_{2k3}\lambda_{2k2}\right|^2}
{m_{\tilde{l_k}}^4}m_\tau^5{}\nonumber\\{}&=&
1.12\cdot
10^{-11}\left|\lambda_{3k2}\lambda_{2k2}+\lambda_{2k3}\lambda_{2k2}\right|^2
\left(\frac{100GeV}{m_{\tilde{l_k}}}\right)^4GeV
\; . \eea Wir beschränken uns nun auf zwei nicht-verschwindende
Kopplungskonstanten und nehmen an, das alle Übrigen verschwinden.
Die Schranken an die Kopplungskonstantenprodukte sind damit:
\beq|\lambda'_{32k}\lambda'_{22k}|<4.0\times
10^{-3}\left(\frac{m_{\tilde{d_k}}}{100GeV}\right)^2\; ,
\eeq\beq|\lambda_{3k2}\lambda'_{k22}|<4.8\times
10^{-3}\left(\frac{m_{\tilde{d_k}}}{100GeV}\right)^2\; ,\eeq mit
einem direkt entsprechenden Ausdruck für
$|\lambda_{2k3}\lambda'_{k22}|$ und mit einem direkt
entsprechenden Ausdruck für $|\lambda_{2k3}\lambda_{2k2}|$ hat
man: \beq|\lambda_{3k2}\lambda_{2k2}|<6.2\times
10^{-4}\left(\frac{m_{\tilde{l_k}}}{100GeV}\right)^2\; .\eeq
\newpage Bei diesen Rechnungen wurde Folgendes verwendet \cite{pdg}:
\begin {itemize}\item $m_\eta =547.30
MeV $ , \item $F_{\eta_8} = 154 MeV$ ,\item
$\frac{\Gamma(\tau\rightarrow\mu\eta)}{\Gamma}<9.6\times 10^{-6}$
und $\frac{\Gamma(\tau\rightarrow 3\mu)}{\Gamma}<1.9\times
10^{-6}$ . \item Die totale Zerfallsrate $\Gamma$ ist:
$$\Gamma=\frac{\hbar}{\tau}=2.265\cdot 10^{-12}GeV\; ,$$ also:
$$\Gamma(\tau\rightarrow\mu\eta)<2.17\cdot 10^{-17}GeV\; \textrm{und}\;
\Gamma(\tau\rightarrow 3\mu)<4.30\cdot 10^{-18}GeV\; .$$
\end{itemize} Die Schranke für
$|\lambda'_{32k}\lambda'_{22k}|$ ist nicht sehr gut bestimmt,
da die experimentelle Unsicherheit in $F_\eta$ relativ groß ist.\\
Es gibt auch andere Wege, diese Schranken zu berechnen:
$$ B\rightarrow \mu\bar{\mu}X \;\; \textrm{und} \;\; B\rightarrow l\nu X \; .$$
Um diese Zerfallsraten zu berechnen müssen einige Näherungen für
die Quarkströme vorgenommen werden (\textit{heavy quark
expansion}, QCD Korrekturen). Die obigen Schranken sind daher genauer.\\ \\
Es gibt einige leptonische $\tau$ Zerfälle, mit e's und $\mu$'s im
Endzustand, die ähnlich dem Zerfall $\tau\rightarrow 3\mu$
berechnet werden können. Es ergeben sich die folgenden oberen
Schranken (in Einheiten von
$\left(\frac{m_{\tilde{l_k}}}{100GeV}\right)^2$):\\
\begin{description}
\item\underline{ \textbf{$\tau\rightarrow e^{-}e^{+}e^{-}$}:
$|\lambda_{3k1}\lambda_{1k1}|<7.7\cdot 10^{-4}$\hspace{280pt}
}\item\underline{ \textbf{$\tau\rightarrow\mu^{-}e^{+}e^{-}$}:
$|\lambda_{3k1}\lambda_{2k1}|<5.9\cdot 10^{-4}$ und
$|\lambda_{3k2}\lambda_{1k1}|<5.9\cdot 10^{-4}$\hspace{142pt}
}\item\underline{ \textbf{$\tau\rightarrow\mu^{+}e^{-}e^{-}$}:
$|\lambda_{3k1}\lambda_{2k1}|<5.5\cdot 10^{-4}$\hspace{280pt} }
\item\underline{ \textbf{$\tau\rightarrow e^{+}\mu^{-}\mu^{-}$}:
$|\lambda_{3k2}\lambda_{1k2}|<5.5\cdot 10^{-4}$\hspace{280pt}
}\item\underline{ \textbf{$\tau\rightarrow e^{-}\mu^{+}\mu^{-}$}:
$|\lambda_{3k2}\lambda_{1k2}|<6.0\cdot 10^{-4}$ und
$|\lambda_{3k1}\lambda_{2k2}|<6.0\cdot 10^{-4}$\hspace{142pt} }
\end {description} Dieselben Schranken gelten unter Vertauschung des $1.$ und $3.$ Index in $\lambda$. Falls
 $i=j$ ist für $\lambda_{ijk}$ die Kopplung gleich Null (Asymmetrie in den ersten beiden
 Indizes).\\ \\ Der Zerfall
 $\tau\rightarrow\mu\gamma$ liefert keine starken Schranken. Die
 aus den entsprechenden Ein-Schleifen-Graphen berechenbaren Werte
 liegen im Bereich von
 $10^{-2}\left(\frac{m_{SUSY}}{100GeV}\right)^2$.
\appendix
\include{appendix}
\addcontentsline{toc}{chapter}{Tabellen}
\section*{\Huge\textbf{Anhang}
\\ \\\huge{Tabellen}}
\pagestyle {plain}
\vspace {1cm}
\begin{table} [h]
\begin{center}
\begin{tabular}{||c|c||}
\hline \textbf{Symbol} & \textbf{Bedeutung}
\\\hline $p$ & Impuls
\\\hline $E$ & Energie
\\\hline $m$ & Masse
\\\hline $\mathcal{L}$ & Lagrangefunktion
\\\hline $\mathcal {M}$ & invariante Amplitude (Matrixelement)
\\\hline $\Gamma$ & Zerfallsrate
\\\hline $\tau$ & Lebensdauer
\\\hline $M$ & Meson
\\\hline $\ell$ & Lepton
\\\hline $C$ & Ladungskonjugationsoperator
\\\hline $P_{L,R}$ & Projektionsoperator
\\\hline $A^\dagger$ & hermitesche Konjugation von $A$
\\\hline $a^*$ & komplexe Konjugation von $a$
\\\hline
\end{tabular}
\end{center}
\caption[Liste der verwendeten Symbole]{\textit{Liste der
verwendeten Symbole}} \label{tab:symbol}
\end{table}
\begin{table} [h]
\begin{center}
\begin{tabular}{||c|c|c|c||}
\hline \multicolumn {4}{||c||}{Konstanten}
\\\hline \multicolumn{2}{||c|}{$\hbar = 6,58211889(26)\times 10^{-22}MeVs$} &
\multicolumn {2}{|c||}{$ G_F= 1,16639(1)\times 10^{-5}GeV^{-2}$}
\\\hline \multicolumn {4}{||c||}{Massen/$MeV$}
\\\hline $m_u=1,5..4,5$ & $m_e=0,510998902(21)$ & $m_{\pi^+}=139,57018(35)$ &
$m_{\pi^0}=134,9766(6)$
\\\hline $m_d=5..8,5$ & $m_\mu =105,658357(5)$ & $m_{K^+}=493,677(16)$ & $m_{K^0}=497,672(31)$
\\\hline $m_s=80..155$ & $m_\tau =1776,99^{+0,29}_{-0,26}$ & $m_{D^+}=1869,3(5)$
& $m_{D^0}=1864,5(5)$
\\\hline $m_c=1000..1400$ & $\Delta m_{K^0}=3,490(6)\cdot 10^{-12}$ & $m_{B^+}=5279,0(5)$ &
$m_{B^0}=5279,4(5)$
\\\hline $m_b=4000..4500$&$\Delta m_{D^0}<5\cdot 10^{-11}$& $m_{D_s^+}=1968,5(6)$ & $m_{B_s^0}=5369,6(24)$
\\\hline $\frac{m_{u}+m_d}{2}=2,5..5,5$&   $\Delta m_{B^0}=3,22(5 )\cdot
10^{-10}$& &
\\\hline \multicolumn {4}{||c||}{Zerfallskonstanten/$MeV$}
\\\hline \multicolumn {2}{||c|}{$f_{\pi^+}=130,7\pm
0,1\pm 0,36$ ($f_{\pi^0}=130\pm 5 $)}& \multicolumn
{2}{|c||}{$f_{K^+}=159,8\pm 1,4\pm 0,44$}
\\\hline \multicolumn
{2}{||c|}{ $f_{D^+}=300^{+180+80}_{-150-40}$}& \multicolumn
{2}{|c||}{ $f_{D_s^+}=285\pm 19\pm 40$}
\\\hline \multicolumn{4}{||c||}{Lebensdauer/$s$}
\\\hline \multicolumn {2}{||c|}{$\tau_{\pi^+}=2,6033(5)\cdot 10^{-8}$}
&\multicolumn {2}{|c||}{ $\tau_{\pi^0}=8,4(6)\cdot
10^{-17}$}
\\\hline \multicolumn{2}{||c|}{$\tau_{K^+}=1,2384(24)\cdot 10^{-8}$} & \multicolumn
{2}{|c||}{$\tau_{K^0_L}=5,17(4)\cdot 10^{-8}$}
\\\hline\multicolumn {2}{||c|}{$\tau_{D^+}=1051(13)\cdot
10^{-15}$}&\multicolumn {2}{|c||}{ $\tau_{D^0}=411,7(27)\cdot
10^{-15}$ }
\\\hline \multicolumn {2}{||c|}{$\tau_{B^+}=1,674(18)\cdot
10^{-12}$}& \multicolumn {2}{|c||}{$\tau_{B^0}=1,542(16)\cdot
10^{-12}$}
\\\hline \multicolumn{2}{||c|}{$\tau_{D_s^+}=490(9)\cdot 10^{-15}$} & \multicolumn
{2}{|c||}{$\tau_{B_s^0}=1,461(57)\cdot 10^{-12}$}
\\\hline
\end{tabular}
\end{center}
\caption[Werte der verwendeten Größen]{\textit{Werte der
verwendeten Größen \cite{pdg}}} \label{tab:symbol}
\end{table}
\begin{table} [t]
\begin{center}
\begin{tabular} {||c||}
\hline $\sigma_i$-Matrizen (Pauli-Matrizen)
\\\hline\makebox(375,70)[b]{$\sigma_{1}=
\begin{pmatrix}0&1\\1&0\\\end{pmatrix} \: \sigma_{2}= \begin{pmatrix}0&-i\\i&0\\
\end{pmatrix} \: \sigma_{3}= \begin{pmatrix}1&0\\0&-1\\\end{pmatrix}$}
\\\hline $\gamma$-Matrizen in Dirac-Darstellung
\\\hline \makebox(375,70)[b]{$\gamma^0=\begin{pmatrix}
  \mathbf{1} & 0 \\
  0 & -\mathbf{1} \\
\end{pmatrix}$ $\gamma^\mu=\begin{pmatrix}
  0 & \sigma^\mu \\
  \sigma^\mu & 0 \\
\end{pmatrix}$ $\gamma^5=\begin{pmatrix}
  0 & \mathbf{1} \\
  \mathbf{1} & 0 \\
\end{pmatrix}$}
\\\hline Gell-Mann-Matrizen\\\hline \makebox(375,100)[b]{$\lambda_0=\sqrt{\frac{2}{3}}\begin{pmatrix}
  1 & 0 & 0 \\
  0 & 1 & 0 \\
  0 & 0 & 1 \\
\end{pmatrix}$ $\lambda_1=\begin{pmatrix}
  0 & 1 & 0 \\
  1 & 0 & 0 \\
  0 & 0 & 0 \\
\end{pmatrix}$
$\lambda_2=\begin{pmatrix}
  0 & -i & 0 \\
  i & 0 & 0 \\
  0 & 0 & 0 \\
\end{pmatrix}$}\\\vspace{0pt}\makebox(375,100)[b]{$\lambda_3=\begin{pmatrix}
  1 & 0 & 0 \\
  0 & -1 & 0 \\
  0 & 0 & 0 \\
\end{pmatrix}$ $\lambda_4=\begin{pmatrix}
  0 & 0 & 1 \\
  0 & 0 & 0 \\
  1 & 0 & 0 \\
\end{pmatrix}$ $\lambda_5=\begin{pmatrix}
  0 & 0 & -i \\
  0 & 0 & 0 \\
  i & 0 & 0 \\
\end{pmatrix}$}\\\vspace{0pt}\makebox(375,100)[b]{ $\lambda_6=\begin{pmatrix}
  0 & 0 & 0 \\
  0 & 0 & 1 \\
  0 & 1 & 0 \\
\end{pmatrix}$ $\lambda_7=\begin{pmatrix}
  0 & 0 & 0 \\
  0 & 0 & -i \\
  0 & i & 0 \\
\end{pmatrix}$ $\lambda_8=\frac{1}{\sqrt{3}}\begin{pmatrix}
  1 & 0 & 0 \\
  0 & 1 & 0 \\
  0 & 0 & -2 \\
\end{pmatrix}$}
\\\hline
\end{tabular}
\end{center}
\caption [Zusammenfassung der verwendeten
Matrizen]{\textit{Zusammenfassung der verwendeten Matrizen}}
\label{tab:matrizen}
\end{table}
\pagebreak
\pagestyle {myheadings}
\markboth {headings} {Tabellen}
\begin{table} [h]
\begin{center}
\begin{tabular}{||c|c|c||}
\hline \textbf{Interaction} & \textbf{4-fermion vertex} &
\textbf{Fierz-transformed vertex}\\
\hline $(\lambda_{LV_{0}}\overline{q}_{L}\gamma_{\mu}\ell_{L}$ &
$\frac{\lambda^{2}_{RV_{0}}}{m^{2}_{0}}
(\overline{d}_{R}\gamma^{\mu}e_{R})
(\overline{e}_{R}\gamma_{\mu}d_{R})$ &
$\frac{\lambda^{2}_{RV_{0}}}{m^{2}_{0}}
(\overline{d}_{R}\gamma^{\mu}d_{R})
(\overline{e}_{R}\gamma_{\mu}e_{R})$\\
$+\lambda_{RV_{0}}\overline{d}_{R}\gamma_\mu
e_{R})V^{\mu\dagger}_{0}$ &
$\frac{\lambda^{2}_{LV_{0}}}{m^{2}_{0}}
(\overline{u}_{L}\gamma^{\mu}\nu_{L})
(\overline{\nu}_{L}\gamma_{\mu}u_{L})$ &
$\frac{\lambda^{2}_{LV_{0}}}{m^{2}_{0}}
(\overline{u}_{L}\gamma^{\mu}u_{L})
(\overline{\nu}_{L}\gamma_{\mu}\nu_{L})$\\ &
$\frac{\lambda^{2}_{LV_{0}}}{m^{2}_{0}}
(\overline{d}_{L}\gamma^{\mu}e_{L})
(\overline{\nu}_{L}\gamma_{\mu}u_{L})$ &
$\frac{\lambda^{2}_{LV_{0}}}{m^{2}_{0}}
(\overline{d}_{L}\gamma^{\mu}u_{L})
(\overline{\nu}_{L}\gamma_{\mu}e_{L})$\\ &
$\frac{\lambda^{2}_{LV_{0}}}{m^{2}_{0}}
(\overline{d}_{L}\gamma^{\mu}e_{L})
(\overline{e}_{L}\gamma_{\mu}d_{L})$ &
$\frac{\lambda^{2}_{LV_{0}}}{m^{2}_{0}}
(\overline{d}_{L}\gamma^{\mu}d_{L})
(\overline{e}_{L}\gamma_{\mu}e_{L})$\\ &
$\frac{\lambda_{LV_{0}}\lambda_{RV_{0}}}{m^{2}_{0}}
(\overline{d}_{R}\gamma^{\mu}e_{R})
(\overline{\ell}_{L}\gamma_{\mu}q_{L})$ &
$\frac{\lambda_{LV_{0}}\lambda_{RV_{0}}}{m^{2}_{0}}
(\overline{d}_{R}u_{L}) (\overline{\nu}_{L}e_{R})$\\ & &
$\frac{\lambda_{LV_{0}}\lambda_{RV_{0}}}{m^{2}_{0}}
(\overline{d}_{R}d_{L}) (\overline{e}_{L}e_{R})$\\\hline
 $\lambda_{R\tilde{V_0}} \bar{u}_R \gamma_{\mu} e_R
\tilde{V}^{\mu \dagger}_o $ &
$\frac{\lambda^{2}_{R\tilde{V_{0}}}}{\tilde{m}^{2}_{0}}
(\overline{u}_{R}\gamma^{\mu}e_{R})
(\overline{e}_{R}\gamma_{\mu}u_{R})$ &
$\frac{\lambda^{2}_{R\tilde{V_{0}}}}{\tilde{m}^{2}_{0}}
(\overline{u}_{R}\gamma^{\mu}u_{R})
(\overline{e}_{R}\gamma_{\mu}e_{R})$\\\hline $(\lambda_{LV_{1/2}}
\bar{d}_R^c \gamma_{\mu}\ell_L$ & $\frac{\lambda^{2}_{R
V_{1/2}}}{m^{2}_{1/2}} (\overline{u}_{L}^{c}\gamma^{\mu}e_{R})
(\overline{e}_{R}\gamma_{\mu}u_{L}^{c})$ & $\frac{\lambda^{2}_{R
V_{1/2}}}{m^{2}_{1/2}} (\overline{u}_{L}^{c}\gamma^{\mu}u_{L}^{c})
(\overline{e}_{R}\gamma_{\mu}e_{R})$\\
$+\lambda_{RV_{1/2}}\bar{q}_L^c \gamma_{\mu} e_R)V_{1/2}^{\mu
\dagger}$ & $\frac{\lambda^{2}_{R V_{1/2}}}{m^{2}_{1/2}}
(\overline{d}_{L}^{c}\gamma^{\mu}e_{R})
(\overline{e}_{R}\gamma_{\mu}d_{L}^{c})$ & $\frac{\lambda^{2}_{R
V_{1/2}}}{m^{2}_{1/2}} (\overline{d}_{L}^{c}\gamma^{\mu}d_{L}^{c})
(\overline{e}_{R}\gamma_{\mu}e_{R})$\\ & $\frac{\lambda^{2}_{L
V_{1/2}}}{m^{2}_{1/2}} (\overline{d}_{R}^{c}\gamma^{\mu}e_{L})
(\overline{e}_{L}\gamma_{\mu}d_{R}^{c})$ & $\frac{\lambda^{2}_{L
V_{1/2}}}{m^{2}_{1/2}} (\overline{d}_{R}^{c}\gamma^{\mu}d_{R}^{c})
(\overline{e}_{L}\gamma_{\mu}e_{L})$\\ & $\frac{\lambda^{2}_{L
V_{1/2}}}{m^{2}_{1/2}} (\overline{d}_{R}^{c}\gamma^{\mu}\nu_{L})
(\overline{\nu}_{L}\gamma_{\mu}d_{R}^{c})$ & $\frac{\lambda^{2}_{L
V_{1/2}}}{m^{2}_{1/2}} (\overline{d}_{R}^{c}\gamma^{\mu}d_{R}^{c})
(\overline{\nu}_{L}\gamma_{\mu}\nu_{L})$\\ & $\frac{\lambda_{L
V_{1/2}}\lambda_{R V_{1/2}}}{m^{2}_{1/2}}
(\overline{d}_{L}^{c}\gamma^{\mu}e_{R})
(\overline{e}_{L}\gamma_{\mu}d_{R}^{c})$ & $\frac{\lambda_{L
V_{1/2}}\lambda_{R V_{1/2}}}{m^{2}_{1/2}}
(\overline{d}_{L}^{c}d_{R}^{c}) (\overline{e}_{L}e_{R})$\\ &
$\frac{\lambda_{L V_{1/2}}\lambda_{R V_{1/2}}}{m^{2}_{1/2}}
(\overline{u}_{L}^{c}\gamma^{\mu}e_{R})
(\overline{\nu}_{L}\gamma_{\mu}d_{R}^{c})$ & $\frac{\lambda_{L
V_{1/2}}\lambda_{R V_{1/2}}}{m^{2}_{1/2}}
(\overline{u}_{L}^{c}d_{R}^{c}) (\overline{\nu}_{L}e_{R})$\\\hline
$ \lambda_{L \tilde{V}_{1/2}}
 \bar{u}_R^c \gamma_{\mu} \ell_L  \tilde{V}^{\mu \dagger}_{1/2}$ &
 $\frac{\lambda^{2}_{L
\tilde{V}_{1/2}}}{\tilde{m}^{2}_{1/2}}
(\overline{u}_{R}^{c}\gamma^{\mu}\nu_{L})
(\overline{\nu}_{L}\gamma_{\mu}u_{R}^{c})$ & $\frac{\lambda^{2}_{L
\tilde{V}_{1/2}}}{\tilde{m}^{2}_{1/2}}
(\overline{u}_{R}^{c}\gamma^{\mu}u_{R}^{c})
(\overline{\nu}_{L}\gamma_{\mu}\nu_{L})$\\ & $\frac{\lambda^{2}_{L
\tilde{V}_{1/2}}}{\tilde{m}^{2}_{1/2}}
(\overline{u}_{R}^{c}\gamma^{\mu}e_{L})
(\overline{e}_{L}\gamma_{\mu}u_{R}^{c})$ & $\frac{\lambda^{2}_{L
\tilde{V}_{1/2}}}{\tilde{m}^{2}_{1/2}}
(\overline{u}_{R}^{c}\gamma^{\mu}u_{R}^{c})
(\overline{e}_{L}\gamma_{\mu}e_{L})$\\\hline $\lambda_{LV_1}
\bar{q}_L \gamma_{\mu} \vec{\sigma} \ell_L \cdot \vec{V}^{\mu
\dagger} _1$ & $\frac{\lambda^{2}_{LV_{1}}}{m^{2}_{1}}
(\overline{u}_{L}\gamma^{\mu}\nu_{L})
(\overline{\nu}_{L}\gamma_{\mu}u_{L})$ &
$\frac{\lambda^{2}_{LV_{1}}}{m^{2}_{1}}
(\overline{u}_{L}\gamma^{\mu}u_{L})
(\overline{\nu}_{L}\gamma_{\mu}\nu_{L})$\\ &
$-\frac{\lambda^{2}_{LV_{1}}}{m^{2}_{1}}
(\overline{u}_{L}\gamma^{\mu}\nu_{L})
(\overline{e}_{L}\gamma_{\mu}d_{L})$ &
$-\frac{\lambda^{2}_{LV_{1}}}{m^{2}_{1}}
(\overline{u}_{L}\gamma^{\mu}d_{L})
(\overline{e}_{L}\gamma_{\mu}\nu_{L})$\\ &
$\frac{\lambda^{2}_{LV_{1}}}{m^{2}_{1}}
(\overline{d}_{L}\gamma^{\mu}e_{L})
(\overline{e}_{L}\gamma_{\mu}d_{L})$ &
$\frac{\lambda^{2}_{LV_{1}}}{m^{2}_{1}}
(\overline{d}_{L}\gamma^{\mu}d_{L})
(\overline{e}_{L}\gamma_{\mu}e_{L})$\\ &
$2\frac{\lambda^{2}_{LV_{1}}}{m^{2}_{1}}
(\overline{u}_{L}\gamma^{\mu}e_{L})
(\overline{e}_{L}\gamma_{\mu}u_{L})$ &
$2\frac{\lambda^{2}_{LV_{1}}}{m^{2}_{1}}
(\overline{u}_{L}\gamma^{\mu}u_{L})
(\overline{e}_{L}\gamma_{\mu}e_{L})$\\ &
$2\frac{\lambda^{2}_{LV_{1}}}{m^{2}_{1}}
(\overline{d}_{L}\gamma^{\mu}\nu_{L})
(\overline{\nu}_{L}\gamma_{\mu}d_{L})$ &
$2\frac{\lambda^{2}_{LV_{1}}}{m^{2}_{1}}
(\overline{d}_{L}\gamma^{\mu}d_{L})
(\overline{\nu}_{L}\gamma_{\mu}\nu_{L})$\\\hline
\end{tabular}
\end{center}
Table 4: \textit{4-Fermion-vertices for vector leptoquarks} \cite
{lepto}. \label{tab:vertizes}\addcontentsline{lot}{table}{4\qquad
4-Fermionen-Vertizes für vektorielle Leptoquarks}
\end{table}
\begin{table}[h]
\begin{center}
\begin{tabular}{||c|c|c||}
\hline \textbf{Interaction} & \textbf{4-fermion vertex} &
\textbf{Fierz-transformed vertex}\\ \hline $\lambda_{LS_0}
\bar{q}^c_L i\sigma_2 \ell_L$ &
$\frac{\lambda^{2}_{RS_{0}}}{m^{2}_{0}} (\overline{u}_{R}^c e_{R})
(\overline{e}_{R}u_{R}^c)$ &
$\frac{\lambda^{2}_{RS_{0}}}{2m^{2}_{0}}
(\overline{u}_{R}^c\gamma^{\mu}u_{R}^c)
(\overline{e}_{R}\gamma_{\mu}e_{R})$\\ $ + \lambda_{RS_o}
\bar{u}^c_R e_R) S_o^{\dagger}$
 & $\frac{\lambda^{2}_{SV_{0}}}{m^{2}_{0}}
(\overline{u}_{L}^c e_{L}) (\overline{e}_{L}u_{L}^c)$ &
$\frac{\lambda^{2}_{LS_{0}}}{2m^{2}_{0}}
(\overline{u}_{L}^c\gamma^{\mu}u_{L}^c)
(\overline{e}_{L}\gamma_{\mu}e_{L})$\\ &
$\frac{\lambda^{2}_{LS_{0}}}{m^{2}_{0}} (\overline{u}_{L}^c e_{L})
(\overline{\nu}_{L}d_{L}^c)$ &
$\frac{\lambda^{2}_{LS_{0}}}{2m^{2}_{0}}
(\overline{u}_{L}^c\gamma^{\mu}d_{L}^c)
(\overline{\nu}_{L}\gamma_{\mu}e_{L})$\\ &
$\frac{\lambda^{2}_{LS_{0}}}{m^{2}_{0}} (\overline{d}_{L}^c
\nu_{L}) (\overline{\nu}_{L}d_{L}^c)$ &
$\frac{\lambda^{2}_{LS_{0}}}{2m^{2}_{0}}
(\overline{d}_{L}^c\gamma^{\mu}d_{L}^c)
(\overline{\nu}_{L}\gamma_{\mu}\nu_{L})$\\ &
$\frac{\lambda_{LS_{0}}\lambda_{RS_{0}}}{m^{2}_{0}}
(\overline{q}_{L}^c i\sigma_2 \ell_{L}) (\overline{e}_{R}u_{R}^c)$
& $\frac{\lambda_{LS_{0}}\lambda_{RS_{0}}}{2m^{2}_{0}}
(\overline{u}_{L}^c u_{R}^c) (\overline{e}_{R}e_{L})$\\ & &
$\frac{\lambda_{LS_{0}}\lambda_{RS_{0}}}{2m^{2}_{0}}
(\overline{d}_{L}^c u_{R}^c) (\overline{e}_{R}\nu_{L})$\\\hline
 $\lambda_{R \tilde{S_o}}
\bar{d}_R^c e_R \tilde{S}_0^{\dagger} $ &
$\frac{\lambda^{2}_{R\tilde{S_{0}}}}{\tilde{m^{2}_{0}}}
(\overline{d}_{R}^c e_{R}) (\overline{e}_{R}d_{R}^c)$ &
$\frac{\lambda^{2}_{R\tilde{S_{0}}}}{\tilde{m^{2}_{0}}}
(\overline{d}_{R}^c\gamma^{\mu}d_{R}^c)
(\overline{e}_{R}\gamma_{\mu}e_{R})$\\\hline $(\lambda_{LS_{1/2}}
\bar{u}_R \ell_L$ & $\frac{\lambda^{2}_{RS_{1/2}}}{m^{2}_{1/2}}
(\overline{u}_{R}\nu_{L}) (\overline{\nu}_{L}u_{R})$ &
$\frac{\lambda^{2}_{R V_{1/2}}}{2m^{2}_{1/2}}
(\overline{u}_{R}\gamma^{\mu}u_{R})
(\overline{\nu}_{L}\gamma_{\mu}\nu_{L})$\\ $
+\lambda_{RS_{1/2}}\bar{q}_L  i\sigma_2  e_R) S_{1/2}^{\dagger}$ &
$\frac{\lambda^{2}_{R S_{1/2}}}{m^{2}_{1/2}} (\overline{u}_{R}
e_{L}) (\overline{e}_{L}u_{R})$ & $\frac{\lambda^{2}_{R
S_{1/2}}}{2m^{2}_{1/2}} (\overline{u}_{R}\gamma^{\mu}u_{R})
(\overline{e}_{L}\gamma_{\mu}e_{L})$\\ & $\frac{\lambda^{2}_{L
S_{1/2}}}{m^{2}_{1/2}} (\overline{u}_{L}e_{R})
(\overline{e}_{R}u_{L})$ & $\frac{\lambda^{2}_{L
S_{1/2}}}{2m^{2}_{1/2}} (\overline{u}_{L}\gamma^{\mu}u_{L})
(\overline{e}_{R}\gamma_{\mu}e_{R})$\\ & $\frac{\lambda^{2}_{L
S_{1/2}}}{m^{2}_{1/2}} (\overline{d}_{L}e_{R})
(\overline{e}_{R}d_{L})$ & $\frac{\lambda^{2}_{L
S_{1/2}}}{2m^{2}_{1/2}} (\overline{d}_{L}\gamma_{\mu}d_{L})
(\overline{e}_{R}\gamma_{\mu}e_{R})$\\ & $\frac{\lambda_{L
S_{1/2}}\lambda_{R S_{1/2}}}{m^{2}_{1/2}}
(\overline{u}_{R}\nu_{L}) (\overline{e}_{R}d_{L})$ &
$\frac{\lambda_{L S_{1/2}}\lambda_{R S_{1/2}}}{2m^{2}_{1/2}}
(\overline{u}_{R}d_{L}) (\overline{e}_{R}\nu_{L})$\\ &
$\frac{\lambda_{L S_{1/2}}\lambda_{R S_{1/2}}}{m^{2}_{1/2}}
(\overline{u}_{R}e_{L}) (\overline{e}_{R}u_{L})$ &
$\frac{\lambda_{L V_{1/2}}\lambda_{R V_{1/2}}}{2m^{2}_{1/2}}
(\overline{u}_{R}u_{L}) (\overline{e}_{R}e_{L})$\\\hline $
\lambda_{L \tilde{S}_{1/2}} \bar{d}_R \ell_L
\tilde{S}_{1/2}^{\dagger}$ &
 $\frac{\lambda^{2}_{L
\tilde{S}_{1/2}}}{\tilde{m}^{2}_{1/2}}
(\overline{d}_{R}^{c}\nu_{L}) (\overline{\nu}_{L}d_{R})$ &
$\frac{\lambda^{2}_{L \tilde{S}_{1/2}}}{2\tilde{m}^{2}_{1/2}}
(\overline{d}_{R}\gamma^{\mu}ud_{R})
(\overline{\nu}_{L}\gamma_{\mu}\nu_{L})$\\ & $\frac{\lambda^{2}_{L
\tilde{S}_{1/2}}}{\tilde{m}^{2}_{1/2}} (\overline{d}_{R}e_{L})
(\overline{e}_{L}d_{R})$ & $\frac{\lambda^{2}_{L
\tilde{S}_{1/2}}}{2\tilde{m}^{2}_{1/2}}
(\overline{d}_{R}\gamma^{\mu}d_{R})
(\overline{e}_{L}\gamma_{\mu}e_{L})$\\\hline $\lambda_{LS_{1}}
\bar{q}^c_L i\sigma_2 \vec{\sigma} \ell_L \cdot
\vec{S}_1^{\dagger}$ & $\frac{\lambda^{2}_{LS_{1}}}{m^{2}_{1}}
(\overline{d}_{L}^c\nu_{L}) (\overline{\nu}_{L}d_{L}^c)$ &
$\frac{\lambda^{2}_{LS_{1}}}{2m^{2}_{1}}
(\overline{d}_{L}^c\gamma^{\mu}d_{L}^c)
(\overline{\nu}_{L}\gamma_{\mu}\nu_{L})$\\ &
$-\frac{\lambda^{2}_{LS_{1}}}{m^{2}_{1}}
(\overline{d}_{L}^c\nu_{L}) (\overline{e}_{L}u_{L}^c)$ &
$-\frac{\lambda^{2}_{LS_{1}}}{2m^{2}_{1}}
(\overline{d}_{L}^c\gamma^{\mu}u_{L}^c)
(\overline{e}_{L}\gamma_{\mu}\nu_{L})$\\ &
$\frac{\lambda^{2}_{LS_{1}}}{m^{2}_{1}} (\overline{u}_{L}^c e_{L})
(\overline{e}_{L}u_{L}^c)$ &
$\frac{\lambda^{2}_{LS_{1}}}{2m^{2}_{1}}
(\overline{u}_{L}^c\gamma^{\mu}u_{L}^c)
(\overline{e}_{L}\gamma_{\mu}e_{L})$\\ &
$2\frac{\lambda^{2}_{LS_{1}}}{m^{2}_{1}} (\overline{d}_{L}^c
e_{L}) (\overline{e}_{L}d_{L}^c)$ &
$\frac{\lambda^{2}_{LS_{1}}}{m^{2}_{1}}
(\overline{d}_{L}^c\gamma^{\mu}d_{L}^c)
(\overline{e}_{L}\gamma_{\mu}e_{L})$\\ &
$2\frac{\lambda^{2}_{LS_{1}}}{m^{2}_{1}}
(\overline{u}_{L}^c\nu_{L}) (\overline{\nu}_{L}u_{L}^c)$ &
$\frac{\lambda^{2}_{LV_{1}}}{m^{2}_{1}}
(\overline{u}_{L}^c\gamma^{\mu}u_{L}^c)
(\overline{\nu}_{L}\gamma_{\mu}\nu_{L})$\\\hline
\end{tabular}
\end{center}
Table 5: \textit{4-Fermion-vertices for scalar leptoquarks} \cite
{lepto}.\label{tab:vertizes}\addcontentsline{lot}{table}{5\qquad
4-Fermionen-Vertizes für skalare Leptoquarks}
\end{table}
\pagebreak
\begin{table} [h]
\begin{center}
\begin{tabular}{||c|c|c|c||}
\hline Product  & Upper
bound/$\left(\frac{m_{LQ}}{100GeV}\right)^2$ & Exchanged particle
& Source of bound\\\hline
 $\Re(\lambda_L^{11}\lambda_L^{*11})$ & $3.4\times 10^{-4}$ & $V_O$ & $R_\pi$\\\hline
 $\Re(\lambda_L^{11}\lambda_L^{*11})$ & $1.7\times 10^{-3}$ & $V_1$ & $R_\pi$\\\hline
 $\Re(\lambda_L^{11}\lambda_R^{*11})$ & $9.8\times 10^{-8}$ & $V_O$ & $R_\pi$\\\hline
 $\Re(\lambda_L^{11}\lambda_R^{*11})$ & $5.0\times 10^{-7}$ & $V_{1/2}$ & $R_\pi$\\\hline
 $\Re(\lambda_L^{21}\lambda_L^{*21})$ & $1.7\times 10^{-3}$ & $V_O$ & $R_\pi$\\\hline
 $\Re(\lambda_L^{21}\lambda_L^{*21})$ & $3.4\times 10^{-4}$ & $V_1$ & $R_\pi$\\\hline
 $\Re(\lambda_L^{21}\lambda_R^{*21})$ & $1.0\times 10^{-4}$ & $V_O$ & $R_\pi$\\\hline
 $\Re(\lambda_L^{21}\lambda_R^{*21})$ & $2.0\times 10^{-5}$ & $V_{1/2}$ & $R_\pi$\\\hline
 $|\lambda_L^{11}\lambda_L^{*21}|$ & $2.9\times 10^{-2}$ & $V_O/V_1$ & $\pi^+\rightarrow\bar{\mu}\nu_e$\\\hline
 $|\lambda_L^{11}\lambda_R^{*21}|$ & $1.7\times 10^{-3}$ & $V_O/V_{1/2}$ & $\pi^+\rightarrow\bar{\mu}\nu_e$\\\hline
 $\Re(\lambda_L^{12}\lambda_L^{*11})$ & $2.8\times 10^{-3}$ & $V_O$ & $R_K$\\\hline
 $\Re(\lambda_L^{12}\lambda_L^{*11})$ & $3.7\times 10^{-3}$ & $V_1$ & $R_K$\\\hline
 $\Re(\lambda_L^{12}\lambda_R^{*11})$ & $9.2\times 10^{-7}$ & $V_O$ & $R_K$\\\hline
 $\Re(\lambda_L^{12}\lambda_R^{*11})$ & $1.2\times 10^{-6}$ & $V_{1/2}$ & $R_K$\\\hline
 $\Re(\lambda_L^{22}\lambda_L^{*21})$ & $3.7\times 10^{-3}$ & $V_O$ & $R_K$\\\hline
 $\Re(\lambda_L^{22}\lambda_L^{*21})$ & $2.8\times 10^{-3}$ & $V_1$ & $R_K$\\\hline
 $\Re(\lambda_L^{22}\lambda_R^{*21})$ & $2.6\times 10^{-4}$ & $V_O$ & $R_K$\\\hline
 $\Re(\lambda_L^{22}\lambda_R^{*21})$ & $1.9\times 10^{-4}$ & $V_{1/2}$ & $R_K$\\\hline
 $|\lambda_L^{12}\lambda_L^{*21}|$ & $5.8\times 10^{-3}$ &
$V_O/V_1$ & $K^+\rightarrow\mu\nu_e$\\\hline
\end{tabular}
\end{center}
Table 6: \textit{Bounds on products of coupling constants for
vector LQs. Bounds with a *) are also valid under the exchange of
the lepton
indices.}\label{tab:vertizes}\addcontentsline{lot}{table}{6\qquad
Schranken an Kopplungskonstantenprodukte für vektorielle LQs}
\end{table}\newpage\begin{table} [h]
\begin{center}
\begin{tabular}{||c|c|c|c||}
\hline Product  & Upper
bound/$\left(\frac{m_{LQ}}{100GeV}\right)^2$ & Exchanged particle
& Source of bound\\\hline $|\lambda_L^{12}\lambda_R^{*21}|$ &
$4.0\times 10^{-4}$ & $V_O/V_{1/2}$ &
$K^+\rightarrow\mu\nu_e$\\\hline $|\lambda_L^{12}\lambda_L^{*11}|$
& $1.2\times 10^{-4}$ & $V_O/V_{1/2}/V_1$ &
$\frac{BR(K^+\rightarrow\pi^+e\bar{e})}{BR(K^+\rightarrow\pi^0\bar{e}\nu_e)}$\\\hline
$|\lambda_L^{12}\lambda_L^{*21}|$ & $1.5\times 10^{-6}$ &
$V_O/V_{1/2}/V_1$ &
$\frac{BR(K^+\rightarrow\pi^+e\bar{\mu})}{BR(K^+\rightarrow\pi^0\bar{\mu}\nu_\mu)}$\\\hline
$|\lambda_L^{22}\lambda_L^{*11}|$ & $6.5\times 10^{-6}$ &
$V_O/V_{1/2}/V_1$ &
$\frac{BR(K^+\rightarrow\pi^+\mu\bar{e})}{BR(K^+\rightarrow\pi^0\bar{\mu}\nu_\mu)}$\\\hline
$|\lambda_L^{i2}\lambda_L^{*m1}|$ & $2.9\times 10^{-6}$ &
$V_{1/2}$ &
$\frac{BR(K^+\rightarrow\pi^+\nu\bar{\nu})}{BR(K^+\rightarrow\pi^0\bar{e}\nu_e)}$\\\hline
$|\lambda_L^{i2}\lambda_L^{*m1}|$ & $1.5\times 10^{-6}$ & $V_1$ &
$\frac{BR(K^+\rightarrow\pi^+\nu\bar{\nu})}{BR(K^+\rightarrow\pi^0\bar{e}\nu_e)}$\\\hline
$|\lambda_L^{22}\lambda_L^{*21}|$ & $7.8\times 10^{-5}$ &
$V_O/V_{1/2}/V_1$ &
$\frac{BR(K^+\rightarrow\pi^+\mu\bar{\mu})}{BR(K^+\rightarrow\pi^0\bar{\mu}\nu_\mu)}$\\\hline
$|\lambda_L^{i2}\lambda_R^{*m1}|$ & $3.8\times 10^{-3}$ &
$V_O/V_{1/2}$ & $\frac{f_S}{f_+(0)}$\\\hline
 $|\lambda_L^{12}\lambda_L^{*11}|$ & $1.9\times 10^{-5}$ & $V_O/V_{1/2}/V_1$ & $K^0_L\rightarrow e\bar{e}$
 \\\hline
 $|\lambda_L^{12}\lambda_L^{*21}|$ & $9.7\times 10^{-8}$ & $V_O/V_{1/2}/V_1$ & $K^0_L\rightarrow\bar{\mu}e$\quad *)
 \\\hline
 $|\lambda_L^{22}\lambda_L^{*21}|$ & $2.7\times 10^{-6}$ & $V_O/V_{1/2}/V_1$ & $K^0_L\rightarrow\bar{\mu}\mu$
 \\\hline
 $|\lambda_L^{12}\lambda_R^{*11}|$ & $9.1\times 10^{-9}$ & $V_O/V_{1/2}$ & $K^0_L\rightarrow e\bar{e}$
 \\\hline
 $|\lambda_L^{12}\lambda_R^{*21}|$ & $6.9\times 10^{-9}$ & $V_O/V_{1/2}$ & $K^0_L\rightarrow\bar{\mu}e$\quad *)
 \\\hline
 $|\lambda_L^{22}\lambda_R^{*21}|$ & $2.8\times 10^{-7}$ & $V_O/V_{1/2}$ & $K^0_L\rightarrow\mu\bar{\mu}$
 \\\hline
  $|\lambda_L^{22}\lambda_L^{*21}|$ & $3.7\times 10^{-1}$ & $V_O/V_{1/2}/V_1$ & $K^0-\bar{K}^0$
  \\\hline
  $|\lambda_L^{32}\lambda_L^{*31}|$ & $2.2\times 10^{-2}$ & $V_O/V_{1/2}/V_1$ & $K^0-\bar{K}^0$
  \\\hline
  $|\lambda_L^{i2}\lambda_L^{*m1}|$ & $7.4\times 10^{-2}$ &
$V_O/V_{1/2}/V_1$ & $V_{cd}$
\\\hline
$|\lambda_L^{i2}\lambda_L^{*m2}|$ & $3.3\times 10^{-1}$ &
$V_O/V_{1/2}/V_1$ & $V_{cs}$
\\\hline
$\Re(\lambda_L^{21}\lambda_L^{*22})$ & $1.0\times 10^{-1}$ &
$V_O/V_1$ & $D^+\rightarrow \mu^+\nu_\mu$
\\\hline
$\Re(\lambda_L^{21}\lambda_R^{*22})$ & $4.3\times 10^{-3}$ &
$V_O/V_{1/2}$ & $D^+\rightarrow \mu^+\nu_\mu$
\\\hline
\end{tabular}\end{center}
\label{tab:vertizes1}Table 6: \textit{Continued.}
\end{table}\newpage\begin{table} [h]
\begin{center}
\begin{tabular}{||c|c|c|c||}
\hline Product  & Upper
bound/$\left(\frac{m_{LQ}}{100GeV}\right)^2$ & Exchanged particle
& Source of bound\\\hline
$\Re(\lambda_L^{22}\lambda_L^{*22})$ & $2.0\times 10^{-2}$ & $V_O$
&
$\frac{BR(D^+_s\rightarrow\mu\nu_\mu)}{BR(D^+_s\rightarrow\tau\nu_\tau)}$\\\hline
$\Re(\lambda_L^{22}\lambda_L^{*22})$ & $9.3\times 10^{-2}$ & $V_1$
&
$\frac{BR(D^+_s\rightarrow\mu\nu_\mu)}{BR(D^+_s\rightarrow\tau\nu_\tau)}$\\\hline
$\Re(\lambda_L^{22}\lambda_R^{*22})$ & $8.3\times 10^{-4}$ & $V_O$
&
$\frac{BR(D^+_s\rightarrow\mu\nu_\mu)}{BR(D^+_s\rightarrow\tau\nu_\tau)}$\\\hline
$\Re(\lambda_L^{22}\lambda_R^{*22})$ & $3.9\times 10^{-3}$ &
$V_{1/2}$ &
$\frac{BR(D^+_s\rightarrow\mu\nu_\mu)}{BR(D^+_s\rightarrow\tau\nu_\tau)}$\\\hline
$\Re(\lambda_L^{32}\lambda_L^{*32})$ & $9.3\times 10^{-2}$ & $V_O$
&
$\frac{BR(D^+_s\rightarrow\mu\nu_\mu)}{BR(D^+_s\rightarrow\tau\nu_\tau)}$\\\hline
$\Re(\lambda_L^{32}\lambda_L^{*32})$ & $2.0\times 10^{-2}$ & $V_1$
&
$\frac{BR(D^+_s\rightarrow\mu\nu_\mu)}{BR(D^+_s\rightarrow\tau\nu_\tau)}$\\\hline
$\Re(\lambda_L^{32}\lambda_R^{*32})$ & $6.6\times 10^{-2}$ & $V_O$
&
$\frac{BR(D^+_s\rightarrow\mu\nu_\mu)}{BR(D^+_s\rightarrow\tau\nu_\tau)}$\\\hline
$\Re(\lambda_L^{32}\lambda_R^{*32})$ & $1.4\times 10^{-2}$ &
$V_{1/2}$ &
$\frac{BR(D^+_s\rightarrow\mu\nu_\mu)}{BR(D^+_s\rightarrow\tau\nu_\tau)}$
\\\hline
$|\lambda_R^{12}\lambda_R^{*11}|$ & $5.6\times 10^{-3}$ &
$\tilde{V}_O/V_{1/2}$ &
$\frac{BR(D^+\rightarrow\pi^+e\bar{e})}{BR(D^0\rightarrow\pi^-\nu_e
\bar{e})}$
\\\hline $|\lambda_R^{22}\lambda_R^{*11}|$ & $4.5\times
10^{-3}$ & $\tilde{V}_O/V_{1/2}$ &
$\frac{BR(D^+\rightarrow\pi^+\mu\bar{e})}{BR(D^0\rightarrow\pi^-\nu_e
\bar{e})}$\quad *)
\\\hline$|\lambda_R^{22}\lambda_R^{*21}|$ &
$3.0\times 10^{-3}$ & $\tilde{V}_O/V_{1/2}$ &
$\frac{BR(D^+\rightarrow\pi^+\mu\bar{\mu})}{BR(D^0\rightarrow\pi^-\nu_e
\bar{e})}$
\\\hline $|\lambda_L^{12}\lambda_L^{*11}|$ & $5.6\times
10^{-3}$ & $\tilde{V}_{1/2}/V_1$**) &
$\frac{BR(D^+\rightarrow\pi^+e\bar{e})}{BR(D^0\rightarrow\pi^-\nu_e
\bar{e})}$
\\\hline $|\lambda_L^{22}\lambda_L^{*11}|$ & $4.5\times
10^{-3}$ & $\tilde{V}_{1/2}/V_1$**) &
$\frac{BR(D^+\rightarrow\pi^+\mu\bar{e})}{BR(D^0\rightarrow\pi^-\nu_e
\bar{e})}$\quad *)
\\\hline$|\lambda_L^{22}\lambda_L^{*21}|$ &
$3.0\times 10^{-3}$ & $\tilde{V}_{1/2}/V_1$**) &
$\frac{BR(D^+\rightarrow\pi^+\mu\bar{\mu})}{BR(D^0\rightarrow\pi^-\nu_e
\bar{e})}$
\\\hline
$|\lambda_R^{12}\lambda_R^{*11}|$ & $1.6$ &
$\tilde{V}_O/V_{1/2}$**) & $D^0\rightarrow e\bar{e}$\\\hline
$|\lambda_R^{12}\lambda_R^{*21}|$ & $1.2\times 10^{-2}$ &
$\tilde{V}_O/V_{1/2}$**) & $D^0\rightarrow\bar{\mu}e$\quad *)
\\\hline
$|\lambda_R^{22}\lambda_R^{*21}|$ & $6.3\times 10^{-3}$ &
$\tilde{V}_O/V_{1/2}$**) & $D^0\rightarrow\bar{\mu}\mu$
\\\hline
    $|\lambda_R^{22}\lambda_R^{*21}|$ & $1.1$ & $\tilde{V}_O/V_{1/2}$**) & $D^0-\bar{D}^0$\\\hline
     $|\lambda_R^{32}\lambda_R^{*31}|$ & $6.4\times 10^{-2}$ & $\tilde{V}_O/V_{1/2}$**) &
  $D^0-\bar{D}^0$
\\\hline
  $|\lambda_L^{i3}\lambda_L^{*m1}|$ & $1.2\times 10^{-3}$ &
$V_O/V_{1/2}/V_1$ & $V_{ub}$
\\\hline
\end{tabular}\end{center}
\label{tab:vertizes1}Table 6: \textit{Continued; **) for $V_1$ a
factor 1/2 has to be employed, LL-couplings for $D^0$ analogous.}
\end{table}\newpage
\begin{table} [h]
\begin{center}
\begin{tabular}{||c|c|c|c||}
\hline Product  & Upper
bound/$\left(\frac{m_{LQ}}{100GeV}\right)^2$ & Exchanged particle
& Source of bound
\\\hline
  $|\lambda_L^{i3}\lambda_L^{*m2}|$ & $1.4\times 10^{-2}$ &
$V_O/V_{1/2}/V_1$ & $V_{cb}$
\\\hline
  $|\lambda_L^{13}\lambda_L^{*11}|$ & $1.5$ &
$V_O/V_1$ & $B^+\rightarrow e^+\nu_e$
\\\hline
$|\lambda_L^{13}\lambda_R^{*11}|$ & $1.2\times 10^{-4}$ &
$V_O/V_{1/2}$ & $B^+\rightarrow e^+\nu_e$
\\\hline
$|\lambda_L^{23}\lambda_L^{*21}|$ & $8.4\times 10^{-3}$ &
$V_O/V_1$ & $B^+\rightarrow \mu^+\nu_\mu$
\\\hline
$|\lambda_L^{23}\lambda_R^{*21}|$ & $1.4\times 10^{-4}$ &
$V_O/V_{1/2}$ & $B^+\rightarrow \mu^+\nu_\mu$
\\\hline
$|\lambda_L^{33}\lambda_L^{*31}|$ & $2.9\times 10^{-3}$ &
$V_O/V_1$ & $B^+\rightarrow \tau^+\nu_\tau$
\\\hline
$|\lambda_L^{33}\lambda_R^{*31}|$ & $8.4\times 10^{-4}$ &
$V_O/V_{1/2}$ & $B^+\rightarrow \tau^+\nu_\tau$
\\\hline
$|\lambda_L^{13}\lambda_L^{*11}|$ & $7.8\times 10^{-3}$ &
$V_O/V_{1/2}/V_1$ & $\frac{BR(B^+\rightarrow
e\bar{e}\pi^+)}{BR(B^+\rightarrow \nu_e\bar{e}\pi^0)}$
\\\hline
$|\lambda_L^{13}\lambda_L^{*21}|$ & $1.0\times 10^{-2}$ &
$V_O/V_{1/2}/V_1$ & $\frac{BR(B^+\rightarrow
e\bar{\mu}\pi^+)}{BR(B^+\rightarrow \nu_e\bar{e}\pi^0)}$\quad *)
\\\hline $|\lambda_L^{23}\lambda_L^{*21}|$ & $1.2\times
10^{-2}$ & $V_O/V_{1/2}/V_1$ & $\frac{BR(B^+\rightarrow
\mu\bar{\mu}\pi^+)}{BR(B^+\rightarrow \nu_e\bar{e}\pi^0)}$
\\\hline
$|\lambda_L^{33}\lambda_L^{*11}|$ & $6.3\times 10^{-3}$ &
$V_O/V_{1/2}/V_1$ & $BR(B^+\rightarrow \tau\bar{e}X)$\quad *)
\\\hline
$|\lambda_L^{33}\lambda_L^{*12}|$ & $7.2\times 10^{-2}$ &
$V_O/V_{1/2}/V_1$ & $BR(B^+\rightarrow \tau\bar{e}X)$\quad *)
\\\hline
$|\lambda_L^{33}\lambda_L^{*21}|$ & $5.0\times 10^{-2}$ &
$V_O/V_{1/2}/V_1$ & $BR(B^+\rightarrow \tau\bar{\mu}X)$\quad *)
\\\hline
$|\lambda_L^{33}\lambda_L^{*22}|$ & $3.9\times 10^{-4}$ &
$V_O/V_{1/2}/V_1$ & $BR(B^+\rightarrow \tau\bar{\mu}X)$\quad
*)\\\hline
 $|\lambda_L^{33}\lambda_L^{*31}|$ & $2.5\times
10^{-1}$ & $V_O/V_{1/2}/V_1$ & $BR(B^+\rightarrow
\tau\bar{\tau}X)$
\\\hline $|\lambda_L^{33}\lambda_L^{*32}|$ &
$3.1\times 10^{-3}$ & $V_O/V_{1/2}/V_1$ & $BR(B^+\rightarrow
\tau\bar{\tau}X)$
\\\hline
$|\lambda_L^{13}\lambda_L^{*12}|$ & $2.2\times 10^{-5}$ &
$V_O/V_{1/2}/V_1$ & $BR(B^+\rightarrow e\bar{e}K)$
\\\hline
$|\lambda_L^{23}\lambda_L^{*12}|$ & $1.5\times 10^{-3}$ &
$V_O/V_{1/2}/V_1$ & $BR(B^+\rightarrow \mu\bar{e}K)$\quad *)
\\\hline
$|\lambda_L^{23}\lambda_L^{*22}|$ & $1.8\times 10^{-5}$ &
$V_O/V_{1/2}/V_1$ & $BR(B^+\rightarrow \mu\bar{\mu}K)$
\\\hline
$|\lambda_L^{13}\lambda_L^{*11}|$ & $3.1\times 10^{-4}$ &
$V_O/V_{1/2}/V_1$ & $BR(B^+\rightarrow e\bar{e}\pi)$
\\\hline
\end{tabular}\end{center}
\label{tab:vertizes1}Table 6: \textit{Continued.}
\end{table}\newpage
\begin{table} [h]
\begin{center}
\begin{tabular}{||c|c|c|c||}
\hline Product  & Upper bound
/$\left(\frac{m_{LQ}}{100GeV}\right)^2$ & Exchanged particle &
Source of bound\\\hline $|\lambda_L^{23}\lambda_L^{*11}|$ &
$1.3\times 10^{-5}$ & $V_O/V_{1/2}/V_1$ & $BR(B^+\rightarrow
\mu\bar{e}\pi)$\quad *)
\\\hline $|\lambda_L^{23}\lambda_L^{*21}|$ & $4.7\times 10^{-4}$
& $V_O/V_{1/2}/V_1$ & $BR(B^+\rightarrow \mu\bar{\mu}\pi)$\\\hline
$|\lambda_L^{13}\lambda_L^{*11}|$ & $2.5\times 10^{-1}$ &
$V_O/V_{1/2}/V_1$ & $B^0\rightarrow e\bar{e}$\\\hline
 $|\lambda_L^{13}\lambda_L^{*21}|$ & $2.3\times 10^{-3}$ & $V_O/V_{1/2}/V_1$ & $B^0\rightarrow\bar{\mu}e$\quad *)
 \\\hline
 $|\lambda_L^{23}\lambda_L^{*21}|$ & $1.1\times 10^{-3}$ & $V_O/V_{1/2}/V_1$ & $B^0\rightarrow\bar{\mu}\mu$
 \\\hline
 $|\lambda_L^{13}\lambda_L^{*31}|$ & $2.9\times 10^{-3}$ & $V_O/V_{1/2}/V_1$ & $B^0\rightarrow\bar{\tau}e$\quad *)
 \\\hline
 $|\lambda_L^{23}\lambda_L^{*31}|$ & $3.7\times 10^{-3}$ & $V_O/V_{1/2}/V_1$ & $B^0\rightarrow\bar{\tau}\mu$\quad *)
 \\\hline
 $|\lambda_L^{13}\lambda_R^{*11}|$ & $3.0\times 10^{-5}$ & $V_O/V_{1/2}$ & $B^0\rightarrow e\bar{e}$
 \\\hline
 $|\lambda_L^{13}\lambda_R^{*21}|$ & $4.0\times 10^{-5}$ & $V_O/V_{1/2}$ & $B^0\rightarrow\bar{\mu}e$\quad *)
 \\\hline
 $|\lambda_L^{23}\lambda_R^{*21}|$ & $2.5\times 10^{-5}$ & $V_O/V_{1/2}$ & $B^0\rightarrow\mu\bar{\mu}$
 \\\hline
 $|\lambda_L^{13}\lambda_R^{*31}|$ & $8.5\times 10^{-4}$ & $V_O/V_{1/2}$ & $B^0\rightarrow\bar{\tau}e$\quad *)
 \\\hline
 $|\lambda_L^{23}\lambda_R^{*31}|$ & $1.1\times 10^{-3}$ & $V_O/V_{1/2}$ & $B^0\rightarrow\bar{\tau}\mu$\quad *)
 \\\hline
 $|\lambda_L^{13}\lambda_L^{*12}|$ & $1.8$ & $V_O/V_{1/2}/V_1$ & $B^0_s\rightarrow e\bar{e}$
 \\\hline
 $|\lambda_L^{13}\lambda_L^{*22}|$ & $4.1\times 10^{-3}$ & $V_O/V_{1/2}/V_1$ & $B^0_s\rightarrow\bar{\mu}e$\quad *)
 \\\hline
 $|\lambda_L^{23}\lambda_L^{*22}|$ & $1.7\times 10^{-3}$ & $V_O/V_{1/2}/V_1$ & $B^0_s\rightarrow\bar{\mu}\mu$
 \\\hline
 $|\lambda_L^{13}\lambda_R^{*12}|$ & $2.0\times 10^{-4}$ & $V_O/V_{1/2}$ & $B^0_s\rightarrow e\bar{e}$
 \\\hline
 $|\lambda_L^{13}\lambda_R^{*22}|$ & $6.8\times 10^{-5}$ & $V_O/V_{1/2}$ & $B^0_s\rightarrow\bar{\mu}e$\quad *)
 \\\hline
 $|\lambda_L^{23}\lambda_R^{*22}|$ & $3.9\times 10^{-5}$ & $V_O/V_{1/2}$ & $B^0_s\rightarrow\mu\bar{\mu}$
 \\\hline
 $|\lambda_L^{23}\lambda_L^{*21}|$ & $2.5$ & $V_O/V_{1/2}/V_1$ & $B^0-\bar{B}^0$
 \\\hline
 $|\lambda_L^{33}\lambda_L^{*31}|$ & $1.5\times 10^{-1}$ & $V_O/V_{1/2}/V_1$ & $B^0-\bar{B}^0$
 \\\hline
\end{tabular}\end{center}
\label{tab:vertizes1}Table 6: \textit{Continued.}
\end{table}
\newpage
\pagebreak
\begin{table} [h]
\begin{center}
\begin{tabular}{||c|c|c|c||}
\hline Product  & Upper
bound/$\left(\frac{m_{SUSY}}{100GeV}\right)^2$ & Exchanged
particle & Source of bound\\\hline
 $\Re(\lambda'_{11k}\lambda'{}_{11k}^*)$ & $6,.8\times 10^{-4}$ & $\tilde{d}^k_R$ & $R_\pi$\\\hline
 $\Re(\lambda_{1k1}\lambda'{}_{k11}^*)$ & $5.0\times 10^{-7}$ & $\tilde{\ell}^k_L$ & $R_\pi$\\\hline
 $\Re(\lambda'_{21k}\lambda'{}_{21k}^*)$ & $3.5\times 10^{-3}$ & $\tilde{d}^k_R$ & $R_\pi$\\\hline
 $\Re(\lambda_{2k2}\lambda'{}_{k11}^*)$ & $2.0\times 10^{-5}$ & $\tilde{\ell}^k_L$ & $R_\pi$\\\hline
 $|\lambda'_{11k}\lambda'{}_{21k}^*|$ & $5.8\times 10^{-2}$ & $\tilde{d}^k_R$ & $\pi^+\rightarrow\bar{\mu}\nu_e$\\\hline
 $|\lambda_{1k2}\lambda'{}_{k11}^*|$ & $1.7\times 10^{-3}$ & $\tilde{\ell}^k_L$ & $\pi^+\rightarrow\bar{\mu}\nu_e$\\\hline
 $\Re(\lambda'_{11k}\lambda'{}_{12k}^*)$ & $5.5\times 10^{-3}$ & $\tilde{d}^k_R$ & $R_K$\\\hline
 $\Re(\lambda_{1k1}\lambda'{}_{k12}^*)$ & $1.2\times 10^{-6}$ & $\tilde{\ell}^k_L$ & $R_K$\\\hline
 $\Re(\lambda'_{21k}\lambda'{}_{22k}^*)$ & $7.4\times 10^{-3}$ & $\tilde{d}^k_R$ & $R_K$\\\hline
 $\Re(\lambda_{2k2}\lambda'{}_{k12}^*)$ & $1.9\times 10^{-4}$ & $\tilde{\ell}^k_L$ & $R_K$\\\hline
 $|\lambda'_{11k}\lambda'{}_{22k}^*|$ & $1.2\times 10^{-2}$ & $\tilde{d}^k_R$ & $K^+\rightarrow\bar{\mu}\nu_e$\\\hline
 $|\lambda_{1k2}\lambda'{}_{k12}^*|$ & $4.0\times 10^{-4}$ & $\tilde{\ell}^k_L$ & $K^+\rightarrow\bar{\mu}\nu_e$\\\hline
 $|\lambda'_{11k}\lambda'{}_{12k}^*|$ & $3.0\times 10^{-4}$ &
$\tilde{d}^k_R$ &
$\frac{BR(K^+\rightarrow\pi^+e\bar{e})}{BR(K^+\rightarrow\pi^0\bar{e}\nu_e)}$\\\hline
$|\lambda'_{11k}\lambda'{}_{22k}^*|$ & $3.0\times 10^{-6}$ &
$\tilde{d}^k_R$ &
$\frac{BR(K^+\rightarrow\pi^+e\bar{\mu})}{BR(K^+\rightarrow\pi^0\bar{\mu}\nu_\mu)}$\\\hline
$|\lambda'_{21k}\lambda'{}_{12k}^*|$ & $1.3\times 10^{-5}$ &
$\tilde{d}^k_R$ &
$\frac{BR(K^+\rightarrow\pi^+\mu\bar{e})}{BR(K^+\rightarrow\pi^0\bar{\mu}\nu_\mu)}$\\\hline
$|\lambda'_{i1k}\lambda'{}_{m2k}^*|$ & $5.9\times 10^{-6}$ &
$\tilde{d}^k_R$ &
$\frac{BR(K^+\rightarrow\pi^+\nu\bar{\nu})}{BR(K^+\rightarrow\pi^0\bar{e}\nu_e)}$\\\hline
$|\lambda'_{21k}\lambda'{}_{22k}^*|$ & $1.6\times 10^{-4}$ &
$\tilde{d}^k_R$ &
$\frac{BR(K^+\rightarrow\pi^+\mu\bar{\mu})}{BR(K^+\rightarrow\pi^0\bar{\mu}\nu_\mu)}$\\\hline
$|\lambda'_{k12}\lambda_{ikj}^*|$ & $3.8\times 10^{-3}$ &
$\tilde{\ell}^k_L$ & $\frac{f_S}{f_+(0)}$\\\hline
 $|\lambda'_{2k1}\lambda'{}_{1k1}^*|$ & $3.7\times 10^{-5}$ & $\tilde{u}^k_L$ & $K^0_L\rightarrow e\bar{e}$\\\hline
\end{tabular}
\end{center}
Table 7: \textit{Bounds on products of SUSY coupling constants.
Bounds with a *) are also valid under the exchange of the lepton
indices.}\addcontentsline{lot}{table}{7\qquad Schranken an
SUSY-Kopplungskonstantenprodukte} \label{tab:vertizes}
\end{table}\newpage\begin{table} [h]
\begin{center}
\begin{tabular}{||c|c|c|c||}
\hline Product  & Upper
bound/$\left(\frac{m_{SUSY}}{100GeV}\right)^2$ & Exchanged
particle & Source of bound\\\hline
 $|\lambda'_{2k1}\lambda'{}_{1k2}^*|$ & $1.9\times 10^{-7}$ & $\tilde{u}^k_L$ & $K^0_L\rightarrow\bar{\mu}e$\quad *)
 \\\hline
 $|\lambda'_{2k2}\lambda'{}_{1k2}^*|$ & $5.4\times 10^{-6}$ & $\tilde{u}^k_L$ & $K^0_L\rightarrow\bar{\mu}\mu$
 \\\hline
 $|\lambda'_{k12}\lambda_{k11}^*|$ & $9.1\times 10^{-9}$ & $\tilde{\nu}^k_L$ & $K^0_L\rightarrow e\bar{e}$
 \\\hline
 $|\lambda'_{k12}\lambda_{k12}^*|$ & $6.9\times 10^{-9}$ & $\tilde{\nu}^k_L$ & $K^0_L\rightarrow\bar{\mu}e$\quad *)
 \\\hline
 $|\lambda'_{k12}\lambda_{k22}^*|$ & $2.8\times 10^{-7}$ & $\tilde{\nu}^k_L$ & $K^0_L\rightarrow\mu\bar{\mu}$
 \\\hline
  $|\lambda'_{1k2}\lambda'{}_{2k2}^*|$ & $7.5\times 10^{-1}$ & $\tilde{u}^k_L$ & $K^0-\bar{K}^0$
  \\\hline
  $|\lambda'_{1k3}\lambda'{}_{2k3}^*|$ & $4.4\times 10^{-2}$ & $\tilde{u}^k_L$ & $K^0-\bar{K}^0$
  \\\hline
  $|\lambda'_{i1k}\lambda'{}_{m2k}^*|$ & $1.5\times 10^{-1}$ &
$\tilde{d}^k_R$ & $V_{cd}$\\\hline
$|\lambda'_{i2k}\lambda'{}_{m2k}^*|$ & $6.6\times 10^{-1}$ &
$\tilde{d}^k_R$ & $V_{cs}$\\\hline
$\Re(\lambda'_{22k}\lambda'{}_{21k}^*)$ & $2.0\times 10^{-1}$ &
$\tilde{d}^k_R$ & $D^+\rightarrow \mu^+\nu_\mu$\\\hline
$\Re(\lambda_{2k2}\lambda'{}_{k12}^*)$ & $4.3\times 10^{-3}$ &
$\tilde{\ell}^k_L$ & $D^+\rightarrow \mu^+\nu_\mu$\\\hline
$\Re(\lambda'_{22k}\lambda'{}_{22k}^*)$ & $1.9\times 10^{-1}$ &
$\tilde{d}^k_R$ &
$\frac{BR(D^+_s\rightarrow\mu\nu_\mu)}{BR(D^+_s\rightarrow\tau\nu_\tau)}$\\\hline
$\Re(\lambda_{2k2}\lambda'{}_{k22}^*)$ & $3.9\times 10^{-3}$ &
$\tilde{\ell}^k_L$ &
$\frac{BR(D^+_s\rightarrow\mu\nu_\mu)}{BR(D^+_s\rightarrow\tau\nu_\tau)}$\\\hline
$\Re(\lambda'_{32k}\lambda'{}_{32k}^*)$ & $4.0\times 10^{-2}$ &
$\tilde{d}^k_R$ &
$\frac{BR(D^+_s\rightarrow\mu\nu_\mu)}{BR(D^+_s\rightarrow\tau\nu_\tau)}$\\\hline
$\Re(\lambda_{3k3}\lambda'{}_{k22}^*)$ & $6.6\times 10^{-2}$ &
$\tilde{\ell}^k_L$ &
$\frac{BR(D^+_s\rightarrow\mu\nu_\mu)}{BR(D^+_s\rightarrow\tau\nu_\tau)}$\\\hline
$|\lambda'_{12k}\lambda'{}_{11k}^*|$ & $1.1\times 10^{-2}$ &
$\tilde{d}_R^k$ &
$\frac{BR(D^+\rightarrow\pi^+e\bar{e})}{BR(D^0\rightarrow\pi^-\nu_e
\bar{e})}$\\\hline $|\lambda'_{22k}\lambda'{}_{11k}^*|$ &
$9.0\times 10^{-3}$ & $\tilde{d}_R^k$ &
$\frac{BR(D^+\rightarrow\pi^+\mu\bar{e})}{BR(D^0\rightarrow\pi^-\nu_e
\bar{e})}$\quad *)\\\hline$|\lambda'_{22k}\lambda'{}_{21k}^*|$ &
$6.0\times 10^{-3}$ & $\tilde{d}_R^k$ &
$\frac{BR(D^+\rightarrow\pi^+\mu\bar{\mu})}{BR(D^0\rightarrow\pi^-\nu_e
\bar{e})}$\\\hline $|\lambda'_{11k}\lambda'{}_{12k}^*|$ & $3.2$ &
$\tilde{d}_R^k$ & $D^0\rightarrow e\bar{e}$\\\hline
 $|\lambda'_{11k}\lambda'{}_{22k}^*|$ & $2.4\times 10^{-2}$ & $\tilde{d}_R^k$ & $D^0\rightarrow\bar{\mu}e$\quad *)
 \\\hline
\end{tabular}\end{center}
\label{tab:vertizes1}Table 7: \textit{Continued.}
\end{table}\newpage
\newpage\begin{table} [h]
\begin{center}
\begin{tabular}{||c|c|c|c||}
\hline Product  & Upper
bound/$\left(\frac{m_{SUSY}}{100GeV}\right)^2$ & Exchanged
particle & Source of bound\\\hline
 $|\lambda'_{21k}\lambda'{}_{22k}^*|$ & $1.3\times 10^{-2}$ & $\tilde{d}_R^k$ & $D^0\rightarrow\bar{\mu}\mu$
 \\\hline
  $|\lambda_{1k1}\lambda_{k21}'{}^*|$ & $4.6\times 10^{-4}$ & $\tilde{\ell}_L^k$ & $D^0\rightarrow e\bar{e}$
  \\\hline
 $|\lambda_{1k2}\lambda_{k21}'{}^*|$ & $5.3\times 10^{-4}$ & $\tilde{\ell}_L^k$ & $D^0\rightarrow\bar{\mu}e$\quad *)
 \\\hline
 $|\lambda_{2k2}\lambda_{k21}'{}^*|$ & $3.8\times 10^{-4}$ & $\tilde{\ell}_L^k$ & $D^0\rightarrow\mu\bar{\mu}$
 \\\hline
  $|\lambda'_{22k}\lambda'{}_{21k}^*|$ & $2.2$ & $\tilde{d}_R^k$ &
 $D^0-\bar{D}^0$\\\hline
 $|\lambda'_{32k}\lambda'{}_{31k}^*|$ & $1.3\times 10^{-1}$ & $\tilde{d}_R^k$ &
  $D^0-\bar{D}^0$\\\hline
    $|\lambda'_{i3k}\lambda'{}_{m1k}^*|$ & $2.4\times 10^{-3}$ &
$\tilde{d}_R^k$ & $V_{ub}$\\\hline
  $|\lambda'_{i3k}\lambda'{}_{m2k}^*|$ & $1.4\times 10^{-2}$ &
$\tilde{d}_R^k$ & $V_{cb}$\\\hline
$|\lambda'_{13k}\lambda'{}_{11k}^*|$ & $3.0$ & $\tilde{d}_R^k$ &
$B^+\rightarrow e^+\nu_e$\\\hline
$|\lambda'_{k13}\lambda_{1k1}^*|$ & $1.2\times 10^{-4}$ &
$\tilde{\ell}_L^k$ & $B^+\rightarrow e^+\nu_e$
\\\hline
$|\lambda'_{23k}\lambda'{}_{21k}^*|$ & $1.7\times 10^{-2}$ &
$\tilde{d}_R^k$ & $B^+\rightarrow \mu^+\nu_\mu$\\\hline

$|\lambda'_{k13}\lambda_{2k2}^*|$ & $1.4\times 10^{-4}$ &
$\tilde{\ell}_L^k$ & $B^+\rightarrow \mu^+\nu_\mu$\\\hline

$|\lambda'_{33k}\lambda'{}_{31k}^*|$ & $5.8\times 10^{-3}$ &
$\tilde{d}_R^k$ & $B^+\rightarrow \tau^+\nu_\tau$\\\hline

$|\lambda'_{k13}\lambda_{3k3}^*|$ & $8.4\times 10^{-4}$ &
$\tilde{\ell}_L^k$ & $B^+\rightarrow \tau^+\nu_\tau$\\\hline
\end{tabular}\end{center}
\label{tab:vertizes1}Table 7: \textit{Continued. Bounds obtained
from semileptonic $B$-decays and decays of the neutral B-meson are
not listed here. They can easily be obtained directly from the
bounds on LQ coupling constants. LL-couplings have to be
multiplied by a factor two, for LR-couplings SUSY and LQ bounds
are the same. Here I refer once more to the Tables 4 and 5: Some
of the possible combinations are forbidden, because the
corresponding 4-fermion vertices do not exist.}
\end{table}
\newpage
\pagestyle {headings}
\include{appendix}

\end {document}

%% file: instituts_deckblatt.tex
\def\sfb#1{\hbox{\setbox0\hbox{#1}\ignorespaces
\dimen0=\wd0 \dimen1=\ht0 \dimen2=\dp0 \advance\dimen0 2pt
\advance\dimen1 1pt \advance\dimen2 1pt \hbox to 0pt{\vrule depth
\dimen2 height \dimen1 \hss}\ignorespaces \hbox to
0pt{\kern1pt\copy0\hss}\ignorespaces \hbox to 0pt{\lower \dimen2
\vbox to 0pt{\hrule width\dimen0}\hss}\ignorespaces \hbox to
0pt{\raise \dimen1 \vbox to 0pt{\hrule
width\dimen0}\hss}\ignorespaces \kern\dimen0 \vrule depth \dimen2
height \dimen1}\ignorespaces\penalty10000}

%
\begin{titlepage}
  \vspace*{-0.5cm}
  \begin{center}
    \font\GIANT=cmr17 scaled\magstep4
    {\GIANT U\kern0.8mm N\kern0.8mm I\kern0.8mm V\kern0.8mm %
    E\kern0.8mm R\kern0.8mm S\kern0.8mm I\kern0.8mm %
    T\kern0.8mm %
    \setbox0=\hbox{A}\setbox1=\hbox{.}%
    \dimen0=\ht0 \advance \dimen0 by -\ht1%
    \makebox[0mm][l]%
    {\raisebox{\dimen0}{.\kern 0.3\wd0 .}}A\kern0.8mm
    T\kern5mm
    B\kern0.8mm O\kern0.8mm N\kern0.8mm N\kern0.8mm
    } \\[8mm]
    {\GIANT
        P\kern0.8mm h\kern0.8mm y\kern0.8mm s\kern0.8mm i\kern0.8mm
    k\kern0.8mm a\kern0.8mm l\kern0.8mm i\kern0.8mm s\kern0.8mm
    c\kern0.8mm h\kern0.8mm e\kern0.8mm s\kern5mm
    I\kern0.8mm n\kern0.8mm s\kern0.8mm t\kern0.8mm
    i\kern0.8mm t\kern0.8mm u\kern0.8mm t\kern0.8mm
    } \\[2.5cm]
    {\Large \bf
Bounds on Leptoquark and Supersymmetric, $R$-parity violating
Interactions from Meson Decays
    }
\vfil
    {\Large
    von \\[1mm]
Margarete Herz}\\

\vfil { \bf
  \parbox[t]{140mm}{\input{abstract}}
} \vfil
%
%
  \end{center}

  \begin{figure}[b]
    \begin{minipage}{\textwidth}
    \begin{raggedright}
    \begin{tabular}{@{}l@{}}
        Post address:  \\
            Nussallee 12   \\
            53115 Bonn   \\
            Germany      \\
    \end{tabular}
    \end{raggedright}
    \hfill
    \parbox{5.5cm}
    {
        \vbox to 2cm
        {
            \vskip -1.4truecm
        \hbox to 5.5cm
        {
            \includegraphics{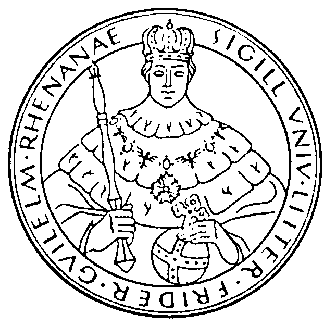}
            \hfill
        }
        \vfill
        \centering
        }
    }
    \hfill
    \begin{raggedleft}
        \begin{tabular}{@{}l@{}}
        BONN-IB-2003-01                   \\      
        Bonn University                 \\
        Januar 2002                   \\     
    \end{tabular}
    \end{raggedleft}
    \end{minipage}
    \vspace*{5mm}
  \end{figure}
\end{titlepage}
%
\begin{titlepage}
\begin{center}
{\Large \bf
\input{Titel1.tex}
}
\\\vspace{20mm}
{\Large
    von         \\
Margarete Herz \\
} \vspace{3cm} {\Large
Diplomarbeit in Physik\\
angefertigt im\\
\vspace{0.5cm}
Physikalischen Institut\\
} \vspace{20mm} {\Large
        vorgelegt der\\
\vspace{0.5cm}
    Mathematisch-Naturwissenschaftlichen Fakult\"at\\
    der\\
    Rheinischen Friedrich-Wilhelms-Universit\"at\\
\vspace{0.5cm}
    Bonn \\

} \vspace{10mm} {\large im November 2002}

\end{center}
\end{titlepage}
%
%
%
\begin{titlepage}
\vspace*{10cm}
\parindent 0pt
\hspace*{\fill} \\[10mm]
\large Ich versichere, dass ich diese Arbeit selbst"andig verfasst
und keine anderen als die angegebenen Quellen und Hilfsmittel
benutzt sowie die Zitate kenntlich gemacht habe.\vspace{50pt}
\begin{tabbing}
xxxxxxxxxxxxxxxxxx \=           \kill
Referent:   \> Prof. Dr. Herbert K. Dreiner \\
Korreferent:    \> Prof. Dr. Hans-Peter Nilles \\
\end{tabbing}
\end{titlepage}
\begin{titlepage}
\hspace{3cm}\\ \vspace{10cm} \\\Huge
\textbf{Dank}\vspace{2cm}\\\normalsize Zunächst möchte ich Prof.
Herbert K. Dreiner für die vielfältige Unterstützung und Hilfe
während meiner Diplomarbeit danken. Prof. Hans-Peter Nilles hat
sich als Korreferent zur Verfügung gestellt. Desweiteren danke ich
den Mitgliedern meiner Gruppe (insbesondere Ulrich Langenfeld,
Christoph Luhn, Marc Thormeier, Akin Wingerter) für ihre
Unterstützung. Meinen Eltern danke ich, dass sie mir das Studium
ermöglicht haben. Nicht zuletzt danke ich Jörg Neuhaus für
Rücksichtnahme, Unterstützung und Hilfe in meiner
Diplomarbeitszeit.
\end{titlepage}

%% file: abstract.tex
We present constraints on products of two leptoquark (LQ) coupling
constants. The bounds are obtained from meson decays, in
particular leptonic $\pi$, $K$, $D$, $D_s$, $B$, $B_s$ decays.
Furthermore semileptonic meson decays and mixing in neutral meson
systems are discussed. We use the Buchmüller-R"uckl-Wyler-model
for scalar and vector LQs. Bounds on $R$-parity violation can be
extracted directly from the corresponding LQ bounds. Our results
are listed in the Tables 6 (for LQs) and 7 (for SUSY particles)
with english captions. The bounds of Davidson/Bailey/Campbell were
updated. The SUSY-bounds of Dreiner/Polesello/Thormeier were
reproduced.

%% file: Titel1.tex
Schranken an die supersymmetrische $R$-Parit"atsverletzung aus
Mesonenzerf"allen durch Leptoquark-Wechselwirkungen